\documentclass[sigconf]{acmart}

\AtBeginDocument{%
  }

\copyrightyear{2025} 
\acmYear{2025} 
\setcopyright{cc}
\setcctype{by}
\acmConference[CHI '25]{CHI Conference on Human Factors in Computing Systems}{April 26-May 1, 2025}{Yokohama, Japan}
\acmBooktitle{CHI Conference on Human Factors in Computing Systems (CHI '25), April 26-May 1, 2025, Yokohama, Japan}\acmDOI{10.1145/3706598.3713275}
\acmISBN{979-8-4007-1394-1/25/04}
\usepackage{makecell}
\usepackage{colortbl}
\usepackage{enumitem}
\usepackage{multirow}
\usepackage{float}
\usepackage{subcaption}
\usepackage{tikz}
\newcommand{\twiceinclude}[2][]{%
  \begin{tikzpicture}%
    \node at (0,0) {\includegraphics[#1]{#2.png}};%
    \node at (0,0) {\includegraphics[#1]{#2.pdf}};%
  \end{tikzpicture}%
}%
\newcommand{\toolname}{\textsc{StoryMate}}

\newcommand\yuling[1]{\textcolor{black}{#1}}
\newcommand\jiaju[1]{\textcolor{black}{#1}}

\begin{document}

\title[Personalizing LLM-Empowered Interactive Story-Reading for Children]{Characterizing LLM-Empowered Personalized Story-Reading and Interaction for Children: \\ Insights from Multi-Stakeholder Perspectives
}

% Characterizing LLM-Empowered Personalized story-reading and Interaction for Children: Insights From Multi-Stakeholders' Perspective

% Towards LLM-Empowered Personalized story-reading and Interaction for Children: Insights From Multi-Stakeholders' Perspective

\author{Jiaju Chen}
\authornote{Work done when visiting Northeastern University.}
\orcid{0009-0005-9598-5838}
\affiliation{%
  \institution{East China Normal University}
  \country{}
}

\author{Minglong Tang}
\affiliation{%
  \institution{East China Normal University}
  \country{}
  }

\author{Yuxuan Lu}
\author{Bingsheng Yao}
\affiliation{%
 \institution{Northeastern University}
 \country{}
}

\author{Elissa Fan}
\affiliation{%
  \institution{Lexington High School}
  \country{}
}

\author{Xiaojuan Ma}
\affiliation{%
  \institution{Hong Kong University of Science and Technology}
  \country{}
}

\author{Ying Xu}
\affiliation{%
  \institution{Harvard University}
  \country{}
}

\author{Dakuo Wang}
\affiliation{%
  \institution{Northeastern University}
  \country{}
}

\author{Yuling Sun}
\authornote{Corresponding author: ylsun@cs.ecnu.edu.cn}
\affiliation{%
  \institution{East China Normal University}
  \country{}
}

\author{Liang He}
\affiliation{%
  \institution{East China Normal University}
  \country{}
}

\renewcommand{\shortauthors}{Jiaju Chen et al.}

\begin{abstract}
  \yuling{Personalized interaction is highly valued by parents in their story-reading activities with children. While AI-empowered story-reading tools have been increasingly used, their abilities to support personalized interaction with children are still limited. Recent advances in large language models (LLMs) show promise in facilitating personalized interactions, but little is known about how to effectively and appropriately use LLMs to enhance children's personalized story-reading experiences. This work explores this question through a design-based study. Drawing on a formative study, we designed and developed \toolname, an LLM-empowered personalized interactive story-reading tool for children, following an empirical study with children, parents, and education experts. Our participants valued the personalized features in \toolname, and also highlighted the need to support personalized content, guiding mechanisms, reading context variations, and interactive interfaces. Based on these findings, we propose a series of design recommendations for better using LLMs to empower children's personalized story reading and interaction.}

\end{abstract}


\begin{CCSXML}
<ccs2012>
   <concept>
       <concept_id>10003120.10003121.10011748</concept_id>
       <concept_desc>Human-centered computing~Empirical studies in HCI</concept_desc>
       <concept_significance>500</concept_significance>
       </concept>
   <concept>
       <concept_id>10003120.10003121.10003129</concept_id>
       <concept_desc>Human-centered computing~Interactive systems and tools</concept_desc>
       <concept_significance>500</concept_significance>
       </concept>
 </ccs2012>
\end{CCSXML}

\ccsdesc[500]{Human-centered computing~Empirical studies in HCI}
\ccsdesc[500]{Human-centered computing~Interactive systems and tools}

\keywords{\yuling{Children, AI, Large Language Model, Story-Reading, Interaction, Personalization, Guided Conversation, Design}}

\begin{teaserfigure}
  \centering
  \twiceinclude[width=0.95\textwidth]{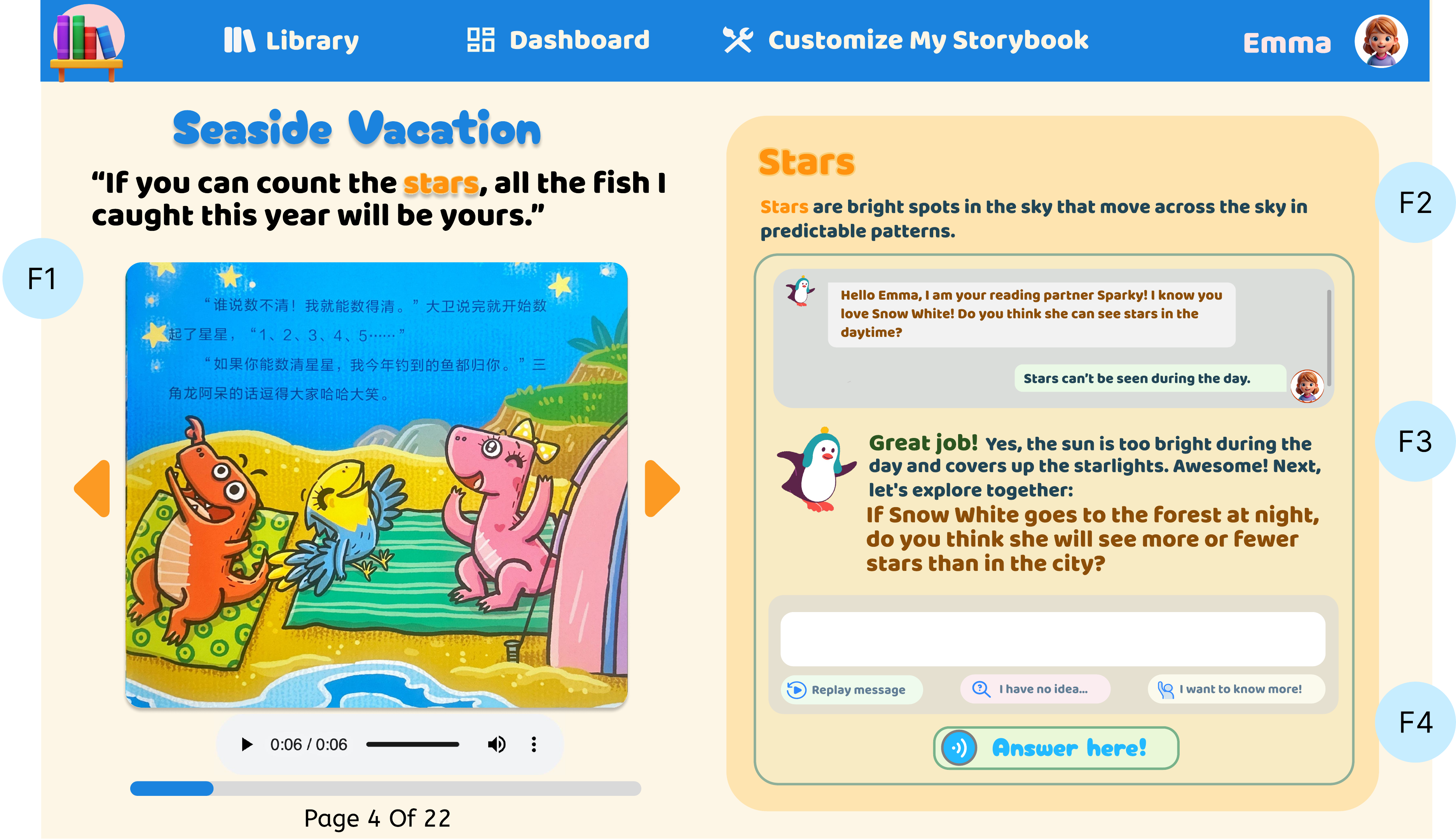}
  \caption{The main interface of \toolname. F1 displays book content with real-time captions and audio narration. F2 shows the concept with its explanation. F3 presents the conversation between the child and the chatbot, extending from the story context to real-world knowledge based on the concept word. F4 provides accessible input options for children.}
  \Description{Main Interface of \toolname}
  \label{fig:teaser}
\end{teaserfigure}


\maketitle

\section{Introduction}

\yuling{
Interactive story-reading is a common parent-child activity in which parents often read a storybook together with children and engage them in \textbf{\textit{personalized, guided conversations and interactions}}~\cite{xuSameBenefitsDifferent2021, sun2024exploring} both within and beyond the story context~\cite{saracho_parents_2017, hu_once_2021, xuRositaReadsMy2023a}. Such personalized, guided conversations and interactions are often shaped by parents' educational philosophy and family environment, and children's personality, cognitive development, and interests \cite{sun2024exploring}.
A large body of pedagogical research has emphasized that these personalized interactions play a beneficial role in, for instance, shaping the way children connect with the stories~\cite{moedt2020effects}, fostering their development of language, communication, and social skills, and enhancing their cultural and emotional awareness~\cite{onuorah_efficacy_2020, kucirkova2017young, kucirkova2021empirical}.
}

\yuling{
Given these benefits, increasing AI-empowered technologies~\cite{fairytaleqa, chen2023fairytalecqa, zhangStoryBuddyHumanAICollaborative2022a, Mathemyths} have been developed and used to support children's interactive story-reading practices. Technologically, these tools ~\cite{zhangStoryBuddyHumanAICollaborative2022a, Elinor} primarily rely on pre-generated questions and intent recognition to drive interactions. Although these approaches have been proven to be effective in generating questions with factually correct answers~\cite{yaoItAITurn2022, yoonDiversityEnhancedNarrative2023}, scholars in Human-Computer Interaction (HCI) and Computer-Supported Collaborative Work (CSCW) have increasingly identified that most existing AI-empowered interactive technologies fail to support personalized conversations and interactions. As \citet{sun2024exploring} insightfully put it, what children practically need are interactions aligning with their personalities, preferences, and situations and often beyond story context to any possible extended content, instead of the ``correct'' answers. }

\yuling{
Recently, Large Language Models (LLMs) have demonstrated more efficient solutions in supporting personalized interactions. Increasing recent HCI attention has been paid to exploring LLM-empowered applications in children-related scenarios, such as vocabulary learning~\cite{osos}, creativity cultivation~\cite{storyprompt}, scientific thinking~\cite{moonstory}, emotional support \cite{chacha}, etc. While these studies have offered valuable insights into using LLMs for child-oriented applications, limited attention has been paid to systematically examining how LLMs can be appropriately used to support children's personalized story-reading activities. In this study, we aim to fill this knowledge gap through a design-based empirical study, addressing the following research questions:
}

\yuling{
\begin{description}
    \item[RQ1] How can LLMs be effectively and appropriately used to support children's personalized story reading and interactions?
    \item[RQ2] How do stakeholders (children, parents, educational experts) experience and perceive LLM-empowered personalized story reading and interaction?
\end{description}
}

\yuling{
 We started with a formative study with parents to understand their experiences and expectations of AI-empowered tools in children's personalized story-reading scenarios. Drawn from the identified design goals, we designed and developed \toolname\ (shown in Figure~\ref{fig:teaser}), an LLM-empowered interactive story-reading tool, which leveraged prompt-engineered, Retrieval-Augmented Generation (RAG)-enhanced LLM (GPT-4) to support child-appropriate, adaptive, and personalized conversation and interaction. Taking \toolname\ as a technology probe, we then conducted an empirical study with 12 children, 14 parents, and 13 experts to explore their experiences and perceptions of LLM-empowered personalized story reading and interaction. 
Our findings suggested that children's personalized requirements for LLM-empowered interactive story-reading were highly complicated and influenced by multiple individual, family, and context factors. We characterized these personalized requirements into four main categories, including conversation content, interactive mechanisms, context, and interactive interfaces.
}

\yuling{
Our study contributes to the HCI community by expanding existing knowledge of LLM-empowered technologies for children. Specifically, based on a formative study, we first proposed \toolname, an LLM-empowered interactive story-reading tool, which involved a prompt-engineered, RAG-enhanced conversation generation model, a child-centered interactive interface, and customizable reading modes to support child-appropriate, adaptive, and personalized interactions. Second, taking \toolname\ as a technology probe, we contribute an empirical understanding of stakeholders' experience and perceptions of \toolname\ as well as LLM-empowered interactive story-reading technologies. Finally, drawing from our findings, we critically discuss the potential opportunities and challenges when using LLMs in children's interactive story-reading scenarios on a large scale, and propose a series of design recommendations for future studies in this area.
}

\section{Related Work}
\yuling{
This section reviews existing work on children's interactive story-reading activities, AI-empowered story reading, as well as personalized interaction for children. These three research lines set up the background for understanding the contributions of our study.}

\subsection{Interactive Story Reading with Children}
\label{rw: reading}

\yuling{
Interactive story reading is a common parent-child activity in which parents often sit together with and read the storybook to and interact with their children. Typically, this process is personalized~\cite{xuSameBenefitsDifferent2021}, that is, parents guide their children in interactive conversations and discussions, according to their educational philosophy as well as children's personality, cognitive development, interests, and so on \cite{sun2024exploring}. 
Existing pedagogical theory~\cite{cobb1999cognitive} has proved that such open-ended, personalized, interactive story-reading activities provide many educational benefits for children, such as enhancing vocabulary acquisition and language development, promoting critical and social thinking, fostering cognitive growth, and so on~\cite{petrieParentchildInteractionStorybook2023, towsonDialogicReadingEarly2017, neuman2016improving}.
}

\yuling{
Given these benefits, researchers have paid significant attention to how to appropriately conduct such kinds of personalized interactive story-reading with children. A series of strategies have been proposed. \citet{whitehurstAcceleratingLanguageDevelopment1988}, for instance, point out that the interactive process and content should be tailored to children's developmental stages. \citet{rubegniGirlWhoWants2021} advise connecting story elements and dialogues to the real world to offer a meaningful context for children. Meanwhile, researchers also foreground parents' significant role in supporting personalized story reading and interactions for children~\cite{kucirkova2017young, kucirkova2021empirical}, and recommended that parents should plan conversation topics and introduce the interactive story-reading activity ahead of reading \cite{WWCPreparingYoung}.
}

\yuling{
Despite these proposed strategies, high-quality personalized story reading and interactions are still challenging for many parents in contemporary society \cite{sun2024exploring, fielding-barnsleyEarlyInterventionHome2003}. On the one hand, many parents lack the necessary knowledge and expertise to guide personalized conversations and interactions, answer children's questions, or provide valuable feedback~\cite{sak2015young, sun2024exploring}. On the other hand, parents in contemporary society often have difficulties fully concentrating on personalized story-reading with children due to the heavy work and housework burden~\cite{sun2024exploring}. These situations are more prominent in families with low socioeconomic status~\cite{buckinghamWhyPoorChildren2023}.}

\yuling{
These challenges have drawn significant attention from researchers in the fields of AI and HCI, and growing interest has been paid to explore technological solutions to address these challenges and support AI-empowered story-reading experiences for children. We now turn to literature in this research area, summarizing and discussing the existing efforts made and the remaining challenges. 
}

\subsection{AI-Empowered Interactive Story Reading for Children}

\yuling{
Existing efforts surrounding AI-empowered interactive story reading for children can be generally divided into two categories. The first category primarily focuses on model and algorithm development, such as constructing story-related datasets \cite{fairytaleqa, chen2023fairytalecqa} and conversation AI models \cite{yaoItAITurn2022, yoonDiversityEnhancedNarrative2023} to support educational-appropriate question generation for children. The main focus of these efforts is to improve the models' technological performance. However, while these AI models have advanced the technical capabilities of AI-empowered story reading and interactions, most have not yet been fully applied to practical applications. It is thus unclear how these technologies work in real-world scenarios. }

\yuling{
The second category, instead, is mainly conducted by scholars in HCI, focusing on the design and evaluation of child-centered interactive technologies and tools. Researchers, for instance, propose StoryBuddy~\cite{zhangStoryBuddyHumanAICollaborative2022a} to support children's story-based reading comprehension, StoryCoder~\cite{storycoder} to foster computational thinking, StoryPrompt~\cite{storyprompt} to enhance literacy and creativity, MatheMyths~\cite{Mathemyths} to promote mathematical language development, and so on. Meanwhile, scholars have also paid attention to deeply evaluating the practical usability and effectiveness of the proposed or existing AI-empowered tools in supporting children's story reading and learning activities \cite{sun2024exploring}. 
}

\yuling{
Generally, existing studies have shown that while AI-empowered tools have introduced innovative technological solutions for children’s interactive story reading, most of the existing tools fall short in supporting \textit{child-centered} and \textit{personalized} interactions. 
Specifically, most existing AI-empowered tools primarily rely on pre-determined, structured conversation patterns~\cite{CAlimit, CAlimit2, zhangStoryBuddyHumanAICollaborative2022a}. While they are effective at offering pre-defined conversations, such approaches often fail to deliver the level of personalized interaction that parents expect \cite{sun2024exploring}.
Meanwhile, many tools~\cite{Elinor, zhangStoryBuddyHumanAICollaborative2022a} mainly focus on conversations within the story narratives, lacking the ability to integrate relevant real-world knowledge—a feature that parents consider practically valuable~\cite{sun2024exploring, kosmas2023storytelling}. 
\citet{sun2024exploring}, through a qualitative study with 17 parents of preschoolers, explored parents' perceptions of various AI-empowered storytelling and reading tools. Their findings demonstrate parents' `dissatisfaction' with these tools, due to their shortage in supporting personalized, adaptive, and child-centered story reading and interactions.
}

\subsection{LLM-Empowered Personalized Interaction for Children}

\yuling{
Compared to traditional AI approaches, the increasingly advanced LLM-empowered technologies have brought revolutionary solutions for personalized interactions, given LLMs' exceptional generation adaptability and versatility \cite{openaiGPT4TechnicalReport2023, Touvron2023Llama2O}. Recent studies have demonstrated LLMs' potential in supporting personalized interaction needs across diverse dimensions (e.g., interaction content~\cite{eapen2023personalization, lyu2023llm}, style~\cite{ha2024clochat}, and more) and scenarios (e.g., education~\cite{hanDesignImplicationsGenerative2023a}, healthcare~\cite{xu2024talk2care}, finance~\cite{easin2024intelligent}, etc.).
}

\yuling{
In child-related scenarios, researchers have explored various LLM-empowered applications, spanning from vocabulary learning~\cite{osos}, creativity cultivation~\cite{storyprompt}, scientific thinking~\cite{moonstory}, to emotional support \cite{chacha}, etc. These studies offer evidence of LLMs' potential and effectiveness in supporting child-oriented, personalized interactions, learning, and development. For instance, \citet{chacha} designed an LLM-empowered chatbot to offer empathetic support for children's emotion sharing, demonstrating LLMs' potential in driving child-oriented emotional interactions. \citet{Mathemyths} designed an LLM-based storytelling agent to foster children's Math language development, demonstrating LLMs' potential as an active learning guide.
Yet, although LLMs' potential has been proved, existing studies haven't deeply examined how LLMs can be appropriately used to support children's personalized interactions. }

\yuling{
Given the specific characteristics of children (e.g., short attention spans, limited self-management abilities, and low AI literacy~\cite{sun2024exploring}), along with the existing challenges of LLMs in generating child-appropriate content~\cite{zhang2023siren, Kasneci2023ChatGPTFG, shen2023chatgpt}, it is urgent to thoroughly examine how to appropriately use LLM to empower children's personalized interactive story reading. Our work contributes to this literature through a design-based study. In particular, we designed and developed \toolname, an LLM-empowered story-reading tool, with features of an LLM-driven chatbot incorporated with RAG and scaffolding techniques to provide child-appropriate, situated, and personalized story reading and interactions. Taking \toolname\ as a technology probe, we explore multiple stakeholders' experiences, perceptions, and expectations of LLM-empowered personalized story reading and interactions for children.
}

\section{Formative Study}
\yuling{
Given children's diverse characteristics and personalized preferences~\cite{sun2024exploring, kucirkova2021empirical}, we first conducted a formative study to learn parents' experiences and perceptions of current AI-empowered story reading tools, and their expectations of LLM-empowered interactive, personalized story-reading tool for children, through which to identify design opportunities that target the existing challenges.
}

\yuling{
We recruited six parents with criteria of 1) with children aged 3-8 and 2) with experience in using AI-empowered story-reading tools, through a snowball sampling approach. Each of them received compensation of 100 CNY for participation. We used semi-structured interviews to collect data, with questions about parents' 1) daily story reading practices, 2) experiences with used AI-empowered story reading tools, including usage scenarios, the benefits and encountered challenges, and 3) perceptions and expectations of LLM-empowered tools for children's personalized story reading. All interviews were audio-recorded with participants' permission and transcribed verbatim into Chinese for analysis. Two authors iteratively read through the data, open-coding and refining the codes based on the thematic analysis method~\cite{ThematicAnalysis1}. 
After several iterations, we identified four Key Insights (KI) of practical parent-child story reading, which aligned with parents' expectations of personalized story reading with children but were not well-supported by existing AI-empowered tools.
}

% \begin{figure*}
%     \centering
%     \includegraphics[width=0.99\linewidth]{imgs/design process.pdf}
%     \caption{Ideation of \toolname}
%     \label{fig:design_process}
%     % \vspace{-16px}
% \end{figure*}

\paragraph{\textbf{KI1:} \yuling{\textbf{Parents and Children Hold Personalized Reading Preferences and Expectations}}}
\label{KI1}

\yuling{
Our formative study indicated that the stakeholders of story reading, including both parents and children, hold different reading preferences and expectations for story-reading activities. For children, our participants told us that children often exhibited diverse and dynamic interests and preferences during story reading. Some children enjoyed asking \textit{``countless whys''} beyond the story itself, while some others preferred to focus more on the storyline. For parents, our participants demonstrated different expectations of their children's story-reading activities. Some hoped to expand children's knowledge by associating elements in the stories with external knowledge (e.g., \textit{``if I read a story about a snake, I want to expand on whether the snake is dangerous, poisonous, etc.''}), while some others preferred their children to think actively by themselves (e.g., \textit{``the story is just a story; what children take away from it depends on themselves.''}). 
}

\paragraph{\textbf{KI2:} \yuling{\textbf{Children Need Personalized Interactions}}}
\label{KI2}
\yuling{
Personalized interaction was highlighted by our participants as a critical feature of children’s story-reading activities. Particularly, such kind of personalized requirements went far beyond the commonly used age- and cognitive-based interaction frameworks. Our participants emphasized that each child held unique preferences for story reading, even if they were in similar contexts or with similar characteristics. Moreover, children's interaction preferences varied from day to day. Yet, most existing AI-empowered story-reading tools couldn't support such kind of personalized interactions very well. One participant described, for instance, the existing tools as \textit{``just working as a story reader''}. Another participant expressed that while the tool they used included the conversation function, the responses were more like \textit{``searching from an online encyclopedia, without considering whether children can adopt and understand them.''}
}

\paragraph{\textbf{KI3:} \yuling{\textbf{Children's Attention is Easily Distracted}}}
\label{KI3}
\yuling{
While AI-based story-reading tools could help alleviate parents' burden by enabling children to read stories with tools, our participants highlighted that it was almost impossible for children to use these tools independently for extended periods. One main reason was \textit{``children’s attention spans are typically short''}, and \textit{``it's hard to sustain children’s attention and motivate them to read a story with it (the tool) for 15-20 minutes steadily.''} Our participants suggested that more child-centered engagement mechanisms should be explored to better maintain children's attention.
}

\begin{figure*}[htbp]
  \centering
  \begin{subfigure}[b]{0.48\textwidth}
    \centering
    \twiceinclude[width=\textwidth,height=2.3in]{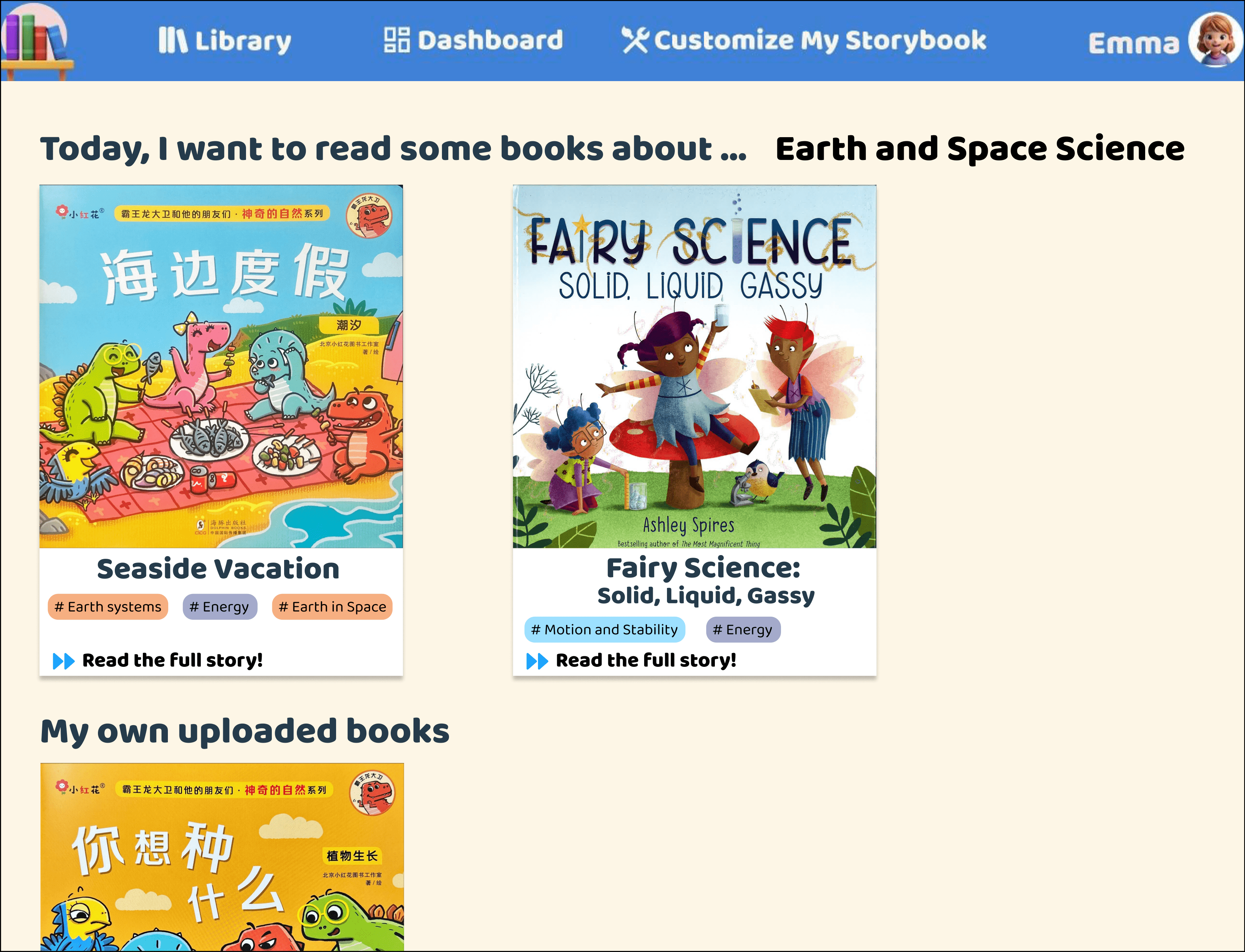}
    \caption{}
    \label{fig:library_interface}
  \end{subfigure}
  \hfill
  \begin{subfigure}[b]{0.48\textwidth}
    \centering
    \twiceinclude[width=\textwidth,height=2.3in]{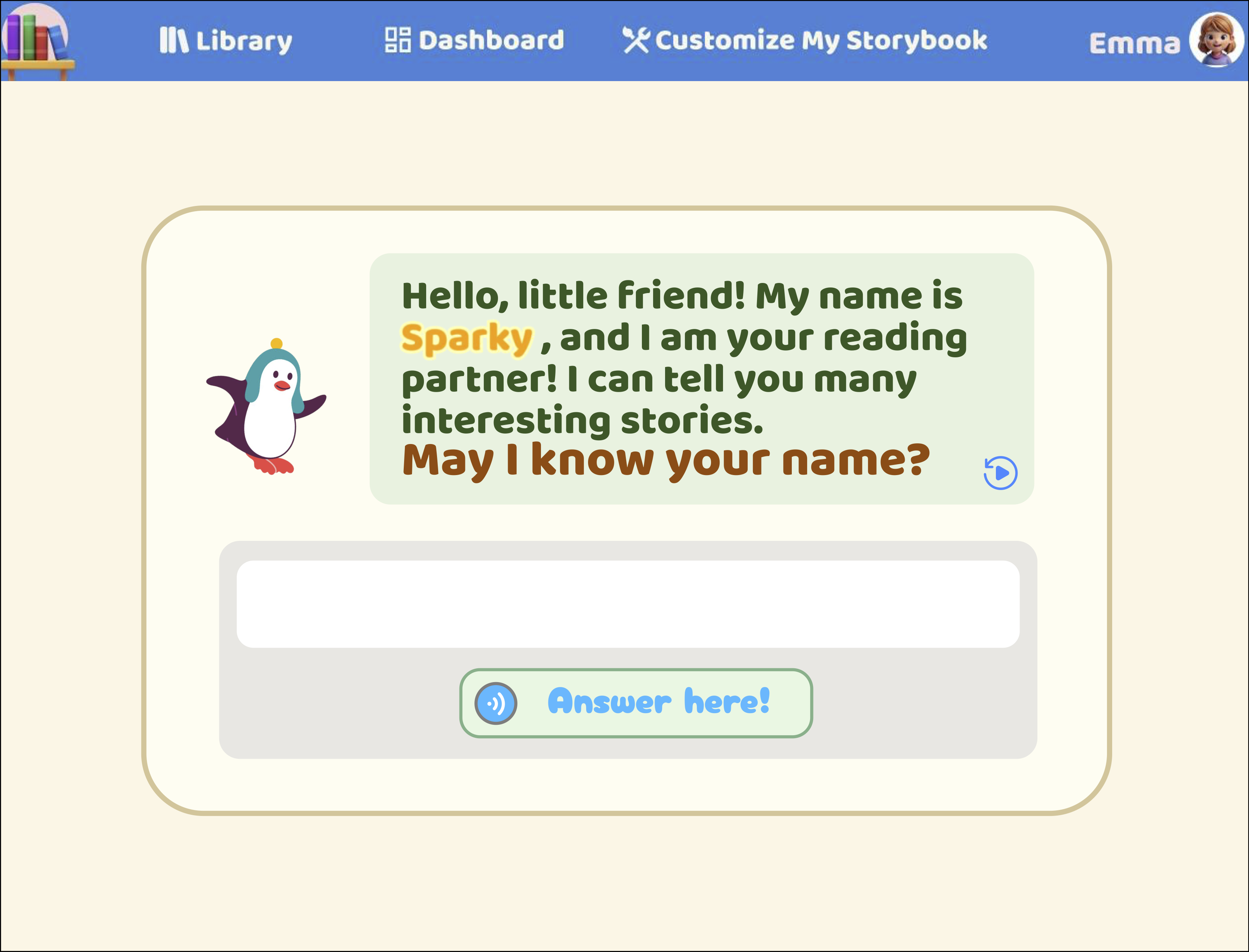}
    \caption{}
    \label{fig:greet_interface}
  \end{subfigure}
\hfill
  \begin{subfigure}[b]{0.48\textwidth}
    \centering
    \twiceinclude[width=\textwidth,height=2.3in]{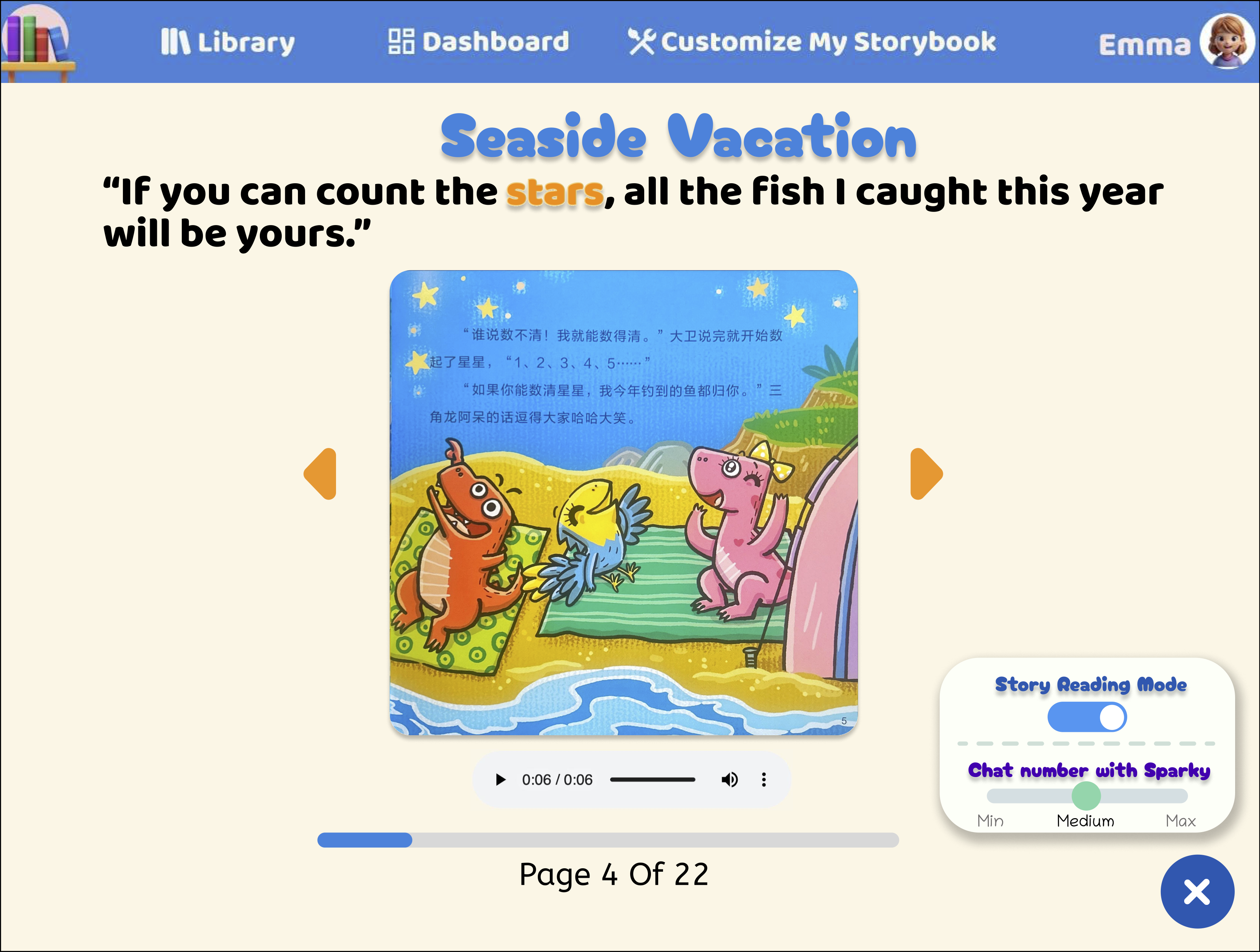}
    \caption{}
    \label{fig:reading_interface}
  \end{subfigure}  
\hfill 
  \begin{subfigure}[b]{0.48\textwidth}
    \centering
    \twiceinclude[width=\textwidth,height=2.3in]{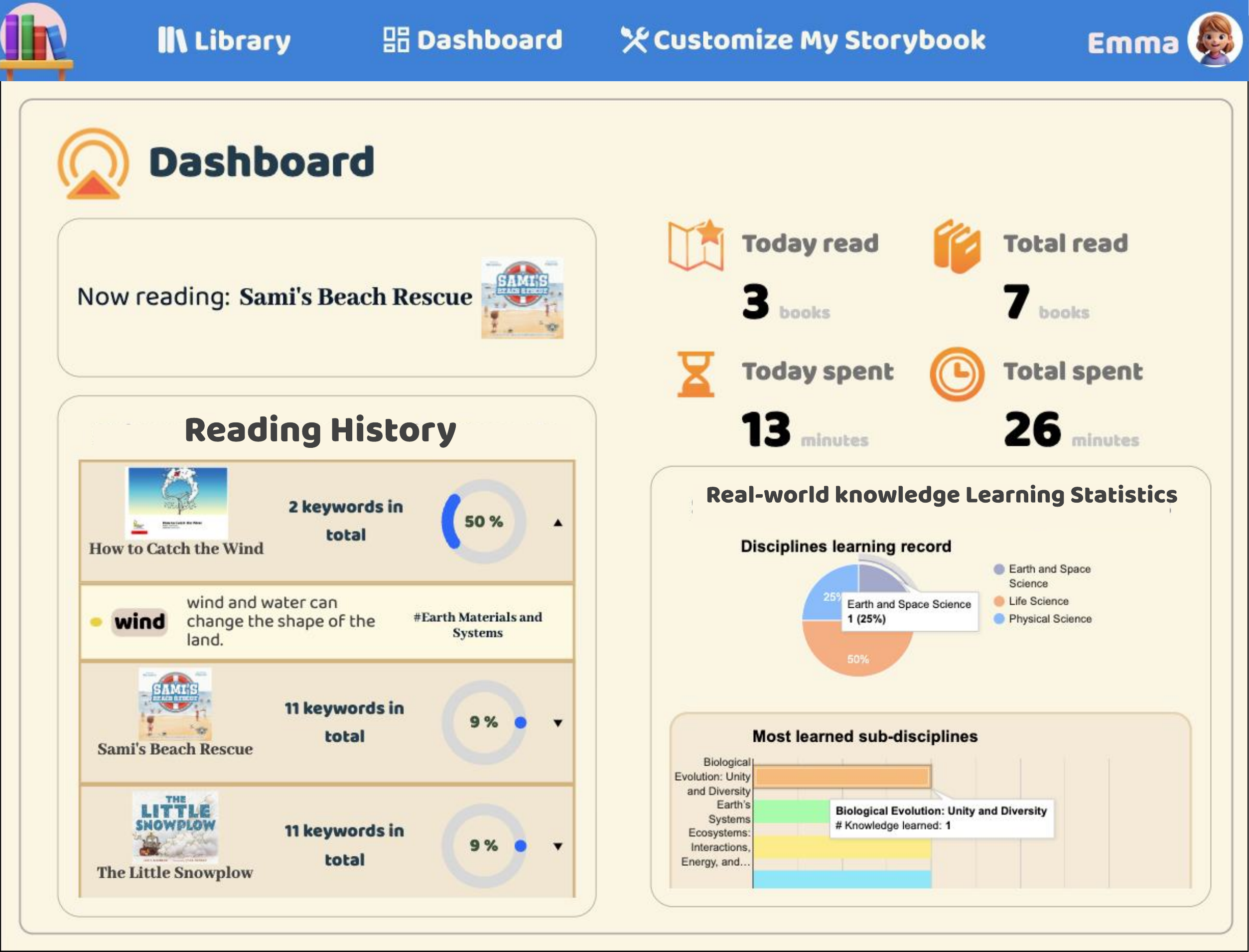}
    \caption{}
    \label{fig:dashboard}
  \end{subfigure}  
  \caption{\toolname's interfaces. \yuling{a) Library interface, which shows customized reading content. b) Greeting interface, wherein a robot guides children's self-introduction. c) Customizable story-reading mode interface, through which children can set read modes. d) Dashboard interface, which records children's reading activities, content, etc.}}
  \label{fig:interface}
\end{figure*}

\paragraph{\textbf{KI4:} \yuling{\textbf{Parents Hope to Encourage Children's Active Thinking}}}
\label{KI4}

\yuling{
Lastly, our participants emphasized the importance of fostering children's active thinking during story-reading. Most of our participants expressed that, when reading with children, they often guided conversations and interactions according to children's interests, hoping to cultivate their active thinking. Yet, most existing tools they had used paid much attention to asking children questions and guiding them toward the correct answers, driven by the underlying education goals. Such kinds of conversation mechanisms limited the opportunities for children's personalized thinking. As one participant pointed out, \textit{``although children's questions often seem strange or silly, these are exactly expressions of their own interests.''}.
}

\section{\toolname}

\yuling{Based on the identified KIs from the formative study, we derived four main Design Goals (\textbf{DGs}) of LLM-empowered personalized story-reading tools for children, including 1) offering personalized reading modes and adjustments to support stakeholders' diverse reading preferences and goals (\textbf{DG1}), 2) supporting child-appropriate, personalized (\textbf{DG2}) and 3) engaging interactions (\textbf{DG3}), and 4) encouraging children to actively think within and beyond the story (\textbf{DG4}).
Based on these design goals, we proposed \toolname, incorporating four Key Features (\textbf{KFs}) to specifically achieve these DGs, including customizable reading modes (\textbf{KF1} for \textbf{DG1}), LLM-empowered personalized interactive chatbot (\textbf{KF2} for \textbf{DG2}), attention-getting mechanisms (\textbf{KF3} for \textbf{DG3}), and RAG-based guided interaction (\textbf{KF4} for \textbf{DG4}). This section will elaborate on the design process and key features of \toolname\ in more detail.}

\subsection{Overview of \toolname}
\label{toolview}
\yuling{
\toolname\ is designed to be a tablet-based tool, employing HTML, JavaScript, and CSS for front-end development, and Flask for back-end development. 
Its main interfaces are shown in Figure~\ref{fig:teaser} and Figure~\ref{fig:interface}. \toolname\ involves three main modules - Library, Dashboard, and Customize My Storybook. Specifically, the Library module (Figure~\ref{fig:library_interface}) primarily supports personalized story reading and interaction through customizable reading modes (Figure~\ref{fig:reading_interface}), story-situated personalized interactions, and external knowledge-enhanced guided interactions (Figure~\ref{fig:teaser}) (key features of \toolname\ for achieving four main DGs. See more details in the following section). The Dashboard module (Figure~\ref{fig:dashboard}) records children's reading process, including reading content, activities, etc. The Customize My Storybook module is designed to support children uploading their own books.
}

\yuling{
As shown in Figure~\ref{fig:teaser} and Figure~\ref{fig:interface}, \toolname\ utilizes graphical user interfaces (GUIs), covering multimodal elements (e.g., graphics, voice, and text) and supporting both language- and text-based interactions. Baidu's text-to-speech model is employed to convert story text and conversations into spoken audio, and GPT-4 is employed to power the chatbot through OpenAI's Chat Completion API. Additionally, we employed various child-friendly components, such as rounded, soft-edged buttons, legible fonts, and a bright color palette, to reduce visual strain and enhance children's readability and accessibility. 
}

\yuling{
To use \toolname\ for interactive story reading, children (or parents) can select a book from the Library or upload a book through ``Customize My Storybook'' (Detailed process is illustrated in the left part of Figure~\ref{fig:interface-upload-dashboard} in Appendix~\ref{app: interface}). If users upload a new storybook, \toolname\ will invoke the embedded OCR technology to detect and extract the text from the photographed pages and generate a digital copy in the Library interface. 
After that, the LLM-empowered chatbot will be activated to greet the children and guide them in introducing themselves (Figure~\ref{fig:greet_interface}), including their nickname, age, interests, favorite storybook or character, etc. 
Meanwhile, the chatbot will guide the children to customize their reading mode (Figure~\ref{fig:reading_interface}), including enabling interactions or not, the frequency of interactions, etc.  
Based on these settings, \toolname\ enters into the interactive story reading phase (Figure~\ref{fig:teaser}), wherein an LLM-powered chatbot is employed to support the child-appropriate, personalized, and engaging interaction with children. Children's reading process, content, and interactions are automatically recorded and displayed on the dashboard (Figure~\ref{fig:dashboard}), helping parents better understand their children's reading behaviors. Figure~\ref{fig:interface-library-read} in Appendix~\ref{app: interface} illustrates the whole interaction process. 
}

\subsection{Key Features of \toolname}
\label{Sec_KeyFeature}

\begin{figure*}
    \centering
    \twiceinclude[width=0.99\linewidth]{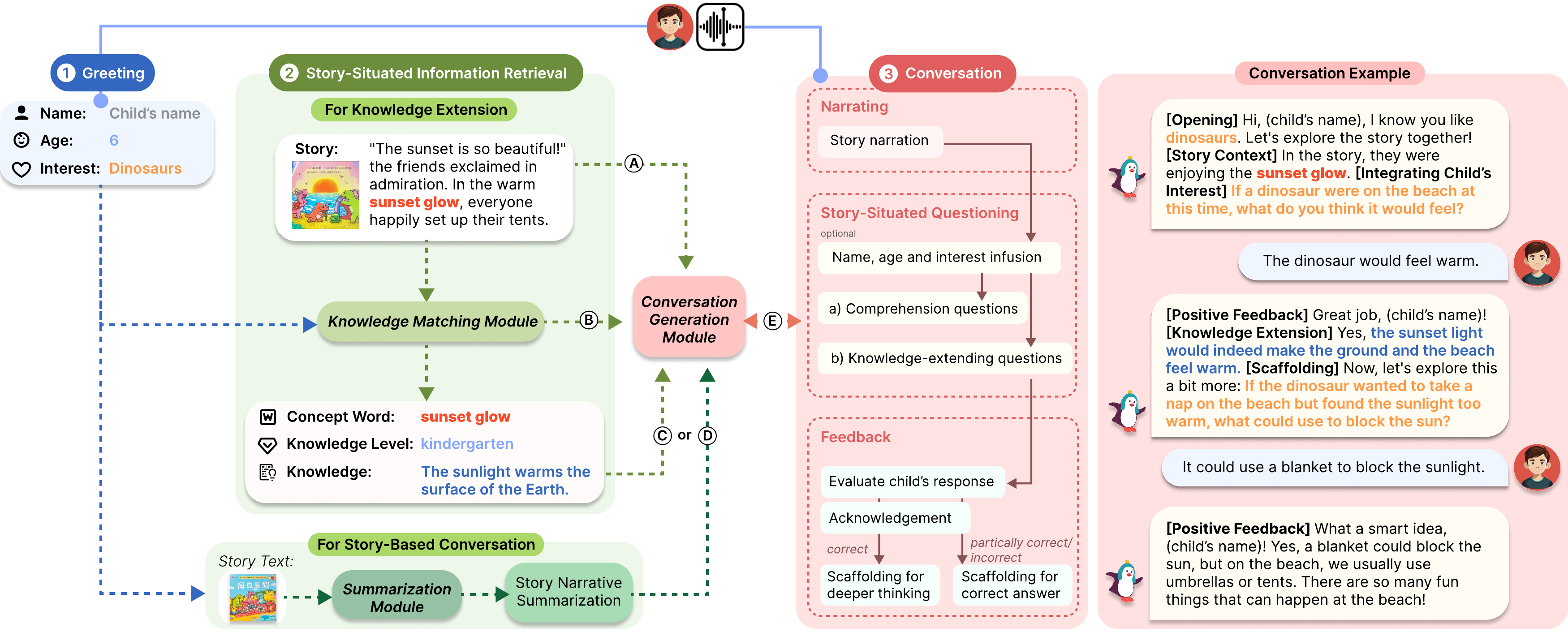}
    \caption{The interaction process of \toolname. Receiving children's information in stage 1, \toolname\ retrieves story-situated information (stage 2). For story-based questions, we use story (A), child's information (B), and summarized story narratives (D) as input. For knowledge-extending questions, we add matched knowledge (C). Then, in stage 3, \toolname\ interacts with children and actively updates conversation status (E). A conversation example is shown at the right end.}
    \label{fig:conversation-eg}
    \vspace{-12px}
\end{figure*}

\yuling{
In \toolname, we specifically designed four key features (\textbf{KFs}), across dimensions of reading and interaction modes, content, and emotional engagement, to achieve four design goals (\textbf{DGs}) respectively. We now elaborated on these four features as well as how they specifically addressed the proposed four design goals. 
}

\paragraph{\textbf{\yuling{KF1: Customizable Read Content and Modes to Support Personalized Reading Preferences and Goals (DG1).}}}
\label{design:readMode}
%上传图书，定制阅读模式

\yuling{
To support children's and parents' personalized reading preferences and goals, we designed customizable reading features for \toolname, allowing children and parents to customize their own library and reading modes. Specifically, \toolname\ allowed users to upload their personal storybooks through the ``Customize My Storybook'' module. F10 and F11 in Figure~\ref{fig:interface-upload-dashboard} (Appendix~\ref{app: interface}) show the detailed process of uploading personal storybooks.
After the storybook was uploaded to the library, users could categorize the storybooks by themes within the Library module according to their preferences. F5 in Figure~\ref{fig:interface-library-read} (Appendix~\ref{app: interface}) illustrates this process.
In addition to customizing the reading content, \toolname\ also provided the function of customizing interaction mode, allowing users to set enabling interactions or not, as well as the interaction frequency. The interface for customizing the interaction mode is shown in F7 and F9 in Figure \ref{fig:interface-library-read} (Appendix~\ref{app: interface}).
Through the customization of reading content and interaction modes, \toolname\ held the potential to meet children's and parents' diverse personalized reading preferences and goals.
}

\paragraph{\textbf{\yuling{KF2: LLM-Empowered Chatbot to Support Child-Appropriate, Adaptive, and Personalized Interaction (DG2).}}}
\label{design:age}
%年龄appropriate，safe，adaptive

\yuling{
To support the personalized interaction requirement, we adopted an interaction mechanism powered by prompt-engineered GPT-4 to guide the interaction style and content to align with the children's general and individual characteristics, making the interactions child-appropriate and personalized. Technically, \toolname\ first obtained children's age and language style from their self-introduction, then integrated this information in prompts to guide GPT-4 in generating conversation with age-appropriate tone and language style. 
Further, to ensure the interaction process was situated and aligned with the children's personal characteristics and preferences, \toolname\ would analyze children's interactions during the real-time interactions, which often reflected their unique thinking, interests, and preferences. Based on the obtained information such as children's engagement level, interactive content, and knowledge accuracy, the chatbot guided the subsequent conversations, ensuring the interaction was adaptive and personalized. Figure \ref{fig:conversation-eg} shows the detailed interaction process, and Appendix \ref{app:prompt} shows the prompts we used. 
Table \ref{tab:chatbot_water_dialogue} and \ref{tab:chatbot_sunlight_dialogue} in Appendix~\ref{example} illustrated two examples of children's conversations with \toolname, demonstrating its potential in guiding personalized conversations.
}

\paragraph{\textbf{\yuling{KF3: Embedding Attention-Getting Mechanisms into Interaction Process to Better Engage Children (DG3).}}}
%peer like interface，Story-situated，name

\yuling{
To make the interaction process more engaging for children, \toolname\ adopted a series of attention-getting mechanisms proposed in the existing literature to enhance relationship-building and emotional bonding with children. Specifically, existing theory \cite{howe2012attachment} suggested that addressing children by name could strengthen the emotional connection between children and interaction partners, making children feel valued and attended to, thus motivating increased engagement. Based on this insight, we extracted the children's names from their self-introductions and used that as the form of address throughout the conversation, fostering a more welcoming and emotionally engaging interaction. 
In the interaction content, we further integrated storylines or characters that children liked into conversations, which could further enhance children's emotional involvement and engagement \cite{Mathemyths}. 
Meanwhile, we designed positive emotional feedback throughout the conversation. For example, when the chatbot received a child's response, it would first acknowledge the child's input and provide encouragement and affirmation, which had been proved as an effective means to promote continued engagement and participation \cite{hardy2020using}. 
Additionally, \toolname's interaction style was designed to resemble a peer-like interaction, combining multiple modalities such as text, images, and sounds. This multi-sensory approach potentially stimulated children's attention and enhanced their multi-sensory engagement \cite{druin1996designing}. 
 The integration of these attention-getting mechanisms allowed \toolname\ to foster engaging and emotionally resonant interactions. Table~\ref{tab:greet_dialogue_1}, \ref{tab:greet_dialogue_2}, \ref{tab:chatbot_water_dialogue} and \ref{tab:chatbot_sunlight_dialogue} in Appendix~\ref{example} illustrate three examples of engaging conversations with \toolname.
}

\paragraph{\textbf{\yuling{KF4: RAG-Based Guided Conversation to Support Children's Active Thinking (DG4).}}}
\label{RAG}

\yuling{
For supporting children's active thinking during story reading, \toolname\ employed a RAG-based guided conversation generation algorithm, which integrated external knowledge that was aligned with children's cognitive characteristics into the conversation generation process, allowing the conversation content to extend within and beyond the story itself. Specifically, to ensure the associated knowledge was age-appropriate and aligned with the children's educational level, we constructed a specialized knowledge graph based on Next Generation Science Standards (NGSS) \cite{national2013next}, a comprehensive, interdisciplinary knowledge base tailored for K12. Drawing on this knowledge graph, we then leveraged a fine-tuned retriever model~\cite{bge_embedding} to match the most relevant knowledge content with the story and integrated the knowledge into the GPT-4-empowered conversation generation process. 
Meanwhile, we incorporated scaffolding techniques into the conversation generation algorithm, guiding GPT-4 to pose a series of tailored questions and provide hints that encourage step-by-step exploration of the story or related knowledge. The technical details of the algorithm are elaborated in Appendix~\ref{app: technical process}. We believe such kinds of external knowledge infusion and guided interaction could potentially foster children's active thinking.
}

% \begin{figure*}
%     \centering
%     \includegraphics[width=0.95\linewidth]{imgs/user journey.pdf}
%     \caption{An example use scenario of \toolname.}
%     \label{fig:user-journey}
%     \vspace{-12px}
% \end{figure*}

\section{User Study}
\begin{table*}[!htbp]
    \centering
    \resizebox{0.99\linewidth}{!}{
    \begin{tabular}{lllll|ll}
    \toprule[1.5pt]
    \rowcolor[HTML]{C0C0C0} 
       Parent ID & Role & Child ID & Child & Reading Habits & Expert ID & Occupation \\
    \midrule
       P1 & Mother & PC1 & 6-year-old girl & Reads with parents & E1 & Primary school Math teacher \\
   \arrayrulecolor[HTML]{778899}     \midrule
       P2 & Mother & PC2 & 8-year-old girl & Reads alone & E2 & Primary school Chinese \\
   \arrayrulecolor[HTML]{778899}     \midrule
       P3 & Mother & \makecell[l]{PC3.a\\PC3.b} & \makecell[l]{6-year-old girl \\ 12-year-old boy} & \makecell[l]{Reads with parents, sometimes asks questions \\ Reads with parents, actively asks questions} & E3 & Primary school Chinese \\
   \arrayrulecolor[HTML]{778899}     \midrule
       P4 & Mother & PC4 & 5-year-old boy & Reads alone & E4 & Primary school Chinese\\
   \arrayrulecolor[HTML]{778899}     \midrule
       P5 & Mother & PC5 & 6-year-old girl & Reads with parents, sometimes asks questions & E5 & Primary school Math\\
   \arrayrulecolor[HTML]{778899}     \midrule
       P6 & Mother & \makecell[l]{-\\-} & \makecell[l]{5-year-old boy \\ 11-year-old girl} & \makecell[l]{Reads with parents, actively asks questions \\ Reads with parents} & E6 & Education expert\\
   \arrayrulecolor[HTML]{778899}     \midrule
       P7 & Father & PC7 & 7-year-old boy & Reads with parents, actively asks questions & E7 & Education expert \\
   \arrayrulecolor[HTML]{778899}     \midrule
       P8 & Mother & PC8 & 6-year-old boy & Reads with parents & E8 & Education expert\\
   \arrayrulecolor[HTML]{778899}     \midrule
       P9 & Mother & PC9 & 7-year-old boy & Reads alone & E9 & Education expert\\
   \arrayrulecolor[HTML]{778899}     \midrule
       P10 & Father & PC10 & 8-year-old girl & Reads with parents, sometimes asks questions & Focus Group & Four primary school teachers \\
   \arrayrulecolor[HTML]{778899}     \midrule
       P11 & Father & \makecell[l]{PC11.a\\PC11.b} & \makecell[l]{4-year-old boy \\ 6-year-old girl} & \makecell[l]{Reads alone with AI-driven storytelling tools \\ Reads alone with AI-driven storytelling tools} \\
   \arrayrulecolor[HTML]{778899}     \midrule
       P12 & Father & - & 3-year-old girl & - \\
   \arrayrulecolor[HTML]{778899}     \midrule
       P13 & Mother & - & 6-year-old boy & Reads alone \\
   \arrayrulecolor[HTML]{778899}     \midrule
       P14 & Mother & - & 4-year-old girl & - \\
    \bottomrule[1.5pt]
    \arrayrulecolor{black}   
    \end{tabular}
    }
    \caption{Participants’ demographic information. P1-P14 are parents, PC1-PC11 are children, and E1-E9 are education experts.}
    \label{tab:parent_demograph}
    \vspace{-16px}
\end{table*}
\yuling{
Taking \toolname\ as a technology probe, we then conducted a qualitative study with 12 children, 14 parents, and 13 experts to explore their experiences and perceptions of \toolname\ and LLM-empowered personalized story reading and interaction. Three authors, all native Mandarin speakers, collected and analyzed data together. This section presents the study methods in detail.
}

\subsection{Participants and Recruitment}

\yuling{
After approval by the institutional review board of the first author's affiliated university, we looked for children aged 6-8 and their parents, as well as education experts—key stakeholders of children's interactive story reading—to experience and evaluate \toolname. The participants were recruited through two authors' social networks.  
We created a poster including an introduction to the research team, study objectives, process, recruitment criteria, data privacy and safety commitments, and contact information. Two authors shared this poster within their social network. Finally, 13 experts and 14 parents contacted us and agreed to participate in the study. Of the 14 parents, 10 brought their children (N=12) to experience \toolname. Of the 13 experts, 9 were educational practitioners (i.e., elementary school teachers), and 4 were researchers who had been extensively involved in early childhood education research.
Their basic information is illustrated in Table~\ref{tab:parent_demograph}. }

\subsection{Data Collection}

\yuling{
The data collection process consisted of three stages: pre-interviews, \toolname\ experience, and post-interviews based on \toolname, with a 100 CNY honorarium provided for participants. We began by introducing our background, research motivation, and procedures to participants. After obtaining informed consent, we conducted the pre-interviews, which mainly focused on participants' daily story-reading scenarios, AI tools they had used, and experiences with such tools. This phase lasted approximately 10-15 minutes.
}

\yuling{
Next, we introduced \toolname\ to participants, including its design motivation, process, and main functions. We particularly introduced the four KFs for personalized interaction. Following our introduction, we invited participants to experience each module. This process lasted about 15 to 20 minutes. For children, with their parent's consent, we observed and recorded their interactions with \toolname, including usage and interaction process, interaction content, engagement level, etc.}

\yuling{
Following the experience, we collected participants' in-depth experiences and perceptions of \toolname\ and broader LLM-empowered story reading and interaction tools for children through interviews and focus groups (Four educational experts participated in the focus group, and the remaining participants were interviewed in the semi-structured format). The interview questions mainly included: 1) your experience with and perceptions of \toolname, the benefits, and challenges you perceived; 2) the potential use scenarios of \toolname; and 3) your perception and expectation to increasing LLM-empowered story reading tools for children. This process lasted 20-30 minutes. All questions were designed to be general to ensure inclusivity and encourage participants to share their varied experiences and perceptions. If participants shared interesting points or prior experiences, we asked for follow-up details and concrete examples.
}

\yuling{
With participants' permission, all interviews were audio-recorded and transcribed verbatim in Chinese for analysis. For children, given their early developmental stages and limited cognitive abilities to be deeply interviewed, we developed a questionnaire on a 7-point Likert scale, with items from Giggle Gauge~\cite{gigglegauge}, to assess their experience. However, we quickly observed that many children struggled to quantify their feelings in this way. We, therefore, shifted to asking simpler, more general questions such as, ``Do you like this tool?'' and ``Would you like to have a partner like this?'' to obtain their experiences (the questionnaire is shown in Appendix~\ref{app:question}). This approach aligned with commonly used methods in existing research involving children~\cite{whatareyoutalkingto, xu2019young, Elinor}.
}

\subsection{Data Analysis}

\yuling{
We applied deductive thematic analysis~\cite{ThematicAnalysis1} for data analysis. We started coding and analyzing while the data was being collected. During the open coding phase, two authors independently reviewed the data, generating codes relevant to our research questions. We iterated this process by generating codes from the collected data, classifying similar codes into themes, regularly discussing with each other to ensure reliability, and refining these codes and themes as new data were obtained. By the end of this phase, we generated an initial code list, including participants' existing story reading practices; children's usage and experiences of \toolname; parents' and experts' perceived benefits in terms of customizable, personalized, intelligent, flexible, knowledge-extended interaction, etc.; encountered obstacles in term of interface design, lack of multimedia interaction, lack of connection with the existing education system, language complexity and difficulty, etc.; as well as their attitudes, perceptions, suggestions, and expectations to future LLM-empowered story reading tools for children.
}

\yuling{
Based on this initial code list, we re-focused our analysis at the broader level of themes, using rounds of discussions to organize and synthesize different codes into potential overarching themes based on our research questions. After several rounds of discussions, we defined the final overarching themes around participants' experiences and perceptions of LLM-empowered \toolname\ in supporting children’s personalized story reading and interactions, including their using experiences to \toolname, and four personalized requirements in terms of interaction content, mechanism, context, and interface. In the following section, we present these themes in detail, using representative quotes from participants translated from Chinese into English to illustrate our findings.
}

\section{Results}
\label{finding}

\yuling{
Our findings were organized to respectively answer two research questions. In Section~\ref{finding: experience}, we demonstrated our participants' experiences with \toolname\ to answer RQ1-how LLMs can be appropriately used to support children’s personalized story reading and interactions. In Section~\ref{finding: requirements}, we elaborated on our participants' personalized requirements to LLM-empowered \toolname, answering RQ2-how stakeholders (children, parents, educational experts) experience and perceive LLM-empowered personalized story reading and interactions. 
}

\subsection{Stakeholders' Experiences to \toolname}
\label{finding: experience}

\yuling{
Overall, participants in our study highly valued \toolname's functions in supporting children's interactive story reading, considering \toolname's conversations with children to be \textit{``interesting''} (P11, E1, E7, E8, focus group), \textit{``intelligent''} (P7, P8, P13, P14), and \textit{``meaningful''} (P3, E1, E4). For children in our study, most of them (N=9/12) engaged smoothly and actively with \toolname, although some (PC1, PC2, PC5) needed parental assistance and guidance. 11 children expressed that they liked \toolname\ and wanted it as a reading companion. For parents, they emphasized \toolname's potential in \textit{``accompanying children''}(P7), \textit{``helping parents pose better questions''}(P5), \textit{``facilitating interest-based personalized and engaging conversations''} (P3, P6, P9, P10). The education experts participating in our study particularly highlighted \toolname’s ability in \textit{``encouraging children's interest in learning''} (E1, E6, E8, E9). At the same time, our participants also reported the challenges in generating conversations that can seamlessly suit every child's reading literacy, cognitive level, and reading preferences. In this section, we elaborate on participants' experiences to \toolname, centered around the four KFs.
}

\subsubsection{\textbf{Customizable Reading Content and Modes (KF1)}}
\label{finding:dg1}

\yuling{
For the customizable reading mode feature (\textbf{KF1}), most of our participants considered it a valuable design in supporting children's diverse reading and interaction preferences. In the practical experience, participants could effectively use this feature to customize their preferred reading modes. For instance, some participants (2 parents, 1 expert, 1 child) preferred immersive reading experiences, prioritizing \textit{``the integrity of the story''} (P9, P13, E1). Others (2 parents, 6 children) favored extending reading to possible external contents through discussions, such as teaching \textit{``moral lessons''} (P8) or addressing \textit{``questions when she (PC5) didn’t understand something''} (P5). E1, a math teacher in primary school, commented on \toolname's customizable reading modes as:
}

\begin{quote}
    \textit{``\yuling{When it comes to picture books, we emphasize integrating the story into the learning process. Since your tool allows customization, students could choose to read the book in its entirety or stop on a specific page for in-depth learning. I think this setup is appropriate for them to adjust reading mode according to their needs.''}}
\end{quote}

\yuling{
Meanwhile, our participants also expressed positive feedback to the ``Customize My Storybook'' feature, considering it allowed children or parents to choose books they enjoyed. P6, a mother of two children, further mentioned that personalized book uploading could align their children's reading activities with school requirements, supporting children to read \textit{``books recommended by the school.''}
In addition to the positive feedback, our participants also suggested that the ``Customize My Storybook'' feature in the Library could benefit from supporting more granular tags, such as book themes, culture, and subjects, then providing children with more diverse choices and helping them explore a wider range of content tailored to their interests and learning needs.
}

\subsubsection{\textbf{LLM-Empowered Chatbot for Personalized and Engaging Interaction (KF2, KF3)}}
\label{finding:dg2}

\yuling{
While many participants reported that existing technologies they used for story-related activities were \textit{``silly and not interactive''} (P8), they considered \toolname's personalized conversations functioned well in supporting children's personalized interaction during story-reading. Educational expert E4 highlighted that \toolname's interactions had the potential to be \textit{``a reading companion''}, and answer children’s `countless whys'. P2, P3, and E3 also thought \toolname' story-situated, adaptive conversations were more engaging for children. As P2 said, \textit{``this approach (situated the interaction into the story) is suitable for her (her daughter) to adopt and engage''}. Further, participants (N=6) highly valued \toolname's use of a peer-like chatbot that engaged in dialogue based on children's names and interests. They considered this an effective way to capture children's attention, while the encouraging and affirming conversation content made the children feel happy and engaged. P10, a father of a 10-year-old girl who loves princesses, commented: \textit{``This book isn't actually about a princess; it's about the seaside. However, the questions were somehow connected to the princess. That's what makes it interesting.''} 
}

\yuling{
Although participants provided positive feedback on \toolname's integration of story contexts and children's personalized interest, they also pointed out challenges regarding the difficulty level and linguistic complexity of the conversation content. Some participants considered that \toolname's conversation difficulty did not always align with their children's age and cognitive level, noting that the chatbot's questions `\textit{`seem to be asked from an adult’s perspective}'' (P8), and sometimes included ``\textit{advanced vocabulary}'' (P1, E2, E7), which made them ``\textit{challenging for children to understand}'' (P1, P8, P12). 
}

\subsubsection{\textbf{RAG-Based Guided Conversation (KF4)}}
\label{finding:dg3}

\yuling{
As for \toolname's RAG-based, external knowledge-enhanced guided conversation, many participants (N=11) considered this function could significantly guide children's active thinking within and beyond the stories. As E6, an educational expert, commented, \textit{``the interactive features play a significant role in stimulating a child's interest, encouraging them to participate actively, which I believe is very important for quality story reading.''} 
Some participants considered the generated scaffolding conversations of \toolname\ even better than school teachers. P10, a father, experienced the guided conversation of \toolname\ and said, \textit{``Many picture book teachers struggle to ask such questions and don't engage in meaningful communication with the children. They tend to just deliver knowledge''}. On the other hand, some parents also felt that \toolname's LLM-guided conversation might not align with every child's preferred way of learning style, as \textit{``some children might not enjoy such (guided conversation) way of receiving knowledge''} (P1, P3).
}

\yuling{
In summary, our participants generally provided relatively positive feedback on \toolname's four KFs in supporting personalized story reading and interaction. At the same time, they also highlighted challenges related to the alignment of the conversation with children's cognitive abilities, multimedia interaction, the relevance of the content to their educational environment, and so on. In particular, our findings suggested that their practical expectations for LLM-empowered personalized reading and interaction exceeded the four key features we designed. We now turn to discuss these personalized requirements in more detail in the sections below.
}

\subsection{\yuling{Personalized Requirements to LLM-Empowered Interactive Story Reading for Children}}
\label{finding: requirements}

\yuling{
We characterized the personalized requirements expressed by our participants for LLM-empowered interactive story reading for children into four main categories, including interactive content, mechanisms, context, and interfaces. 
}

\subsubsection{\textbf{Personalized Requirements in Conversation Content}}
\label{finding: content}

\yuling{
The personalized requirement regarding conversation content was the first major requirement from our participants, which was specifically reflected in the conversation topics, the language complexity, and the incorporation degree of extended knowledge.}

\paragraph{\yuling{Conversation Topics}}
\yuling{
Throughout our study, the conversation topic between children and \toolname\ displayed obvious personalized differences—children showed varying interests even when engaging with the same story and interaction. For example, some children were more interested in the plot and narrative details, while others focused on different aspects, such as the color of the pictures, the character's gender, and other story elements.
Table~\ref{tab:eg_content} showed two distinct pieces of feedback from two children, reflecting completely different perspectives. 
}

\begin{table}[h]
    \centering
    \vspace{-8px}
    \begin{tabular}{c|p{6cm}}
    \toprule
       \parbox{1.5cm}{\centering\includegraphics[width=0.8cm]{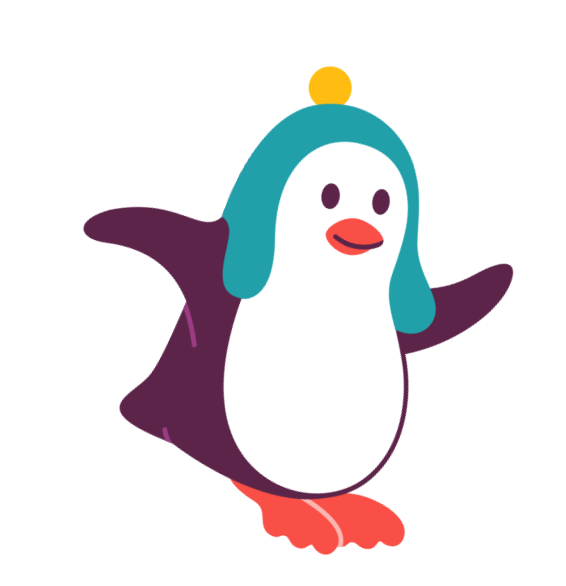}} & \textit{\jiaju{... On this beach where everyone is setting up tents, if there were a small animal you like, what effect would the sunset light have on its body?}}\\
    \arrayrulecolor[HTML]{778899}
    \midrule
    \arrayrulecolor{black}
       PC2 & \textit{\jiaju{Turn to other colors, like red.}} \\
    \arrayrulecolor[HTML]{778899}
    \midrule
    \arrayrulecolor{black}
        PC4 & \textit{The dinosaurs would probably feel warm and cozy.} \\
    \bottomrule
    \end{tabular}
    \caption{The conversation between \toolname\ and PC2, PC4.}
    \label{tab:eg_content}
    \vspace{-12px}
\end{table}

\paragraph{\yuling{The Complexity of Used Language}}
\label{finding: question difficulty}
\yuling{
The personalized requirement in language complexity was mainly reflected in the children's ability to comprehend the conversation content of \toolname. Even within the same age group, children's understanding of the same questions varied significantly. In our study, most children (N=9/12) were able to interact smoothly with the chatbot and answer fluently, and many parents and experts (N=6) agreed that the conversation content was appropriate for children's developmental stages. However, the other children (N=3/12) felt challenging to understand. Their parents explained that the guided interaction content, such as, ``designing a structure to block out the sun,'' and inquiry methods like ``What do you think of XX?'' was a little difficult for children to answer, requiring further explanation from parents. They, as well as two experts, suggested adopting a more conversational style of interaction. 
}

\paragraph{\yuling{The Incorporation of Extended Knowledge}}
\yuling{The incorporation of extended knowledge was primarily reflected in the differentiated requirement of participants regarding whether and how to integrate external knowledge into story-based interactions. During our study, while many children (N=10/12) enjoyed conversations that integrated real-world knowledge, some, like PC1, felt it overwhelming, commenting on the using process as ``\textit{taking 100 Math questions}''. 
The parents and experts in our study also shared varying opinions on whether and how to extend knowledge during children's story reading. Some participants (e.g., E6, P12) believed that rather than introducing additional knowledge, young children's reading should focus solely on the story itself. In contrast, many parents (P2, P4, P5, P7, P8, P10) and experts (E1, E5, E8, E9, focus group) were enthusiastic about and supportive of extending real-world knowledge. Particularly, compared with existing storytelling tools, which were commented by them as \textit{``stupid and only knows how to read the words in the book''} (P8), they appreciated how \toolname\ integrated real-world knowledge into story contexts. As P4 noted,}

\begin{quote}
    \textit{``\yuling{The knowledge in books is limited. Through this tool, a lot of knowledge can be extended. For instance, he (PC4) has recently been learning about stars and the galaxy. This might only be introduced in a one-sided way in books, but this tool can provide more information. He can also actively ask the chatbot questions.}''} 
\end{quote}

\yuling{
Some parents also suggested that not only knowledge but also real-life events should be integrated into the interactions with children, which would make the conversation more engaging. For example, P6 suggested that, if a typhoon occurred near the child's home, the chatbot could embed such an event in conversations and \textit{``teach children how to securely seal windows with tapes.''} These personalized requirements in interaction content highlighted a need for supporting adaptive interaction with children, and balancing the depth of external knowledge with the child's cognitive and emotional engagement, to ensure the interaction remains enjoyable and accessible.
}

\subsubsection{\textbf{Personalized Requirements in Interactive Mechanism}}
\label{finding: machanism}

\yuling{
The interactive mechanism here specifically referred to when and how to guide the interaction, as well as how to motivate children to interact during the story-reading process. In our study, we found that even though children and parents had already customized their reading modes and interaction requirements before starting using \toolname, they still exhibited varying preferences in interactive mechanisms during the actual reading process. 
}

\paragraph{Guiding Mechanism}
\yuling{
Although most parents and experts (N=12) appreciated \toolname's customizable interaction mode design, they expressed differing opinions regarding the granularity of interaction timing. Some parents, for instance, preferred providing children with a more integral reading experience, believing that interrupting the story for interaction—regardless of frequency—could disrupt the story's coherence and negatively impact the child's reading experience. P3, a mother of a 6-year-old girl and a 12-year-old boy, shared her children's different preferences in guiding mechanism: \textit{``This approach (uninterrupted reading) might suit some children, while others, like my son, might not find it as fitting because he likes to ask questions anytime.''} 
P9, a mother of an 8-year-old boy, suggested \textit{``waiting until the end allows the child to better understand the story.''} P13 shared a similar perspective, suggesting that \textit{``it's better not to stop and delve into the interactions immediately. Instead, continue reading and let the child grasp the overall story.''} E8, an educational expert, commented positively on the conversation timing of \toolname, suggesting that \textit{``not having a question on every page is good. If there were a question on every page, children might feel like they are taking a test.''}
}

\yuling{Both parents and experts shared personalized opinions regarding the questioning methods during the conversation. Many participants (N=7) considered \toolname's questions to be effective in assisting children to comprehend the story and encourage their active thinking. Some experts further suggested improved approaches for posing questions. For instance, E7 recommended \textit{``asking some simple questions first, then gradually increasing the difficulty.''} E1 and E3 suggested \textit{asking one question at a time''} to let children focus on understanding each question fully. P14, a mother of a 4-year-old girl, proposed to \textit{``stratify questions by age, with 1-2 questions for 3-4-year-olds and more questions for 5-6-year-olds.''}}

\paragraph{Incentive Mechanisms}
\label{finding: incentive mechanisms}

\yuling{
As introduced in Section~\ref{Sec_KeyFeature}, we incorporated encouragement and incentive mechanisms in \toolname's conversation to better engage children. During our study, participants provided positive feedback on these incentives. Some participants expressed a desire for even more diverse incentive mechanisms, such as \textit{``giving a little firework when the child answers correctly''} (P14) and \textit{``designing coins or badges for children to earn''} (E4), to improve children's \textit{``sense of achievement''} (P13). 
However, not all participants favored the addition of these incentive mechanisms. Some pointed out that such kinds of design might make children read for rewards or for comparing with others, rather than enjoying the reading itself. 
}

\yuling{
This difference in perceptions toward the incentive mechanism was also evidenced in how participants perceived the dashboard (Figure~\ref{fig:dashboard}). Some participants valued the dashboard as a useful tool to help understand what children read and what challenges they faced while answering questions. Some even suggested adding a child-friendly version of the dashboard, which could serve as a more positive incentive, offering children a sense of accomplishment as they saw their progress visually updated. However, many others expressed concerns that this approach focused much on quantitative performance metrics, suggesting that parents and teachers should take a more holistic approach to children's reading. As P8 mentioned, ``\textit{An overly quantitative approach might not have been suitable for young children. It shifted the focus too much toward measurable outcomes rather than children's curiosity, creativity, and intrinsic motivation}''.
These requirements suggested the delicate balance between motivating children through feedback and ensuring that the emphasis remained on fostering a genuine, long-term love for reading rather than striving for external rewards or achievements.
}

\subsubsection{\textbf{Personalized Requirements Caused by Reading Context}}
\label{finding: context}

\yuling{
The third major personalized requirement our participants expressed for LLM-empowered interactive story reading for children was the one caused by children's reading context. Our findings suggested that children's reading activities were not merely interactions between children and storybooks, but rather shaped by children's family and education context. 
Such kind of situated interaction was influenced by factors such as the parents' knowledge level, parenting philosophy, and children's education context, which in turn led to varying interaction requirements. 
}

\paragraph{Parent's Daily Companionship Pattern}
\yuling{Parent was the most critical role in children's story-reading behaviors. Their involvement level and type greatly influenced children's reading behaviors and experiences, as well as how they interacted and engaged with \toolname. Specifically, some children were able to use \toolname\ independently, while others needed their parents to repeat or explain the questions posed by \toolname. One primary reason for these differences was the role parents play during their daily reading process.
P3's child, PC3.a, for instance, was able to read and interact with \toolname\ independently. P3 explained, ``\textit{At home, I'm the same. I don't emphasize things deliberately. I just answer when she asks.}'' In contrast, P1's daughter needed P1 to repeat the questions of \toolname\ because ``\textit{I spend more time with her, so she's more dependent on me}''. Additionally, parents' knowledge level and daily life circumstances also indirectly influenced their children's interactions with \toolname. Some parents (e.g., P2, P3, P7) in our study reported having limited time to spend with their children, which often led to children reading alone. As a result, their children showed strong independent reading and interaction with \toolname.}

\paragraph{Parent's Education Philosophies}
\label{finding: education philosopies}
\yuling{Our study also revealed that parents' various educational philosophies would influence the types of knowledge they wanted their children to learn and how they expected children to engage in learning, which in turn impacted their interaction needs with \toolname. For example, some parents preferred to follow their children's interests and allowed them to read ``\textit{interesting, funny, and humorous books}'' (P12), while others aimed to teach their children knowledge as much as possible, such as ``\textit{character recognition, pinyin, math, and English vocabulary}'' (P5, P7). P11 also expressed a desire to foster humanistic qualities in children. This diversity in educational philosophies led to different interactions with \toolname.}

\paragraph{Children's Education Environment.}
\label{sec: school env}
\yuling{
Children's educational environment, including curriculum guidelines, course arrangements, and the degree of the school's digitalization, also shaped their reading practices and capabilities with \toolname. As reported by participants, different schools adopted various reading approaches, such as \textit{``organizing reading classes where children could discuss and share their thoughts''} (E3), and \textit{``assigning psychological readings to foster confident, brave, and resilient traits in children''} (E5). These different reading activities indirectly influenced children's interaction needs with \toolname. Schools' varying course arrangements, such as the pinyin course, also affected children's literacy when interacting with \toolname. Some children (N=4/12) relied on pinyin to understand the text, while others, benefiting from more advanced pinyin education, were able to comprehend without it. Additionally, the digitalization degree of schools also contributed to children's varied literacy. As E4 noted, \textit{``In schools with strong AI integration, students often engaged in AI tools, such as using AI to create comic strips, making them more familiar with smart tools compared to children in ordinary schools.''} These differences in literacy levels further influenced how children interacted with \toolname. }

\subsubsection{\textbf{Personalized Requirements in Interactive Interface}}
\label{finding: interface}

\yuling{
The last personalized requirement for LLM-empowered story reading and interaction for children, which was also the most frequently mentioned by our participants, was interactive interfaces. As introduced in Section \ref{toolview}, \toolname\ featured a multimodal GUI, integrating voice, graphics, and text to engage with children. While many children expressed interest and engagement when interacting with \toolname, parents and experts expressed diverse opinions and requirements for the interfaces.
}

\yuling{
First, our participants expressed differing experiences and opinions regarding the use of multi-modal interfaces. While many participants (N=7) recommended ``\textit{telling stories vividly and expressively, with accompanying background music}'' (P6, E3), some others expressed concerns that overly rich interfaces might distract children, making them focus more on the animation rather than the story itself. While many children (N=8/12) enjoyed reading picture books, some children (e.g., PC2, PC3.b) preferred books with more text. E7, an educational expert, noted that this differentiation was based on children's reading development level. She explained, ``\textit{If a child’s reading development is relatively advanced, they will want to move beyond picture books and read text, as the text contains more information. Conversely, if a child's reading development is less advanced, they will prefer picture books because they find text more challenging}.'' }

\yuling{
Meanwhile, participants had different opinions regarding the use of GUI. Some parents (P2, P9, P11) believed that VUI was better suited for children, as it was safer for their eyesight and encouraged imaginative thinking. As P11 stated, ``\textit{Listening is a great method. A purely dialogic process, where only auditory input is involved, can significantly enhance children's imagination.}'' However, many others (N=15/26) emphasized the importance of GUI for better engagement, with E2 noting, ``\textit{Children's visual perception is the most developed sense in their early years.}'' Additionally, some parents (P2, P3, P7) concerned the potential harm of digital media to children's eyesight and believed that the paper book could better protect children's eyesight, as well as support a better reading experience for children through ``\textit{flipping pages back and forth}'' (P3). 
Other participants (P5, P4, P7), instead, did not share this cautious stance towards digital media. They believed that children's eyesight could be protected through practices like ``\textit{restricting screen time}''. 
}

\section{Discussion}
\yuling{
In the Finding section, we illustrated whether and how LLM-empowered \toolname\ was used to support children’s personalized story reading and interactions (\textbf{RQ1}) and how stakeholders (children, parents, educational experts) experience and perceive it (\textbf{RQ2}). We demonstrated participants' relatively positive feedback on \toolname\ in supporting children’s personalized reading and interaction. At the same time, as key stakeholders, they also expressed a series of personalized requirements for LLM-empowered personalized story reading and interaction tools. We categorized these personalized requirements into four interactive content, mechanism, context, and interface, and elaborated on them in detail. In what follows, we provide an in-depth discussion of our two research questions based on these findings.
}

\subsection{How Can LLMs Be Appropriately Used to Support Children's Personalized Story Reading and Interaction?}
\yuling{
To address \textbf{RQ1}, we conducted a formative study to understand parents' requirements for interactive story reading. Based on the identified key insights (\textbf{KIs}) and design goals (\textbf{DGs}), we designed and deployed \toolname, leveraging key features (\textbf{KFs}) of customizable reading content and modes to support personalized reading preferences, LLM-empowered chatbot to provide child-appropriate, adaptive, and personalized interaction, embedding attention-getting mechanisms into the interaction process to better engage children, and RAG-based guided conversation to support children's active thinking. 
As shown in Section \ref{finding: experience}, our participants provided generally positive feedback on \toolname, emphasizing its potential in supporting children’s adaptive, flexible, and personalized reading experiences. Meanwhile, our study also observed that children's personalized needs in practical story-reading scenarios were complex and far beyond the current functionalities of \toolname. Some of these needs, we believe, could be addressed by integrating advanced technologies, while others require more child-centered attention and discussion. According to our findings, we now elaborate on the opportunities and challenges of using LLMs to support children’s personalized story-reading, offering insights for future LLM-empowered story-reading technologies for children.
}

\paragraph{\textbf{LLM's Potentials in Supporting Personalized and Adaptive Story Reading and Interaction}}

\yuling{Our findings in Section~\ref{finding: experience} revealed that, compared with existing AI-based story reading tools (e.g., StoryBuddy~\cite{zhangStoryBuddyHumanAICollaborative2022a}, Rosita~\cite{xuRositaReadsMy2023a}, Elinor~\cite{Elinor}), which primarily relied on pre-determined intent recognition, rule-based syntactic parsing (e.g., using ``slot fillers'' to process user responses within predefined templates), or pre-generated static Question-Answer pairs (QA pairs) to drive conversations~\cite{qu2019user, Elinor}, LLM-empowered \toolname\ hold greater potential in supporting the adaptive and engaging conversations that both children and parents require.}

\yuling{Specifically, our study suggests that LLMs have great potential in guiding interaction based on children's cognitive abilities and interests. As shown in the conversations in Table~\ref{tab:chatbot_summary_dialogue} and noted by our participants, the questions generated by \toolname\ exhibit a balanced level of difficulty. For the challenging questions, \toolname\ can effectively scaffold children's active thinking. Meanwhile, LLMs can effectively integrate children's unique interests in diverse conversation situations and respond adaptively to their personalized answers (see Section~\ref{finding:dg2}). Such capability significantly enhances the attractiveness of the interaction process for children. Moreover, LLMs can effectively support guided, expansive conversations by integrating external knowledge into the conversation, which in turn inspires children's active thinking. As shown in Table~\ref{tab:chatbot_sunlight_dialogue} and~\ref{tab:chatbot_water_dialogue}, \toolname's chatbot connects the story context with relevant real-world knowledge, which our participants consider a great way to encourage children to think actively both within and beyond the story. This capability meets parents' expectations of guiding children to think proactively through story reading (see Section~\ref{finding:dg3}).
}

\yuling{
Technically, \toolname\ primarily uses prompt-engineering to design child-appropriate, adaptive, and engaging conversations. The prompts used can be found in Appendix~\ref{app:prompt}. For guided conversations enriched with external knowledge, we adopt a lightweight approach by leveraging RAG to embed educational content from NGSS into GPT-4 through prompts. Technical details can be found in Appendix~\ref{app: technical process}. These technologies can be generalized to other children's interaction scenarios, contributing to future studies in this research area.}

\yuling{
Through our study, there is no doubt that LLMs have immense potential in supporting more situated, personalized, and engaging conversations and interactions with children. We also believe that, as technologies become increasingly intelligent, the capabilities of LLMs in this regard will be more pronounced. Yet, we have to acknowledge that although these LLM-empowered approaches we employed significantly enhance the personalization and engagement of story reading and interaction in our experimental study, they still face certain practical challenges in meeting the real conversational needs of children. For example, although \toolname\ instructs GPT-4 to dynamically adapt to children's overall response literacy and leverage RAG to incorporate age-appropriate knowledge, our participants still considered it challenging to generate conversations that can seamlessly suit every child's reading literacy, cognitive level, and reading preferences. Drawing from these identified practical challenges, we proposed a series of design recommendations (see Section \ref{Discussion2-designImplication}). Meanwhile, our study also suggests that in LLM-empowered story reading and interaction scenarios, there are issues that go beyond technical challenges and require a more socio-technological perspective to fully address, which will be deeply discussed in the following section.
}

\paragraph{\textbf{LLM’s Challenges in Supporting Child-Centered Interactive Story Reading}}

\yuling{
During our research, we observed that the relationship between children and \toolname\ in the guidance process is quite subtle. Often, we assume that AI guides children through interaction and conversation. However, in our observations of children interacting with \toolname, we found that children frequently exhibited smarter feedback and, to some extent, took on the role of a guide. Educational experts we interviewed also pointed out that, in many ways, children can be considered ``smarter'' than LLMs due to their creative and unconventional thinking.
Our findings echo ongoing discussions in the field of intelligent education—as educational tools become increasingly intelligent, how should the teacher-student relationship mediated by AI tools evolve \cite{zhang2021ai, extance2023chatgpt}? A consistent view suggests that the educational paradigm reshaped by LLMs should focus more on using LLMs to inspire thinking \cite{extance2023chatgpt}. In the scenario of children's story reading, fostering children's thinking and creativity becomes even more crucial, as it plays a key role in their personalized development.
}

\yuling{
Drawing on this perspective, we reexamined LLM-empowered \toolname. We found that, although the GPT-4-empowered chatbot can scaffold children's thinking step by step, the guided conversations inadvertently shift the focus toward the instructed agenda (i.e., story comprehension or knowledge enrichment), potentially sidelining children’s curiosity and imagination. For example, when asking about the effects of sunshine on animals, PC2 responded with ``\textit{animals will turn red}''. While such a response may not align with conventional explanations, it illustrates the child's imaginative exploration of the world. However, LLMs often interpreted such responses in a more literal or conventional manner, limiting the understanding or appreciation of children's imaginative thoughts. Existing HCI studies have also pointed out similar limitations of LLMs, such as, in comprehensively understanding children's expressions and thinking perspectives~\cite{Mathemyths, chacha}.
}

\yuling{
Based on our findings as well as existing discussion, we suggest further studies in this area should carefully consider the balance between children's digital autonomy with the guidance provided by LLMs. Specifically, LLM-empowered tools should grant children more agency and autonomy during the interactions~\cite{wang2023DigitalAutonomy}, such as allowing them to initiate interactions and ask questions based on their interests and needs. In addition, in the education domain, there are several classic pedagogical approaches focusing on the teaching-learning relationship, such as `learning by doing'~\cite{anzai1979theory}, `learning by teaching'~\cite{goodlad1989peer}, and `inquiry-based learning'~\cite{pedaste2015phases}. They hold great potential for improving child-centered or children-guided education experiences, but are often difficult to fully personalize in traditional learning environments (i.e., classroom setting)~\cite{Bingham2018AheadOT, personalize_challenge, amro2019exploring}. LLMs have great potential to support such educational approaches, enabling more child-centered story reading and interactions. 
}

\subsection{How Do Stakeholders Perceive LLM-Empowered Personalized Story Reading and Interaction?}
\label{Discussion2-designImplication}

\yuling{To address \textbf{RQ2}, we recruited three kinds of stakeholders in children's story reading and interaction scenarios—children, parents, and education expert—to experience and evaluate LLM-empowered \toolname\ as well as this technical trend. Our findings suggested that although stakeholders experienced relatively positive feedback to \toolname, their practical personalized requirements were multi-dimensional and far beyond the designed personalized features of \toolname. We specifically characterized these requirements into interactive content, mechanisms, context, and interfaces, and elaborate them respectively in Section~\ref{finding: requirements}. Drawing from these findings, we now discuss design recommendations for addressing these requirements and bettering LLMs to empower children's personalized story reading and interaction. }

\yuling{
It is noteworthy here that we are not arguing for the increased development and use of advanced, smarter technologies in the scenario of children's story reading—what some may refer to as \textit{Technicism} in education~\cite{halliday1998technicism}. Instead, we express concerns regarding the potential pitfalls of technicism in this particular scenario. From technicism's perspective, typical solutions for those challenges might involve extensive child-related data collection to calibrate better models, allowing them to become more attuned to children's language patterns, emotions, cognitive styles, etc~\cite{talaat2023real, van-dijk-etal-2023-chiscor, sagae2021tracking}. However, such approaches raise significant risks regarding data privacy and ethics, especially for vulnerable children~\cite{jacobs2006development, wang2019children}. Furthermore, this kind of technicism may also limit children's free development and parent-child relationships \cite{sun2024exploring}. We thus advocate for a more integrated development direction that combines technology with thoughtful interaction design—one that better supports child-centered story reading and interaction experiences.
}

\paragraph{\textbf{Adapting Conversation Content to Children's Individual Capability and Preferences}}
\yuling{
Leveraging GPT-4, \toolname\ was designed to support personalized conversation by adapting to children of different ages (Section \ref{design:age}), interests, knowledge scope (Section \ref{RAG}), and reading needs (Section \ref{design:readMode}). Yet, our findings (Section~\ref{finding: content}) revealed that, due to children's varied characteristics, personalized knowledge scopes, cognitive development levels, and growth environment, stakeholders’ requirements for personalized conversation content are more complicated in practice, reflecting in the personalized requirements in conversation topic, language complexity, the incorporation of extended knowledge, etc. From a technical perspective, incorporating prompting-based strategies (e.g., In-Context Learning~\cite{dong2022surveyincontext} and Chain-of-Thought~\cite{wei2022chainofthought} reasoning), along with expert-annotated demonstrations (e.g., question-answer pairs with difficulty ratings), could better calibrate LLM-generated content to children's needs. 
From a design perspective, future tools could offer flexible options for book types and knowledge extension, empowering children to engage with content that resonates with their interests and learning goals. Additionally, future designs could maintain dynamic profiles for each child that track their progress and preferences over time~\cite{zhang2023memory, zhong2024memorybank} to enable long-term adaptability and more personalized interactions.
}

\paragraph{\textbf{Children-Driven Interactive Mechanism}}

\yuling{
Existing literature has pointed out the importance of aligning interaction styles with individual needs, which is the key to enhancing children's engagement, learning outcomes, and cognitive development~\cite{FuturegazingPotentialDigital, chen2023adaptive, whalley2021towards, BUNTING2021100236}. Our findings extend this perspective, revealing that stakeholders have personalized requirements in guiding and incentive mechanisms.  
Although prior research on AI-empowered story-related tools has explored interaction strategies (e.g., story customization~\cite{storydrawer, storyprompt}, collaborative storytelling~\cite{Mathemyths}, and gamification~\cite{storycoder}), these interaction patterns are often limited to AI-guided interactions, which may undermine children's autonomy and initiative. We suggest future conversation-based interaction studies could leverage engagement-promoting techniques in pedagogical research, such as responsive interaction~\cite{kaiser1996effects}, incidental teaching~\cite{hart1975incidental}, Zone of Proximal Development~\cite{vygotsky1978mind}, to craft both engaging and educational-suitable conversations.
Meanwhile, we also suggest future designs pay attention to enhancing children's digital autonomy~\cite{wang2023DigitalAutonomy} by, for instance, enabling proactive question-asking and allowing them to skip unwanted interactions, and ensuring greater flexibility for children to choose their preferred interaction style~\cite{Kucirkova02012022}.
}

\paragraph{\textbf{Supporting Environmentally Sensitive Interactions}}

\yuling{
Our findings (Section~\ref{finding: context}), along with existing educational research, revealed that children's story reading and interaction were a typically situated activity. That is, children's reading context, their parents' involvement, knowledge levels, and education philosophy, their schools' education environment and teaching programs, significantly influence children's reading behaviors and preferences \cite{DC1_familycontext_1, DC1_familycontext_2_and_parentedu_1, DC1_school_1}, which indirectly influence their adoption, use, and experience to \toolname. We, therefore, suggest that future HCI researchers and designers in this area should not only consider the personalized characteristics of children but also take into account the personalized features of the reading environments in which they are situated, including family and school environments, immediate surroundings~\cite{suardi2021curiosity}, etc. The human-in-the-loop approach \cite{yaoItAITurn2022} could also be adopted to allow parents or teachers to participate in the interactive generation algorithms and guide the algorithms to be more sensitive to children's reading environment, knowledge level, and cognitive abilities. 
}

\paragraph{\textbf{Versatile and Customizable Interactive Interfaces}}
\yuling{
\toolname\ leveraged a graphical user interface, integrating text, audio, and graphics to support children's interaction. However, our participants expressed highly personalized requirements for the interactive interface, in terms of the use and the richness of the multi-modal interface (see Section~\ref{finding: interface}). These personalized requirements highlighted the need for more versatile and customizable interactive interfaces. On the one hand, the tools should incorporate more multi-modal interaction methods, providing children with a richer and more engaging user experience. Technically, prior research has shown that combining interactions with diverse formats—such as visuals (images, animations, layouts) and audio (sound, music, voice commands)—can significantly boost children's engagement and enhance their interaction experience~\cite{roseli2023affective, educsci14060604}. Recently emerging text-to-image generation~\cite{zhang2023text} and text-to-video generation~\cite{cho2024sora} technologies can also be used to enrich the interaction experience, further stimulating children's curiosity and participation. On the other hand, the tools should include customizable interactive design features, allowing users to select interaction modalities based on their personal preferences and ensuring that the interaction is tailored to the child's individual reading style and interests.
}

\subsection{Limitation and Future Work}

\yuling{
This work is our preliminary investigation about whether and how to effectively utilize LLMs to support children's interactive story-reading activities. We employed a probe-based qualitative method to gather insights. Several limitations should be acknowledged. First, we primarily collected user feedback through observation and qualitative interviews. Given the limited study duration, children were not given sufficient time to deeply familiarize themselves with the tool and its features and build an emotional connection with \toolname. This might impact their experience.
Second, the qualitative study may introduce possible sample bias, affecting the generalizability of our findings. Specifically, our participants were primarily from Shanghai, one of China's most economically developed cities. This demographic characteristic caused some important factors, such as family income levels and AI literacy, to be not included in our analysis. 
Third, our research took \toolname\ as a technical probe. Its specific design features and interaction methods might influence participants' behaviors and experiences. 
In future work, we aim to optimize \toolname\ according to participants' feedback and then recruit a more demographically diverse group of parents and children for long-term experience and behavioral data collection. This approach will allow us to draw more universal conclusions and refine the tool to better suit a broader range of users.
}

\section{Conclusion}

\yuling{
This study presents a probe-based empirical study to explore how to better use LLMs to support children’s personalized story-reading activities. Based on key design insights from formative interviews with parents, we developed and implemented \toolname, an LLM-empowered personalized story reading and interactive tool for children. We then take \toolname\ as the technical probe, and recruit children, parents, and education experts to experience and evaluate it as well as this increasing technology trend. Our findings elaborate on their experiences and perceptions of \toolname, and summarize their personalized requirements. We conclude by proposing a series of critical discussions and design recommendations for future LLM-empowered personalized story-reading technologies.
}

\begin{acks}
\yuling{We sincerely appropriate East China Normal University (ECNU), Lab of Artificial Intelligence for Education at ECNU, and Shanghai Institute of Artificial Intelligence for Education for supporting our study.
}We are also deeply grateful to the children, parents, and educational experts for their participation in our study and to Shihan Fu for her valuable feedback.
\end{acks}
\bibliographystyle{ACM-Reference-Format}
\bibliography{main}


\begin{thebibliography}{94}

%%% the Web address.

\ifx \showCODEN    \undefined \def \showCODEN     #1{\unskip}     \fi
\ifx \showDOI      \undefined \def \showDOI       #1{#1}\fi
\ifx \showISBNx    \undefined \def \showISBNx     #1{\unskip}     \fi
\ifx \showISBNxiii \undefined \def \showISBNxiii  #1{\unskip}     \fi
\ifx \showISSN     \undefined \def \showISSN      #1{\unskip}     \fi
\ifx \showLCCN     \undefined \def \showLCCN      #1{\unskip}     \fi
\ifx \shownote     \undefined \def \shownote      #1{#1}          \fi
\ifx \showarticletitle \undefined \def \showarticletitle #1{#1}   \fi
\ifx \showURL      \undefined \def \showURL       {\relax}        \fi
% invisible to TeX
\providecommand\bibfield[2]{#2}
\providecommand\bibinfo[2]{#2}
\providecommand\natexlab[1]{#1}
\providecommand\showeprint[2][]{arXiv:#2}

\bibitem[Amro and Borup(2019)]%
        {amro2019exploring}
\bibfield{author}{\bibinfo{person}{Falah Amro} {and} \bibinfo{person}{Jered Borup}.} \bibinfo{year}{2019}\natexlab{}.
\newblock \showarticletitle{Exploring blended teacher roles and obstacles to success when using personalized learning software}.
\newblock \bibinfo{journal}{\emph{Journal of Online Learning Research}} \bibinfo{volume}{5}, \bibinfo{number}{3} (\bibinfo{year}{2019}), \bibinfo{pages}{229--250}.
\newblock


\bibitem[Anzai and Simon(1979)]%
        {anzai1979theory}
\bibfield{author}{\bibinfo{person}{Yuichiro Anzai} {and} \bibinfo{person}{Herbert~A Simon}.} \bibinfo{year}{1979}\natexlab{}.
\newblock \showarticletitle{The theory of learning by doing.}
\newblock \bibinfo{journal}{\emph{Psychological review}} \bibinfo{volume}{86}, \bibinfo{number}{2} (\bibinfo{year}{1979}), \bibinfo{pages}{124}.
\newblock


\bibitem[Aronson(1994)]%
        {ThematicAnalysis1}
\bibfield{author}{\bibinfo{person}{Jodi Aronson}.} \bibinfo{year}{1994}\natexlab{}.
\newblock \showarticletitle{A pragmatic view of thematic analysis}.
\newblock \bibinfo{journal}{\emph{The qualitative report}} \bibinfo{volume}{2}, \bibinfo{number}{1} (\bibinfo{year}{1994}), \bibinfo{pages}{1--3}.
\newblock


\bibitem[Bingham et~al\mbox{.}(2018)]%
        {Bingham2018AheadOT}
\bibfield{author}{\bibinfo{person}{Andrea~J. Bingham}, \bibinfo{person}{John~F. Pane}, \bibinfo{person}{Elizabeth~D. Steiner}, {and} \bibinfo{person}{Laura~S. Hamilton}.} \bibinfo{year}{2018}\natexlab{}.
\newblock \showarticletitle{Ahead of the Curve: Implementation Challenges in Personalized Learning School Models}.
\newblock \bibinfo{journal}{\emph{Educational Policy}}  \bibinfo{volume}{32} (\bibinfo{year}{2018}), \bibinfo{pages}{454 -- 489}.
\newblock
\urldef\tempurl%
\url{https://api.semanticscholar.org/CorpusID:147628309}
\showURL{%
\tempurl}


\bibitem[Buckingham et~al\mbox{.}(2023)]%
        {buckinghamWhyPoorChildren2023}
\bibfield{author}{\bibinfo{person}{Jennifer Buckingham}, \bibinfo{person}{Robyn Beaman}, {and} \bibinfo{person}{Kevin Wheldall}.} \bibinfo{year}{2023}\natexlab{}.
\newblock \showarticletitle{Why Poor Children Are More Likely to Become Poor Readers: The Early Years}.
\newblock In \bibinfo{booktitle}{\emph{Mapping the {{Field}}}}. \bibinfo{publisher}{Routledge}.
\newblock
\showISBNx{978-1-00-340372-2}


\bibitem[Bunting et~al\mbox{.}(2021)]%
        {BUNTING2021100236}
\bibfield{author}{\bibinfo{person}{Leona Bunting}, \bibinfo{person}{Ylva~Hård {af Segerstad}}, {and} \bibinfo{person}{Wolmet Barendregt}.} \bibinfo{year}{2021}\natexlab{}.
\newblock \showarticletitle{Swedish teachers’ views on the use of personalised learning technologies for teaching children reading in the English classroom}.
\newblock \bibinfo{journal}{\emph{International Journal of Child-Computer Interaction}}  \bibinfo{volume}{27} (\bibinfo{year}{2021}), \bibinfo{pages}{100236}.
\newblock
\showISSN{2212-8689}
\urldef\tempurl%
\url{https://doi.org/10.1016/j.ijcci.2020.100236}
\showDOI{\tempurl}


\bibitem[Chen et~al\mbox{.}(2023)]%
        {chen2023adaptive}
\bibfield{author}{\bibinfo{person}{Ching-Huei Chen}, \bibinfo{person}{Victor Law}, {and} \bibinfo{person}{Kun Huang}.} \bibinfo{year}{2023}\natexlab{}.
\newblock \showarticletitle{Adaptive scaffolding and engagement in digital game-based learning}.
\newblock \bibinfo{journal}{\emph{Educational technology research and development}} \bibinfo{volume}{71}, \bibinfo{number}{4} (\bibinfo{year}{2023}), \bibinfo{pages}{1785--1798}.
\newblock


\bibitem[Chen et~al\mbox{.}(2024)]%
        {chen2023fairytalecqa}
\bibfield{author}{\bibinfo{person}{Jiaju Chen}, \bibinfo{person}{Yuxuan Lu}, \bibinfo{person}{Shao Zhang}, \bibinfo{person}{Bingsheng Yao}, \bibinfo{person}{Yuanzhe Dong}, \bibinfo{person}{Ying Xu}, \bibinfo{person}{Yunyao Li}, \bibinfo{person}{Qianwen Wang}, \bibinfo{person}{Dakuo Wang}, {and} \bibinfo{person}{Yuling Sun}.} \bibinfo{year}{2024}\natexlab{}.
\newblock \showarticletitle{{S}tory{S}park{QA}: Expert-Annotated {QA} Pairs with Real-World Knowledge for Children{'}s Story-Based Learning}. In \bibinfo{booktitle}{\emph{Proceedings of the 2024 Conference on Empirical Methods in Natural Language Processing}}, \bibfield{editor}{\bibinfo{person}{Yaser Al-Onaizan}, \bibinfo{person}{Mohit Bansal}, {and} \bibinfo{person}{Yun-Nung Chen}} (Eds.). \bibinfo{publisher}{Association for Computational Linguistics}, \bibinfo{address}{Miami, Florida, USA}, \bibinfo{pages}{17351--17370}.
\newblock
\urldef\tempurl%
\url{https://doi.org/10.18653/v1/2024.emnlp-main.961}
\showDOI{\tempurl}


\bibitem[Cheng et~al\mbox{.}(2024)]%
        {moonstory}
\bibfield{author}{\bibinfo{person}{Alan~Y Cheng}, \bibinfo{person}{Meng Guo}, \bibinfo{person}{Melissa Ran}, \bibinfo{person}{Arpit Ranasaria}, \bibinfo{person}{Arjun Sharma}, \bibinfo{person}{Anthony Xie}, \bibinfo{person}{Khuyen~N Le}, \bibinfo{person}{Bala Vinaithirthan}, \bibinfo{person}{Shihe Luan}, \bibinfo{person}{David Thomas~Henry Wright}, {et~al\mbox{.}}} \bibinfo{year}{2024}\natexlab{}.
\newblock \showarticletitle{Scientific and Fantastical: Creating Immersive, Culturally Relevant Learning Experiences with Augmented Reality and Large Language Models}. In \bibinfo{booktitle}{\emph{Proceedings of the CHI Conference on Human Factors in Computing Systems}}. \bibinfo{pages}{1--23}.
\newblock


\bibitem[Cho et~al\mbox{.}(2024)]%
        {cho2024sora}
\bibfield{author}{\bibinfo{person}{Joseph Cho}, \bibinfo{person}{Fachrina~Dewi Puspitasari}, \bibinfo{person}{Sheng Zheng}, \bibinfo{person}{Jingyao Zheng}, \bibinfo{person}{Lik-Hang Lee}, \bibinfo{person}{Tae-Ho Kim}, \bibinfo{person}{Choong~Seon Hong}, {and} \bibinfo{person}{Chaoning Zhang}.} \bibinfo{year}{2024}\natexlab{}.
\newblock \showarticletitle{Sora as an agi world model? a complete survey on text-to-video generation}.
\newblock \bibinfo{journal}{\emph{arXiv preprint arXiv:2403.05131}} (\bibinfo{year}{2024}).
\newblock


\bibitem[Cobb and Bowers(1999)]%
        {cobb1999cognitive}
\bibfield{author}{\bibinfo{person}{Paul Cobb} {and} \bibinfo{person}{Janet Bowers}.} \bibinfo{year}{1999}\natexlab{}.
\newblock \showarticletitle{Cognitive and situated learning perspectives in theory and practice}.
\newblock \bibinfo{journal}{\emph{Educational researcher}} \bibinfo{volume}{28}, \bibinfo{number}{2} (\bibinfo{year}{1999}), \bibinfo{pages}{4--15}.
\newblock


\bibitem[Dietz et~al\mbox{.}(2021)]%
        {storycoder}
\bibfield{author}{\bibinfo{person}{Griffin Dietz}, \bibinfo{person}{Jimmy~K Le}, \bibinfo{person}{Nadin Tamer}, \bibinfo{person}{Jenny Han}, \bibinfo{person}{Hyowon Gweon}, \bibinfo{person}{Elizabeth~L Murnane}, {and} \bibinfo{person}{James~A. Landay}.} \bibinfo{year}{2021}\natexlab{}.
\newblock \showarticletitle{StoryCoder: Teaching Computational Thinking Concepts Through Storytelling in a Voice-Guided App for Children}. In \bibinfo{booktitle}{\emph{Proceedings of the 2021 CHI Conference on Human Factors in Computing Systems}} (Yokohama, Japan) \emph{(\bibinfo{series}{CHI '21})}. \bibinfo{publisher}{Association for Computing Machinery}, \bibinfo{address}{New York, NY, USA}, Article \bibinfo{articleno}{54}, \bibinfo{numpages}{15}~pages.
\newblock
\showISBNx{9781450380966}
\urldef\tempurl%
\url{https://doi.org/10.1145/3411764.3445039}
\showDOI{\tempurl}


\bibitem[Dietz et~al\mbox{.}(2020)]%
        {gigglegauge}
\bibfield{author}{\bibinfo{person}{Griffin Dietz}, \bibinfo{person}{Zachary Pease}, \bibinfo{person}{Brenna McNally}, {and} \bibinfo{person}{Elizabeth Foss}.} \bibinfo{year}{2020}\natexlab{}.
\newblock \showarticletitle{Giggle gauge: a self-report instrument for evaluating children's engagement with technology}. In \bibinfo{booktitle}{\emph{Proceedings of the Interaction Design and Children Conference}} (London, United Kingdom) \emph{(\bibinfo{series}{IDC '20})}. \bibinfo{publisher}{Association for Computing Machinery}, \bibinfo{address}{New York, NY, USA}, \bibinfo{pages}{614–623}.
\newblock
\showISBNx{9781450379816}
\urldef\tempurl%
\url{https://doi.org/10.1145/3392063.3394393}
\showDOI{\tempurl}


\bibitem[Dong et~al\mbox{.}(2022)]%
        {dong2022surveyincontext}
\bibfield{author}{\bibinfo{person}{Qingxiu Dong}, \bibinfo{person}{Lei Li}, \bibinfo{person}{Damai Dai}, \bibinfo{person}{Ce Zheng}, \bibinfo{person}{Jingyuan Ma}, \bibinfo{person}{Rui Li}, \bibinfo{person}{Heming Xia}, \bibinfo{person}{Jingjing Xu}, \bibinfo{person}{Zhiyong Wu}, \bibinfo{person}{Tianyu Liu}, {et~al\mbox{.}}} \bibinfo{year}{2022}\natexlab{}.
\newblock \showarticletitle{A survey on in-context learning}.
\newblock \bibinfo{journal}{\emph{arXiv preprint arXiv:2301.00234}} (\bibinfo{year}{2022}).
\newblock


\bibitem[Druin and Solomon(1996)]%
        {druin1996designing}
\bibfield{author}{\bibinfo{person}{Allison Druin} {and} \bibinfo{person}{Cynthia Solomon}.} \bibinfo{year}{1996}\natexlab{}.
\newblock \bibinfo{booktitle}{\emph{Designing multimedia environments for children: Computers, creativity, and kids}}.
\newblock \bibinfo{publisher}{John Wiley \& Sons, Inc.}
\newblock


\bibitem[Eapen and Adhithyan(2023)]%
        {eapen2023personalization}
\bibfield{author}{\bibinfo{person}{Joel Eapen} {and} \bibinfo{person}{VS Adhithyan}.} \bibinfo{year}{2023}\natexlab{}.
\newblock \showarticletitle{Personalization and customization of llm responses}.
\newblock \bibinfo{journal}{\emph{International Journal of Research Publication and Reviews}} \bibinfo{volume}{4}, \bibinfo{number}{12} (\bibinfo{year}{2023}), \bibinfo{pages}{2617--2627}.
\newblock


\bibitem[Easin et~al\mbox{.}(2024)]%
        {easin2024intelligent}
\bibfield{author}{\bibinfo{person}{Arafat~Md Easin}, \bibinfo{person}{Saha Sourav}, {and} \bibinfo{person}{Orosz Tam{\'a}s}.} \bibinfo{year}{2024}\natexlab{}.
\newblock \showarticletitle{An intelligent llm-powered personalized assistant for digital banking using langgraph and chain of thoughts}. In \bibinfo{booktitle}{\emph{2024 IEEE 22nd Jubilee International Symposium on Intelligent Systems and Informatics (SISY)}}. IEEE, \bibinfo{pages}{625--630}.
\newblock


\bibitem[Extance(2023)]%
        {extance2023chatgpt}
\bibfield{author}{\bibinfo{person}{Andy Extance}.} \bibinfo{year}{2023}\natexlab{}.
\newblock \showarticletitle{ChatGPT has entered the classroom: how LLMs could transform education}.
\newblock \bibinfo{journal}{\emph{Nature}} \bibinfo{volume}{623}, \bibinfo{number}{7987} (\bibinfo{year}{2023}), \bibinfo{pages}{474--477}.
\newblock


\bibitem[Ezeokoli and Patience(2016)]%
        {DC1_school_1}
\bibfield{author}{\bibinfo{person}{FO Ezeokoli} {and} \bibinfo{person}{Igubor Patience}.} \bibinfo{year}{2016}\natexlab{}.
\newblock \showarticletitle{Effects of two methods of instruction on students’ critical response to prose literature text in English in some secondary schools in Benin city}.
\newblock \bibinfo{journal}{\emph{International Journal of Education and Practice}} \bibinfo{volume}{4}, \bibinfo{number}{5} (\bibinfo{year}{2016}), \bibinfo{pages}{167--182}.
\newblock


\bibitem[Fan et~al\mbox{.}(2024)]%
        {storyprompt}
\bibfield{author}{\bibinfo{person}{Min Fan}, \bibinfo{person}{Xinyue Cui}, \bibinfo{person}{Jing Hao}, \bibinfo{person}{Renxuan Ye}, \bibinfo{person}{Wanqing Ma}, \bibinfo{person}{Xin Tong}, {and} \bibinfo{person}{Meng Li}.} \bibinfo{year}{2024}\natexlab{}.
\newblock \showarticletitle{StoryPrompt: Exploring the Design Space of an AI-Empowered Creative Storytelling System for Elementary Children}. In \bibinfo{booktitle}{\emph{Extended Abstracts of the 2024 CHI Conference on Human Factors in Computing Systems}} \emph{(\bibinfo{series}{CHI EA '24})}. \bibinfo{publisher}{Association for Computing Machinery}, \bibinfo{address}{New York, NY, USA}, Article \bibinfo{articleno}{303}, \bibinfo{numpages}{8}~pages.
\newblock
\showISBNx{9798400703317}
\urldef\tempurl%
\url{https://doi.org/10.1145/3613905.3651118}
\showDOI{\tempurl}


\bibitem[{Fielding-Barnsley} and Purdie(2003)]%
        {fielding-barnsleyEarlyInterventionHome2003}
\bibfield{author}{\bibinfo{person}{Ruth {Fielding-Barnsley}} {and} \bibinfo{person}{Nola Purdie}.} \bibinfo{year}{2003}\natexlab{}.
\newblock \showarticletitle{Early Intervention in the Home for Children at Risk of Reading Failure}.
\newblock \bibinfo{journal}{\emph{Support for Learning}} \bibinfo{volume}{18}, \bibinfo{number}{2} (\bibinfo{year}{2003}), \bibinfo{pages}{77--82}.
\newblock
\showISSN{1467-9604}
\urldef\tempurl%
\url{https://doi.org/10.1111/1467-9604.00284}
\showDOI{\tempurl}


\bibitem[Goodlad(1989)]%
        {goodlad1989peer}
\bibfield{author}{\bibinfo{person}{S Goodlad}.} \bibinfo{year}{1989}\natexlab{}.
\newblock \showarticletitle{Peer tutoring: A guide to learning by teaching}.
\newblock \bibinfo{journal}{\emph{Nichols Pub}} (\bibinfo{year}{1989}).
\newblock


\bibitem[Ha et~al\mbox{.}(2024)]%
        {ha2024clochat}
\bibfield{author}{\bibinfo{person}{Juhye Ha}, \bibinfo{person}{Hyeon Jeon}, \bibinfo{person}{Daeun Han}, \bibinfo{person}{Jinwook Seo}, {and} \bibinfo{person}{Changhoon Oh}.} \bibinfo{year}{2024}\natexlab{}.
\newblock \showarticletitle{CloChat: Understanding How People Customize, Interact, and Experience Personas in Large Language Models}. In \bibinfo{booktitle}{\emph{Proceedings of the CHI Conference on Human Factors in Computing Systems}}. \bibinfo{pages}{1--24}.
\newblock


\bibitem[Halliday(1998)]%
        {halliday1998technicism}
\bibfield{author}{\bibinfo{person}{John Halliday}.} \bibinfo{year}{1998}\natexlab{}.
\newblock \showarticletitle{Technicism, reflective practice and authenticity in teacher education}.
\newblock \bibinfo{journal}{\emph{Teaching and teacher education}} \bibinfo{volume}{14}, \bibinfo{number}{6} (\bibinfo{year}{1998}), \bibinfo{pages}{597--605}.
\newblock


\bibitem[Han and Cai(2023)]%
        {hanDesignImplicationsGenerative2023a}
\bibfield{author}{\bibinfo{person}{Ariel Han} {and} \bibinfo{person}{Zhenyao Cai}.} \bibinfo{year}{2023}\natexlab{}.
\newblock \showarticletitle{Design Implications of Generative {{AI}} Systems for Visual Storytelling for Young Learners}. In \bibinfo{booktitle}{\emph{Proceedings of the 22nd {{Annual ACM Interaction Design}} and {{Children Conference}}}}. \bibinfo{publisher}{ACM}, \bibinfo{address}{Chicago IL USA}, \bibinfo{pages}{470--474}.
\newblock
\showISBNx{9798400701313}
\urldef\tempurl%
\url{https://doi.org/10.1145/3585088.3593867}
\showDOI{\tempurl}


\bibitem[Hardy and McLeod(2020)]%
        {hardy2020using}
\bibfield{author}{\bibinfo{person}{Jessica~K Hardy} {and} \bibinfo{person}{Ragan~H McLeod}.} \bibinfo{year}{2020}\natexlab{}.
\newblock \showarticletitle{Using positive reinforcement with young children}.
\newblock \bibinfo{journal}{\emph{Beyond Behavior}} \bibinfo{volume}{29}, \bibinfo{number}{2} (\bibinfo{year}{2020}), \bibinfo{pages}{95--107}.
\newblock


\bibitem[Hart and Risley(1975)]%
        {hart1975incidental}
\bibfield{author}{\bibinfo{person}{Betty Hart} {and} \bibinfo{person}{Todd~R Risley}.} \bibinfo{year}{1975}\natexlab{}.
\newblock \showarticletitle{Incidental teaching of language in the preschool 1}.
\newblock \bibinfo{journal}{\emph{Journal of applied behavior analysis}} \bibinfo{volume}{8}, \bibinfo{number}{4} (\bibinfo{year}{1975}), \bibinfo{pages}{411--420}.
\newblock


\bibitem[Howe(2012)]%
        {howe2012attachment}
\bibfield{author}{\bibinfo{person}{David Howe}.} \bibinfo{year}{2012}\natexlab{}.
\newblock \showarticletitle{Attachment theory}.
\newblock \bibinfo{journal}{\emph{Social work theories and methods}}  \bibinfo{volume}{75} (\bibinfo{year}{2012}).
\newblock


\bibitem[Hu et~al\mbox{.}(2021)]%
        {hu_once_2021}
\bibfield{author}{\bibinfo{person}{Jiangbo Hu}, \bibinfo{person}{Camilla Gordon}, \bibinfo{person}{Ning Yang}, {and} \bibinfo{person}{Yonggang Ren}.} \bibinfo{year}{2021}\natexlab{}.
\newblock \showarticletitle{“{Once} {Upon} {A} {Star}”: {A} {Science} {Education} {Program} {Based} on {Personification} {Storytelling} in {Promoting} {Preschool} {Children}’s {Understanding} of {Astronomy} {Concepts}}.
\newblock \bibinfo{journal}{\emph{Early Education and Development}} \bibinfo{volume}{32}, \bibinfo{number}{1} (\bibinfo{date}{Jan.} \bibinfo{year}{2021}), \bibinfo{pages}{7--25}.
\newblock
\showISSN{1040-9289, 1556-6935}
\urldef\tempurl%
\url{https://doi.org/10.1080/10409289.2020.1759011}
\showDOI{\tempurl}


\bibitem[Jacobs and Klaczynski(2006)]%
        {jacobs2006development}
\bibfield{author}{\bibinfo{person}{Janis~E Jacobs} {and} \bibinfo{person}{Paul~A Klaczynski}.} \bibinfo{year}{2006}\natexlab{}.
\newblock \bibinfo{booktitle}{\emph{The development of judgment and decision making in children and adolescents}}.
\newblock \bibinfo{publisher}{Psychology Press}.
\newblock


\bibitem[Kaiser et~al\mbox{.}(1996)]%
        {kaiser1996effects}
\bibfield{author}{\bibinfo{person}{Ann~P Kaiser}, \bibinfo{person}{Mary~Louise Hemmeter}, \bibinfo{person}{Michaelene~M Ostrosky}, \bibinfo{person}{Rebecca Fischer}, \bibinfo{person}{Paul Yoder}, {and} \bibinfo{person}{Maureen Keefer}.} \bibinfo{year}{1996}\natexlab{}.
\newblock \showarticletitle{The effects of teaching parents to use responsive interaction strategies}.
\newblock \bibinfo{journal}{\emph{Topics in early childhood special education}} \bibinfo{volume}{16}, \bibinfo{number}{3} (\bibinfo{year}{1996}), \bibinfo{pages}{375--406}.
\newblock


\bibitem[Kasneci et~al\mbox{.}(2023)]%
        {Kasneci2023ChatGPTFG}
\bibfield{author}{\bibinfo{person}{Enkelejda Kasneci}, \bibinfo{person}{Kathrin Se{\ss}ler}, \bibinfo{person}{Stefan K{\"u}chemann}, \bibinfo{person}{Maria Bannert}, \bibinfo{person}{Daryna Dementieva}, \bibinfo{person}{Frank Fischer}, \bibinfo{person}{Urs Gasser}, \bibinfo{person}{George~Louis Groh}, \bibinfo{person}{Stephan G{\"u}nnemann}, \bibinfo{person}{Eyke H{\"u}llermeier}, \bibinfo{person}{Stephan Krusche}, \bibinfo{person}{Gitta Kutyniok}, \bibinfo{person}{Tilman Michaeli}, \bibinfo{person}{Claudia Nerdel}, \bibinfo{person}{J{\"u}rgen Pfeffer}, \bibinfo{person}{Oleksandra Poquet}, \bibinfo{person}{Michael Sailer}, \bibinfo{person}{Albrecht Schmidt}, \bibinfo{person}{Tina Seidel}, \bibinfo{person}{Matthias Stadler}, \bibinfo{person}{Jochen Weller}, \bibinfo{person}{Jochen Kuhn}, {and} \bibinfo{person}{Gjergji Kasneci}.} \bibinfo{year}{2023}\natexlab{}.
\newblock \showarticletitle{ChatGPT for good? On opportunities and challenges of large language models for education}.
\newblock \bibinfo{journal}{\emph{Learning and Individual Differences}} (\bibinfo{year}{2023}).
\newblock
\urldef\tempurl%
\url{https://api.semanticscholar.org/CorpusID:257445349}
\showURL{%
\tempurl}


\bibitem[Kewalramani et~al\mbox{.}(2024)]%
        {educsci14060604}
\bibfield{author}{\bibinfo{person}{Sarika Kewalramani}, \bibinfo{person}{George Aranda}, \bibinfo{person}{Jiqing Sun}, \bibinfo{person}{Gerarda Richards}, \bibinfo{person}{Linda Hobbs}, \bibinfo{person}{Lihua Xu}, \bibinfo{person}{Victoria Millar}, \bibinfo{person}{Belinda Dealy}, {and} \bibinfo{person}{Bridgette Van~Leuven}.} \bibinfo{year}{2024}\natexlab{}.
\newblock \showarticletitle{A Systematic Review of the Role of Multimodal Resources for Inclusive STEM Engagement in Early-Childhood Education}.
\newblock \bibinfo{journal}{\emph{Education Sciences}} \bibinfo{volume}{14}, \bibinfo{number}{6} (\bibinfo{year}{2024}).
\newblock
\showISSN{2227-7102}
\urldef\tempurl%
\url{https://doi.org/10.3390/educsci14060604}
\showDOI{\tempurl}


\bibitem[Kosmas et~al\mbox{.}(2023)]%
        {kosmas2023storytelling}
\bibfield{author}{\bibinfo{person}{Chatziioannidis Kosmas}, \bibinfo{person}{Antoniadis Ioannis}, {and} \bibinfo{person}{Fragulis George}.} \bibinfo{year}{2023}\natexlab{}.
\newblock \showarticletitle{Storytelling archetypes in gamification}. In \bibinfo{booktitle}{\emph{AIP Conference Proceedings}}, Vol.~\bibinfo{volume}{2909}. AIP Publishing.
\newblock


\bibitem[Kucirkova(2022)]%
        {Kucirkova02012022}
\bibfield{author}{\bibinfo{person}{Natalia Kucirkova}.} \bibinfo{year}{2022}\natexlab{}.
\newblock \showarticletitle{Children’s agency and reading with story-apps: considerations of design, behavioural and social dimensions}.
\newblock \bibinfo{journal}{\emph{Qualitative Research in Psychology}} \bibinfo{volume}{19}, \bibinfo{number}{1} (\bibinfo{year}{2022}), \bibinfo{pages}{66--90}.
\newblock
\urldef\tempurl%
\url{https://doi.org/10.1080/14780887.2018.1545065}
\showDOI{\tempurl}
\showeprint{https://doi.org/10.1080/14780887.2018.1545065}


\bibitem[Kucirkova and Flewitt(2020)]%
        {FuturegazingPotentialDigital}
\bibfield{author}{\bibinfo{person}{Natalia Kucirkova} {and} \bibinfo{person}{Rosie Flewitt}.} \bibinfo{year}{2020}\natexlab{}.
\newblock \showarticletitle{The future-gazing potential of digital personalization in young children’s reading: views from education professionals and app designers}.
\newblock \bibinfo{journal}{\emph{Early Child Development and Care}} \bibinfo{volume}{190}, \bibinfo{number}{2} (\bibinfo{year}{2020}), \bibinfo{pages}{135--149}.
\newblock
\urldef\tempurl%
\url{https://doi.org/10.1080/03004430.2018.1458718}
\showDOI{\tempurl}
\showeprint{https://doi.org/10.1080/03004430.2018.1458718}


\bibitem[Kucirkova et~al\mbox{.}(2021)]%
        {kucirkova2021empirical}
\bibfield{author}{\bibinfo{person}{Natalia Kucirkova}, \bibinfo{person}{Merideth Gattis}, \bibinfo{person}{Thomas~P Spargo}, \bibinfo{person}{Beatriz~Seisdedos de Vega}, {and} \bibinfo{person}{Rosie Flewitt}.} \bibinfo{year}{2021}\natexlab{}.
\newblock \showarticletitle{An empirical investigation of parent-child shared reading of digital personalized books}.
\newblock \bibinfo{journal}{\emph{International Journal of Educational Research}}  \bibinfo{volume}{105} (\bibinfo{year}{2021}), \bibinfo{pages}{101710}.
\newblock


\bibitem[Kucirkova et~al\mbox{.}(2017)]%
        {kucirkova2017young}
\bibfield{author}{\bibinfo{person}{Natalia Kucirkova}, \bibinfo{person}{Karen Littleton}, {and} \bibinfo{person}{Teresa Cremin}.} \bibinfo{year}{2017}\natexlab{}.
\newblock \showarticletitle{Young children’s reading for pleasure with digital books: six key facets of engagement}.
\newblock \bibinfo{journal}{\emph{Cambridge Journal of Education}} \bibinfo{volume}{47}, \bibinfo{number}{1} (\bibinfo{year}{2017}), \bibinfo{pages}{67--84}.
\newblock


\bibitem[Lee et~al\mbox{.}(2024)]%
        {osos}
\bibfield{author}{\bibinfo{person}{Jungeun Lee}, \bibinfo{person}{Suwon Yoon}, \bibinfo{person}{Kyoosik Lee}, \bibinfo{person}{Eunae Jeong}, \bibinfo{person}{Jae-Eun Cho}, \bibinfo{person}{Wonjeong Park}, \bibinfo{person}{Dongsun Yim}, {and} \bibinfo{person}{Inseok Hwang}.} \bibinfo{year}{2024}\natexlab{}.
\newblock \showarticletitle{Open Sesame? Open Salami! Personalizing Vocabulary Assessment-Intervention for Children via Pervasive Profiling and Bespoke Storybook Generation}. In \bibinfo{booktitle}{\emph{Proceedings of the CHI Conference on Human Factors in Computing Systems}}. \bibinfo{pages}{1--32}.
\newblock


\bibitem[Lee and Bowen(2006)]%
        {DC1_familycontext_2_and_parentedu_1}
\bibfield{author}{\bibinfo{person}{Jung-Sook Lee} {and} \bibinfo{person}{Natasha~K Bowen}.} \bibinfo{year}{2006}\natexlab{}.
\newblock \showarticletitle{Parent involvement, cultural capital, and the achievement gap among elementary school children}.
\newblock \bibinfo{journal}{\emph{American educational research journal}} \bibinfo{volume}{43}, \bibinfo{number}{2} (\bibinfo{year}{2006}), \bibinfo{pages}{193--218}.
\newblock


\bibitem[Lyu et~al\mbox{.}(2023)]%
        {lyu2023llm}
\bibfield{author}{\bibinfo{person}{Hanjia Lyu}, \bibinfo{person}{Song Jiang}, \bibinfo{person}{Hanqing Zeng}, \bibinfo{person}{Yinglong Xia}, \bibinfo{person}{Qifan Wang}, \bibinfo{person}{Si Zhang}, \bibinfo{person}{Ren Chen}, \bibinfo{person}{Christopher Leung}, \bibinfo{person}{Jiajie Tang}, {and} \bibinfo{person}{Jiebo Luo}.} \bibinfo{year}{2023}\natexlab{}.
\newblock \showarticletitle{Llm-rec: Personalized recommendation via prompting large language models}.
\newblock \bibinfo{journal}{\emph{arXiv preprint arXiv:2307.15780}} (\bibinfo{year}{2023}).
\newblock


\bibitem[Moedt and Holmes(2020)]%
        {moedt2020effects}
\bibfield{author}{\bibinfo{person}{Kelly Moedt} {and} \bibinfo{person}{Robyn~M Holmes}.} \bibinfo{year}{2020}\natexlab{}.
\newblock \showarticletitle{The effects of purposeful play after shared storybook readings on kindergarten children's reading comprehension, creativity, and language skills and abilities}.
\newblock \bibinfo{journal}{\emph{Early Child Development and Care}} \bibinfo{volume}{190}, \bibinfo{number}{6} (\bibinfo{year}{2020}), \bibinfo{pages}{839--854}.
\newblock


\bibitem[Neuman et~al\mbox{.}(2016)]%
        {neuman2016improving}
\bibfield{author}{\bibinfo{person}{Susan~B Neuman}, \bibinfo{person}{Tanya Kaefer}, {and} \bibinfo{person}{Ashley~M Pinkham}.} \bibinfo{year}{2016}\natexlab{}.
\newblock \showarticletitle{Improving low-income preschoolers’ word and world knowledge: The effects of content-rich instruction}.
\newblock \bibinfo{journal}{\emph{The Elementary School Journal}} \bibinfo{volume}{116}, \bibinfo{number}{4} (\bibinfo{year}{2016}), \bibinfo{pages}{652--674}.
\newblock


\bibitem[Onuorah(2020)]%
        {onuorah_efficacy_2020}
\bibfield{author}{\bibinfo{person}{Adaorah~R Onuorah}.} \bibinfo{year}{2020}\natexlab{}.
\newblock \showarticletitle{Efficacy of digital storytelling intervention on social skills acquisition among primary school children}.
\newblock \bibinfo{journal}{\emph{Journal of Engineering and Applied Sciences}} \bibinfo{volume}{15}, \bibinfo{number}{12} (\bibinfo{year}{2020}), \bibinfo{pages}{2562--2567}.
\newblock


\bibitem[OpenAI(2023)]%
        {openaiGPT4TechnicalReport2023}
\bibfield{author}{\bibinfo{person}{OpenAI}.} \bibinfo{year}{2023}\natexlab{}.
\newblock \bibinfo{title}{{{GPT-4 Technical Report}}}.
\newblock
\newblock
\urldef\tempurl%
\url{https://doi.org/10.48550/arXiv.2303.08774}
\showDOI{\tempurl}
\showeprint[arxiv]{2303.08774}~[cs]


\bibitem[Pedaste et~al\mbox{.}(2015)]%
        {pedaste2015phases}
\bibfield{author}{\bibinfo{person}{Margus Pedaste}, \bibinfo{person}{Mario M{\"a}eots}, \bibinfo{person}{Leo~A Siiman}, \bibinfo{person}{Ton De~Jong}, \bibinfo{person}{Siswa~AN Van~Riesen}, \bibinfo{person}{Ellen~T Kamp}, \bibinfo{person}{Constantinos~C Manoli}, \bibinfo{person}{Zacharias~C Zacharia}, {and} \bibinfo{person}{Eleftheria Tsourlidaki}.} \bibinfo{year}{2015}\natexlab{}.
\newblock \showarticletitle{Phases of inquiry-based learning: Definitions and the inquiry cycle}.
\newblock \bibinfo{journal}{\emph{Educational research review}}  \bibinfo{volume}{14} (\bibinfo{year}{2015}), \bibinfo{pages}{47--61}.
\newblock


\bibitem[Petrie et~al\mbox{.}(2023)]%
        {petrieParentchildInteractionStorybook2023}
\bibfield{author}{\bibinfo{person}{Abigail Petrie}, \bibinfo{person}{Robert Mayr}, \bibinfo{person}{Fei Zhao}, {and} \bibinfo{person}{Simona Montanari}.} \bibinfo{year}{2023}\natexlab{}.
\newblock \showarticletitle{Parent-Child Interaction during Storybook Reading: Wordless Narrative Books versus Books with Text}.
\newblock \bibinfo{journal}{\emph{Journal of Child Language}} \bibinfo{volume}{50}, \bibinfo{number}{1} (\bibinfo{date}{Jan.} \bibinfo{year}{2023}), \bibinfo{pages}{104--131}.
\newblock
\showISSN{0305-0009, 1469-7602}
\urldef\tempurl%
\url{https://doi.org/10.1017/S0305000921000763}
\showDOI{\tempurl}


\bibitem[plc(2017)]%
        {hsbc2017value}
\bibfield{author}{\bibinfo{person}{HSBC~Holdings plc}.} \bibinfo{year}{2017}\natexlab{}.
\newblock \bibinfo{booktitle}{\emph{The Value of Education: Higher and Higher}}.
\newblock
\urldef\tempurl%
\url{https://www.hsbc.com/-/files/hsbc/media/media-release/2017/170628-the-value-of-education-higher-and-higher-global-report.pdf}
\showURL{%
Retrieved December 10, 2024 from \tempurl}


\bibitem[Qu et~al\mbox{.}(2019)]%
        {qu2019user}
\bibfield{author}{\bibinfo{person}{Chen Qu}, \bibinfo{person}{Liu Yang}, \bibinfo{person}{W~Bruce Croft}, \bibinfo{person}{Yongfeng Zhang}, \bibinfo{person}{Johanne~R Trippas}, {and} \bibinfo{person}{Minghui Qiu}.} \bibinfo{year}{2019}\natexlab{}.
\newblock \showarticletitle{User intent prediction in information-seeking conversations}. In \bibinfo{booktitle}{\emph{Proceedings of the 2019 Conference on Human Information Interaction and Retrieval}}. \bibinfo{pages}{25--33}.
\newblock


\bibitem[Roseli and Aziz(2023)]%
        {roseli2023affective}
\bibfield{author}{\bibinfo{person}{Nur Hazwani~Mohamad Roseli} {and} \bibinfo{person}{Nurulnadwan Aziz}.} \bibinfo{year}{2023}\natexlab{}.
\newblock \showarticletitle{Affective Design Components of Educational Application for Children}.
\newblock \bibinfo{journal}{\emph{Information Management and Business Review}} \bibinfo{volume}{15}, \bibinfo{number}{4 (SI) I} (\bibinfo{year}{2023}), \bibinfo{pages}{191--199}.
\newblock


\bibitem[Rubegni et~al\mbox{.}(2021)]%
        {rubegniGirlWhoWants2021}
\bibfield{author}{\bibinfo{person}{Elisa Rubegni}, \bibinfo{person}{Rebecca Dore}, \bibinfo{person}{Monica Landoni}, {and} \bibinfo{person}{Ling Kan}.} \bibinfo{year}{2021}\natexlab{}.
\newblock \showarticletitle{``{{The}} Girl Who Wants to Fly'': {{Exploring}} the Role of Digital Technology in Enhancing Dialogic Reading}.
\newblock \bibinfo{journal}{\emph{International Journal of Child-Computer Interaction}}  \bibinfo{volume}{30} (\bibinfo{year}{2021}).
\newblock
\showISSN{2212-8697}
\urldef\tempurl%
\url{https://doi.org/10.1016/j.ijcci.2020.100239}
\showDOI{\tempurl}


\bibitem[Sagae(2021)]%
        {sagae2021tracking}
\bibfield{author}{\bibinfo{person}{Kenji Sagae}.} \bibinfo{year}{2021}\natexlab{}.
\newblock \showarticletitle{Tracking child language development with neural network language models}.
\newblock \bibinfo{journal}{\emph{Frontiers in Psychology}}  \bibinfo{volume}{12} (\bibinfo{year}{2021}), \bibinfo{pages}{674402}.
\newblock


\bibitem[Sak(2015)]%
        {sak2015young}
\bibfield{author}{\bibinfo{person}{Ramazan Sak}.} \bibinfo{year}{2015}\natexlab{}.
\newblock \showarticletitle{Young children’s difficult questions and adults’ answers}.
\newblock \bibinfo{journal}{\emph{The Anthropologist}} \bibinfo{volume}{22}, \bibinfo{number}{2} (\bibinfo{year}{2015}), \bibinfo{pages}{293--300}.
\newblock


\bibitem[Saracho(2017)]%
        {saracho_parents_2017}
\bibfield{author}{\bibinfo{person}{Olivia~N. Saracho}.} \bibinfo{year}{2017}\natexlab{}.
\newblock \showarticletitle{Parents’ shared storybook reading – learning to read}.
\newblock \bibinfo{journal}{\emph{Early Child Development and Care}} \bibinfo{volume}{187}, \bibinfo{number}{3-4} (\bibinfo{date}{April} \bibinfo{year}{2017}), \bibinfo{pages}{554--567}.
\newblock
\showISSN{0300-4430, 1476-8275}
\urldef\tempurl%
\url{https://doi.org/10.1080/03004430.2016.1261514}
\showDOI{\tempurl}


\bibitem[Seo et~al\mbox{.}(2024)]%
        {chacha}
\bibfield{author}{\bibinfo{person}{Woosuk Seo}, \bibinfo{person}{Chanmo Yang}, {and} \bibinfo{person}{Young-Ho Kim}.} \bibinfo{year}{2024}\natexlab{}.
\newblock \showarticletitle{ChaCha: Leveraging Large Language Models to Prompt Children to Share Their Emotions about Personal Events}. In \bibinfo{booktitle}{\emph{Proceedings of the CHI Conference on Human Factors in Computing Systems}} (Honolulu, HI, USA) \emph{(\bibinfo{series}{CHI '24})}. \bibinfo{publisher}{Association for Computing Machinery}, \bibinfo{address}{New York, NY, USA}, Article \bibinfo{articleno}{903}, \bibinfo{numpages}{20}~pages.
\newblock
\showISBNx{9798400703300}
\urldef\tempurl%
\url{https://doi.org/10.1145/3613904.3642152}
\showDOI{\tempurl}


\bibitem[Shen et~al\mbox{.}(2023)]%
        {shen2023chatgpt}
\bibfield{author}{\bibinfo{person}{Xinyue Shen}, \bibinfo{person}{Zeyuan Chen}, \bibinfo{person}{Michael Backes}, {and} \bibinfo{person}{Yang Zhang}.} \bibinfo{year}{2023}\natexlab{}.
\newblock \showarticletitle{In chatgpt we trust? measuring and characterizing the reliability of chatgpt}.
\newblock \bibinfo{journal}{\emph{arXiv preprint arXiv:2304.08979}} (\bibinfo{year}{2023}).
\newblock


\bibitem[Sperry and Sperry(2021)]%
        {DC1_familycontext_1}
\bibfield{author}{\bibinfo{person}{Linda~L Sperry} {and} \bibinfo{person}{Douglas~E Sperry}.} \bibinfo{year}{2021}\natexlab{}.
\newblock \showarticletitle{Entering into the story: Implications for emergent literacy}.
\newblock \bibinfo{journal}{\emph{Frontiers in Psychology}}  \bibinfo{volume}{12} (\bibinfo{year}{2021}), \bibinfo{pages}{665092}.
\newblock


\bibitem[States(2013)]%
        {national2013next}
\bibfield{author}{\bibinfo{person}{NGSS~Lead States}.} \bibinfo{year}{2013}\natexlab{}.
\newblock \bibinfo{booktitle}{\emph{Next generation science standards: For states, by states}}.
\newblock \bibinfo{publisher}{National Academies Press}.
\newblock


\bibitem[Suardi et~al\mbox{.}(2021)]%
        {suardi2021curiosity}
\bibfield{author}{\bibinfo{person}{Mahara Suardi}, \bibinfo{person}{Leli Kurniawati}, {and} \bibinfo{person}{Yeni Rachmawati}.} \bibinfo{year}{2021}\natexlab{}.
\newblock \showarticletitle{Curiosity in young children}. In \bibinfo{booktitle}{\emph{5th International Conference on Early Childhood Education (ICECE 2020)}}. Atlantis Press, \bibinfo{pages}{224--228}.
\newblock


\bibitem[Sun et~al\mbox{.}(2024)]%
        {sun2024exploring}
\bibfield{author}{\bibinfo{person}{Yuling Sun}, \bibinfo{person}{Jiaju Chen}, \bibinfo{person}{Bingsheng Yao}, \bibinfo{person}{Jiali Liu}, \bibinfo{person}{Dakuo Wang}, \bibinfo{person}{Xiaojuan Ma}, \bibinfo{person}{Yuxuan Lu}, \bibinfo{person}{Ying Xu}, {and} \bibinfo{person}{Liang He}.} \bibinfo{year}{2024}\natexlab{}.
\newblock \showarticletitle{Exploring Parent's Needs for Children-Centered AI to Support Preschoolers' Interactive Storytelling and Reading Activities}.
\newblock \bibinfo{journal}{\emph{Proc. ACM Hum.-Comput. Interact.}} \bibinfo{volume}{8}, \bibinfo{number}{CSCW2}, Article \bibinfo{articleno}{496} (\bibinfo{date}{Nov.} \bibinfo{year}{2024}), \bibinfo{numpages}{25}~pages.
\newblock
\urldef\tempurl%
\url{https://doi.org/10.1145/3687035}
\showDOI{\tempurl}


\bibitem[Talaat(2023)]%
        {talaat2023real}
\bibfield{author}{\bibinfo{person}{Fatma~M Talaat}.} \bibinfo{year}{2023}\natexlab{}.
\newblock \showarticletitle{Real-time facial emotion recognition system among children with autism based on deep learning and IoT}.
\newblock \bibinfo{journal}{\emph{Neural Computing and Applications}} \bibinfo{volume}{35}, \bibinfo{number}{17} (\bibinfo{year}{2023}), \bibinfo{pages}{12717--12728}.
\newblock


\bibitem[Touvron et~al\mbox{.}(2023)]%
        {Touvron2023Llama2O}
\bibfield{author}{\bibinfo{person}{Hugo Touvron}, \bibinfo{person}{Louis Martin}, \bibinfo{person}{Kevin~R. Stone}, \bibinfo{person}{Peter Albert}, \bibinfo{person}{Amjad Almahairi}, \bibinfo{person}{Yasmine Babaei}, \bibinfo{person}{Nikolay Bashlykov}, \bibinfo{person}{Soumya Batra}, \bibinfo{person}{Prajjwal Bhargava}, \bibinfo{person}{Shruti Bhosale}, \bibinfo{person}{Daniel~M. Bikel}, \bibinfo{person}{Lukas Blecher}, \bibinfo{person}{Cristian~Cant{\'o}n Ferrer}, \bibinfo{person}{Moya Chen}, \bibinfo{person}{Guillem Cucurull}, \bibinfo{person}{David Esiobu}, \bibinfo{person}{Jude Fernandes}, \bibinfo{person}{Jeremy Fu}, \bibinfo{person}{Wenyin Fu}, \bibinfo{person}{Brian Fuller}, \bibinfo{person}{Cynthia Gao}, \bibinfo{person}{Vedanuj Goswami}, \bibinfo{person}{Naman Goyal}, \bibinfo{person}{Anthony~S. Hartshorn}, \bibinfo{person}{Saghar Hosseini}, \bibinfo{person}{Rui Hou}, \bibinfo{person}{Hakan Inan}, \bibinfo{person}{Marcin Kardas}, \bibinfo{person}{Viktor Kerkez}, \bibinfo{person}{Madian Khabsa},
  \bibinfo{person}{Isabel~M. Kloumann}, \bibinfo{person}{A.~V. Korenev}, \bibinfo{person}{Punit~Singh Koura}, \bibinfo{person}{Marie-Anne Lachaux}, \bibinfo{person}{Thibaut Lavril}, \bibinfo{person}{Jenya Lee}, \bibinfo{person}{Diana Liskovich}, \bibinfo{person}{Yinghai Lu}, \bibinfo{person}{Yuning Mao}, \bibinfo{person}{Xavier Martinet}, \bibinfo{person}{Todor Mihaylov}, \bibinfo{person}{Pushkar Mishra}, \bibinfo{person}{Igor Molybog}, \bibinfo{person}{Yixin Nie}, \bibinfo{person}{Andrew Poulton}, \bibinfo{person}{Jeremy Reizenstein}, \bibinfo{person}{Rashi Rungta}, \bibinfo{person}{Kalyan Saladi}, \bibinfo{person}{Alan Schelten}, \bibinfo{person}{Ruan Silva}, \bibinfo{person}{Eric~Michael Smith}, \bibinfo{person}{R. Subramanian}, \bibinfo{person}{Xia Tan}, \bibinfo{person}{Binh Tang}, \bibinfo{person}{Ross Taylor}, \bibinfo{person}{Adina Williams}, \bibinfo{person}{Jian~Xiang Kuan}, \bibinfo{person}{Puxin Xu}, \bibinfo{person}{Zhengxu Yan}, \bibinfo{person}{Iliyan Zarov}, \bibinfo{person}{Yuchen Zhang},
  \bibinfo{person}{Angela Fan}, \bibinfo{person}{Melanie Kambadur}, \bibinfo{person}{Sharan Narang}, \bibinfo{person}{Aurelien Rodriguez}, \bibinfo{person}{Robert Stojnic}, \bibinfo{person}{Sergey Edunov}, {and} \bibinfo{person}{Thomas Scialom}.} \bibinfo{year}{2023}\natexlab{}.
\newblock \showarticletitle{Llama 2: Open Foundation and Fine-Tuned Chat Models}.
\newblock \bibinfo{journal}{\emph{ArXiv}}  \bibinfo{volume}{abs/2307.09288} (\bibinfo{year}{2023}).
\newblock
\urldef\tempurl%
\url{https://api.semanticscholar.org/CorpusID:259950998}
\showURL{%
\tempurl}


\bibitem[Towson et~al\mbox{.}(2017)]%
        {towsonDialogicReadingEarly2017}
\bibfield{author}{\bibinfo{person}{Jacqueline~A. Towson}, \bibinfo{person}{Angel Fettig}, \bibinfo{person}{Veronica~P. Fleury}, {and} \bibinfo{person}{Diana~L. Abarca}.} \bibinfo{year}{2017}\natexlab{}.
\newblock \showarticletitle{Dialogic {{Reading}} in {{Early Childhood Settings}}: {{A Summary}} of the {{Evidence Base}}}.
\newblock \bibinfo{journal}{\emph{Topics in Early Childhood Special Education}} (\bibinfo{date}{Sept.} \bibinfo{year}{2017}).
\newblock
\urldef\tempurl%
\url{https://doi.org/10.1177/0271121417724875}
\showDOI{\tempurl}


\bibitem[van Dijk et~al\mbox{.}(2023)]%
        {van-dijk-etal-2023-chiscor}
\bibfield{author}{\bibinfo{person}{Bram van Dijk}, \bibinfo{person}{Max van Duijn}, \bibinfo{person}{Suzan Verberne}, {and} \bibinfo{person}{Marco Spruit}.} \bibinfo{year}{2023}\natexlab{}.
\newblock \showarticletitle{{C}hi{SC}or: A Corpus of Freely-Told Fantasy Stories by {D}utch Children for Computational Linguistics and Cognitive Science}. In \bibinfo{booktitle}{\emph{Proceedings of the 27th Conference on Computational Natural Language Learning (CoNLL)}}, \bibfield{editor}{\bibinfo{person}{Jing Jiang}, \bibinfo{person}{David Reitter}, {and} \bibinfo{person}{Shumin Deng}} (Eds.). \bibinfo{publisher}{Association for Computational Linguistics}, \bibinfo{address}{Singapore}, \bibinfo{pages}{352--363}.
\newblock
\urldef\tempurl%
\url{https://doi.org/10.18653/v1/2023.conll-1.23}
\showDOI{\tempurl}


\bibitem[Vermette et~al\mbox{.}(2019)]%
        {personalize_challenge}
\bibfield{author}{\bibinfo{person}{Laton Vermette}, \bibinfo{person}{Joanna McGrenere}, \bibinfo{person}{Colin Birge}, \bibinfo{person}{Adam Kelly}, {and} \bibinfo{person}{Parmit~K. Chilana}.} \bibinfo{year}{2019}\natexlab{}.
\newblock \showarticletitle{Freedom to Personalize My Digital Classroom: Understanding Teachers' Practices and Motivations}. In \bibinfo{booktitle}{\emph{Proceedings of the 2019 CHI Conference on Human Factors in Computing Systems}} (Glasgow, Scotland Uk) \emph{(\bibinfo{series}{CHI '19})}. \bibinfo{publisher}{Association for Computing Machinery}, \bibinfo{address}{New York, NY, USA}, \bibinfo{pages}{1–14}.
\newblock
\showISBNx{9781450359702}
\urldef\tempurl%
\url{https://doi.org/10.1145/3290605.3300548}
\showDOI{\tempurl}


\bibitem[Vygotsky(1978)]%
        {vygotsky1978mind}
\bibfield{author}{\bibinfo{person}{Lev~Semenovich Vygotsky}.} \bibinfo{year}{1978}\natexlab{}.
\newblock \bibinfo{booktitle}{\emph{Mind in society: The development of higher psychological processes}}. Vol.~\bibinfo{volume}{86}.
\newblock \bibinfo{publisher}{Harvard university press}.
\newblock


\bibitem[Wang et~al\mbox{.}(2019)]%
        {wang2019children}
\bibfield{author}{\bibinfo{person}{Ge Wang}, \bibinfo{person}{Jun Zhao}, {and} \bibinfo{person}{Nigel Shadbolt}.} \bibinfo{year}{2019}\natexlab{}.
\newblock \showarticletitle{Are Children Fully Aware of Online Privacy Risks and How Can We Improve Their Coping Ability?}
\newblock \bibinfo{journal}{\emph{arXiv preprint arXiv:1902.02635}} (\bibinfo{year}{2019}).
\newblock


\bibitem[Wang et~al\mbox{.}(2023)]%
        {wang2023DigitalAutonomy}
\bibfield{author}{\bibinfo{person}{Ge Wang}, \bibinfo{person}{Jun Zhao}, \bibinfo{person}{Max Van~Kleek}, {and} \bibinfo{person}{Nigel Shadbolt}.} \bibinfo{year}{2023}\natexlab{}.
\newblock \showarticletitle{12 Ways to Empower: Designing for Children’s Digital Autonomy}. In \bibinfo{booktitle}{\emph{Proceedings of the 2023 CHI Conference on Human Factors in Computing Systems}}. \bibinfo{pages}{1--27}.
\newblock


\bibitem[Wei et~al\mbox{.}(2022)]%
        {wei2022chainofthought}
\bibfield{author}{\bibinfo{person}{Jason Wei}, \bibinfo{person}{Xuezhi Wang}, \bibinfo{person}{Dale Schuurmans}, \bibinfo{person}{Maarten Bosma}, \bibinfo{person}{Fei Xia}, \bibinfo{person}{Ed Chi}, \bibinfo{person}{Quoc~V Le}, \bibinfo{person}{Denny Zhou}, {et~al\mbox{.}}} \bibinfo{year}{2022}\natexlab{}.
\newblock \showarticletitle{Chain-of-thought prompting elicits reasoning in large language models}.
\newblock \bibinfo{journal}{\emph{Advances in neural information processing systems}}  \bibinfo{volume}{35} (\bibinfo{year}{2022}), \bibinfo{pages}{24824--24837}.
\newblock


\bibitem[Westlund and Breazeal(2015)]%
        {CAlimit}
\bibfield{author}{\bibinfo{person}{Jacqueline~Kory Westlund} {and} \bibinfo{person}{Cynthia Breazeal}.} \bibinfo{year}{2015}\natexlab{}.
\newblock \showarticletitle{The Interplay of Robot Language Level with Children's Language Learning during Storytelling}. In \bibinfo{booktitle}{\emph{Proceedings of the Tenth Annual ACM/IEEE International Conference on Human-Robot Interaction Extended Abstracts}} (Portland, Oregon, USA) \emph{(\bibinfo{series}{HRI'15 Extended Abstracts})}. \bibinfo{publisher}{Association for Computing Machinery}, \bibinfo{address}{New York, NY, USA}, \bibinfo{pages}{65–66}.
\newblock
\showISBNx{9781450333184}
\urldef\tempurl%
\url{https://doi.org/10.1145/2701973.2701989}
\showDOI{\tempurl}


\bibitem[Whalley et~al\mbox{.}(2021)]%
        {whalley2021towards}
\bibfield{author}{\bibinfo{person}{Brian Whalley}, \bibinfo{person}{Derek France}, \bibinfo{person}{Julian Park}, \bibinfo{person}{Alice Mauchline}, {and} \bibinfo{person}{Katharine Welsh}.} \bibinfo{year}{2021}\natexlab{}.
\newblock \showarticletitle{Towards flexible personalized learning and the future educational system in the fourth industrial revolution in the wake of Covid-19}.
\newblock \bibinfo{journal}{\emph{Higher Education Pedagogies}} \bibinfo{volume}{6}, \bibinfo{number}{1} (\bibinfo{year}{2021}), \bibinfo{pages}{79--99}.
\newblock


\bibitem[What-Works-Clearinghouse({[n.\,d.]})]%
        {WWCPreparingYoung}
\bibfield{author}{\bibinfo{person}{What-Works-Clearinghouse}.} \bibinfo{year}{[n.\,d.]}\natexlab{}.
\newblock \bibinfo{booktitle}{\emph{{{WWC}} {\textbar} {{Preparing Young Children}} for {{School}}}}.
\newblock
\urldef\tempurl%
\url{https://ies.ed.gov/ncee/wwc/PracticeGuide/30}
\showURL{%
\tempurl}


\bibitem[Whitehurst et~al\mbox{.}(1988)]%
        {whitehurstAcceleratingLanguageDevelopment1988}
\bibfield{author}{\bibinfo{person}{G.~J. Whitehurst}, \bibinfo{person}{F.~L. Falco}, \bibinfo{person}{C.~J. Lonigan}, \bibinfo{person}{J.~E. Fischel}, \bibinfo{person}{B.~D. DeBaryshe}, \bibinfo{person}{M.~C. {Valdez-Menchaca}}, {and} \bibinfo{person}{M. Caulfield}.} \bibinfo{year}{1988}\natexlab{}.
\newblock \showarticletitle{Accelerating Language Development through Picture Book Reading.}
\newblock \bibinfo{journal}{\emph{Developmental Psychology}} \bibinfo{volume}{24}, \bibinfo{number}{4} (\bibinfo{year}{1988}), \bibinfo{pages}{552--559}.
\newblock
\showISSN{1939-0599(Electronic),0012-1649(Print)}
\urldef\tempurl%
\url{https://doi.org/10.1037/0012-1649.24.4.552}
\showDOI{\tempurl}


\bibitem[Xiao et~al\mbox{.}(2023)]%
        {bge_embedding}
\bibfield{author}{\bibinfo{person}{Shitao Xiao}, \bibinfo{person}{Zheng Liu}, \bibinfo{person}{Peitian Zhang}, {and} \bibinfo{person}{Niklas Muennighoff}.} \bibinfo{year}{2023}\natexlab{}.
\newblock \bibinfo{title}{C-Pack: Packaged Resources To Advance General Chinese Embedding}.
\newblock
\newblock
\showeprint[arxiv]{2309.07597}~[cs.CL]


\bibitem[Xu et~al\mbox{.}(2024)]%
        {xu2024talk2care}
\bibfield{author}{\bibinfo{person}{Xuhai Xu}, \bibinfo{person}{Bingsheng Yao}, \bibinfo{person}{Ziqi Yang}, \bibinfo{person}{Shao Zhang}, \bibinfo{person}{Ethan Rogers}, \bibinfo{person}{Stephen Intille}, \bibinfo{person}{Nawar Shara}, \bibinfo{person}{Guodong Gao}, {and} \bibinfo{person}{Dakuo Wang}.} \bibinfo{year}{2024}\natexlab{}.
\newblock \showarticletitle{Talk2Care: Facilitating Asynchronous Patient-Provider Communication with Large-Language-Model}. In \bibinfo{booktitle}{\emph{Proceedings of the AAAI Symposium Series}}, Vol.~\bibinfo{volume}{4}. \bibinfo{pages}{146--151}.
\newblock


\bibitem[Xu et~al\mbox{.}(2021a)]%
        {CAlimit2}
\bibfield{author}{\bibinfo{person}{Ying Xu}, \bibinfo{person}{Stacy Branham}, \bibinfo{person}{Xinwei Deng}, \bibinfo{person}{Penelope Collins}, {and} \bibinfo{person}{Mark Warschauer}.} \bibinfo{year}{2021}\natexlab{a}.
\newblock \showarticletitle{Are Current Voice Interfaces Designed to Support Children’s Language Development?}. In \bibinfo{booktitle}{\emph{Proceedings of the 2021 CHI Conference on Human Factors in Computing Systems}} (Yokohama, Japan) \emph{(\bibinfo{series}{CHI '21})}. \bibinfo{publisher}{Association for Computing Machinery}, \bibinfo{address}{New York, NY, USA}, Article \bibinfo{articleno}{633}, \bibinfo{numpages}{12}~pages.
\newblock
\showISBNx{9781450380966}
\urldef\tempurl%
\url{https://doi.org/10.1145/3411764.3445271}
\showDOI{\tempurl}


\bibitem[Xu et~al\mbox{.}(2023)]%
        {xuRositaReadsMy2023a}
\bibfield{author}{\bibinfo{person}{Ying Xu}, \bibinfo{person}{Kunlei He}, \bibinfo{person}{Valery Vigil}, \bibinfo{person}{Santiago {Ojeda-Ramirez}}, \bibinfo{person}{Xuechen Liu}, \bibinfo{person}{Julian Levine}, \bibinfo{person}{Kelsyann Cervera}, {and} \bibinfo{person}{Mark Warschauer}.} \bibinfo{year}{2023}\natexlab{}.
\newblock \showarticletitle{``{{Rosita Reads With My Family}}'': {{Developing A Bilingual Conversational Agent}} to {{Support Parent-Child Shared Reading}}}. In \bibinfo{booktitle}{\emph{Proceedings of the 22nd {{Annual ACM Interaction Design}} and {{Children Conference}}}}. \bibinfo{publisher}{ACM}, \bibinfo{address}{Chicago IL USA}, \bibinfo{pages}{160--172}.
\newblock
\showISBNx{9798400701313}
\urldef\tempurl%
\url{https://doi.org/10.1145/3585088.3589354}
\showDOI{\tempurl}


\bibitem[Xu et~al\mbox{.}(2022a)]%
        {Elinor}
\bibfield{author}{\bibinfo{person}{Ying Xu}, \bibinfo{person}{Valery Vigil}, \bibinfo{person}{Andres~S. Bustamante}, {and} \bibinfo{person}{Mark Warschauer}.} \bibinfo{year}{2022}\natexlab{a}.
\newblock \showarticletitle{“Elinor’s Talking to Me!”:Integrating Conversational AI into Children’s Narrative Science Programming}. In \bibinfo{booktitle}{\emph{Proceedings of the 2022 CHI Conference on Human Factors in Computing Systems}} (New Orleans, LA, USA) \emph{(\bibinfo{series}{CHI '22})}. \bibinfo{publisher}{Association for Computing Machinery}, \bibinfo{address}{New York, NY, USA}, Article \bibinfo{articleno}{166}, \bibinfo{numpages}{16}~pages.
\newblock
\showISBNx{9781450391573}
\urldef\tempurl%
\url{https://doi.org/10.1145/3491102.3502050}
\showDOI{\tempurl}


\bibitem[Xu et~al\mbox{.}(2021b)]%
        {xuSameBenefitsDifferent2021}
\bibfield{author}{\bibinfo{person}{Ying Xu}, \bibinfo{person}{Dakuo Wang}, \bibinfo{person}{Penelope Collins}, \bibinfo{person}{Hyelim Lee}, {and} \bibinfo{person}{Mark Warschauer}.} \bibinfo{year}{2021}\natexlab{b}.
\newblock \showarticletitle{Same Benefits, Different Communication Patterns: {{Comparing Children}}'s Reading with a Conversational Agent vs. a Human Partner}.
\newblock \bibinfo{journal}{\emph{Computers \& Education}}  \bibinfo{volume}{161} (\bibinfo{date}{Feb.} \bibinfo{year}{2021}), \bibinfo{pages}{104059}.
\newblock
\showISSN{0360-1315}
\urldef\tempurl%
\url{https://doi.org/10.1016/j.compedu.2020.104059}
\showDOI{\tempurl}


\bibitem[Xu et~al\mbox{.}(2022b)]%
        {fairytaleqa}
\bibfield{author}{\bibinfo{person}{Ying Xu}, \bibinfo{person}{Dakuo Wang}, \bibinfo{person}{Mo Yu}, \bibinfo{person}{Daniel Ritchie}, \bibinfo{person}{Bingsheng Yao}, \bibinfo{person}{Tongshuang Wu}, \bibinfo{person}{Zheng Zhang}, \bibinfo{person}{Toby Li}, \bibinfo{person}{Nora Bradford}, \bibinfo{person}{Branda Sun}, \bibinfo{person}{Tran Hoang}, \bibinfo{person}{Yisi Sang}, \bibinfo{person}{Yufang Hou}, \bibinfo{person}{Xiaojuan Ma}, \bibinfo{person}{Diyi Yang}, \bibinfo{person}{Nanyun Peng}, \bibinfo{person}{Zhou Yu}, {and} \bibinfo{person}{Mark Warschauer}.} \bibinfo{year}{2022}\natexlab{b}.
\newblock \showarticletitle{Fantastic Questions and Where to Find Them: {F}airytale{QA} {--} An Authentic Dataset for Narrative Comprehension}. In \bibinfo{booktitle}{\emph{Proceedings of the 60th Annual Meeting of the Association for Computational Linguistics (Volume 1: Long Papers)}}, \bibfield{editor}{\bibinfo{person}{Smaranda Muresan}, \bibinfo{person}{Preslav Nakov}, {and} \bibinfo{person}{Aline Villavicencio}} (Eds.). \bibinfo{publisher}{Association for Computational Linguistics}, \bibinfo{address}{Dublin, Ireland}, \bibinfo{pages}{447--460}.
\newblock
\urldef\tempurl%
\url{https://doi.org/10.18653/v1/2022.acl-long.34}
\showDOI{\tempurl}


\bibitem[Xu and Warschauer(2019)]%
        {xu2019young}
\bibfield{author}{\bibinfo{person}{Ying Xu} {and} \bibinfo{person}{Mark Warschauer}.} \bibinfo{year}{2019}\natexlab{}.
\newblock \showarticletitle{Young children's reading and learning with conversational agents}. In \bibinfo{booktitle}{\emph{Extended abstracts of the 2019 CHI conference on human factors in computing systems}}. \bibinfo{pages}{1--8}.
\newblock


\bibitem[Xu and Warschauer(2020)]%
        {whatareyoutalkingto}
\bibfield{author}{\bibinfo{person}{Ying Xu} {and} \bibinfo{person}{Mark Warschauer}.} \bibinfo{year}{2020}\natexlab{}.
\newblock \showarticletitle{What Are You Talking To?: Understanding Children's Perceptions of Conversational Agents}. In \bibinfo{booktitle}{\emph{Proceedings of the 2020 CHI Conference on Human Factors in Computing Systems}} (Honolulu, HI, USA) \emph{(\bibinfo{series}{CHI '20})}. \bibinfo{publisher}{Association for Computing Machinery}, \bibinfo{address}{New York, NY, USA}, \bibinfo{pages}{1–13}.
\newblock
\showISBNx{9781450367080}
\urldef\tempurl%
\url{https://doi.org/10.1145/3313831.3376416}
\showDOI{\tempurl}


\bibitem[Yao et~al\mbox{.}(2022)]%
        {yaoItAITurn2022}
\bibfield{author}{\bibinfo{person}{Bingsheng Yao}, \bibinfo{person}{Dakuo Wang}, \bibinfo{person}{Tongshuang Wu}, \bibinfo{person}{Zheng Zhang}, \bibinfo{person}{Toby Li}, \bibinfo{person}{Mo Yu}, {and} \bibinfo{person}{Ying Xu}.} \bibinfo{year}{2022}\natexlab{}.
\newblock \showarticletitle{It Is {{AI}}'s {{Turn}} to {{Ask Humans}} a {{Question}}: {{Question-Answer Pair Generation}} for {{Children}}'s {{Story Books}}}. In \bibinfo{booktitle}{\emph{Proceedings of the 60th {{Annual Meeting}} of the {{Association}} for {{Computational Linguistics}} ({{Volume}} 1: {{Long Papers}})}}, \bibfield{editor}{\bibinfo{person}{Smaranda Muresan}, \bibinfo{person}{Preslav Nakov}, {and} \bibinfo{person}{Aline Villavicencio}} (Eds.). \bibinfo{publisher}{Association for Computational Linguistics}, \bibinfo{address}{Dublin, Ireland}, \bibinfo{pages}{731--744}.
\newblock
\urldef\tempurl%
\url{https://doi.org/10.18653/v1/2022.acl-long.54}
\showDOI{\tempurl}


\bibitem[Yoon and Bak(2023)]%
        {yoonDiversityEnhancedNarrative2023}
\bibfield{author}{\bibinfo{person}{Hokeun Yoon} {and} \bibinfo{person}{JinYeong Bak}.} \bibinfo{year}{2023}\natexlab{}.
\newblock \showarticletitle{Diversity {{Enhanced Narrative Question Generation}} for {{Storybooks}}}. In \bibinfo{booktitle}{\emph{Proceedings of the 2023 {{Conference}} on {{Empirical Methods}} in {{Natural Language Processing}}}}, \bibfield{editor}{\bibinfo{person}{Houda Bouamor}, \bibinfo{person}{Juan Pino}, {and} \bibinfo{person}{Kalika Bali}} (Eds.). \bibinfo{publisher}{Association for Computational Linguistics}, \bibinfo{address}{Singapore}, \bibinfo{pages}{465--482}.
\newblock
\urldef\tempurl%
\url{https://doi.org/10.18653/v1/2023.emnlp-main.31}
\showDOI{\tempurl}


\bibitem[Zhang et~al\mbox{.}(2024)]%
        {Mathemyths}
\bibfield{author}{\bibinfo{person}{Chao Zhang}, \bibinfo{person}{Xuechen Liu}, \bibinfo{person}{Katherine Ziska}, \bibinfo{person}{Soobin Jeon}, \bibinfo{person}{Chi-Lin Yu}, {and} \bibinfo{person}{Ying Xu}.} \bibinfo{year}{2024}\natexlab{}.
\newblock \showarticletitle{Mathemyths: Leveraging Large Language Models to Teach Mathematical Language through Child-AI Co-Creative Storytelling}. In \bibinfo{booktitle}{\emph{Proceedings of the CHI Conference on Human Factors in Computing Systems}} (Honolulu, HI, USA) \emph{(\bibinfo{series}{CHI '24})}. \bibinfo{publisher}{Association for Computing Machinery}, \bibinfo{address}{New York, NY, USA}, Article \bibinfo{articleno}{274}, \bibinfo{numpages}{23}~pages.
\newblock
\showISBNx{9798400703300}
\urldef\tempurl%
\url{https://doi.org/10.1145/3613904.3642647}
\showDOI{\tempurl}


\bibitem[Zhang et~al\mbox{.}(2022b)]%
        {storydrawer}
\bibfield{author}{\bibinfo{person}{Chao Zhang}, \bibinfo{person}{Cheng Yao}, \bibinfo{person}{Jiayi Wu}, \bibinfo{person}{Weijia Lin}, \bibinfo{person}{Lijuan Liu}, \bibinfo{person}{Ge Yan}, {and} \bibinfo{person}{Fangtian Ying}.} \bibinfo{year}{2022}\natexlab{b}.
\newblock \showarticletitle{StoryDrawer: a child--AI collaborative drawing system to support children's creative visual storytelling}. In \bibinfo{booktitle}{\emph{Proceedings of the 2022 CHI conference on human factors in computing systems}}. \bibinfo{pages}{1--15}.
\newblock


\bibitem[Zhang et~al\mbox{.}(2023c)]%
        {zhang2023text}
\bibfield{author}{\bibinfo{person}{Chenshuang Zhang}, \bibinfo{person}{Chaoning Zhang}, \bibinfo{person}{Mengchun Zhang}, {and} \bibinfo{person}{In~So Kweon}.} \bibinfo{year}{2023}\natexlab{c}.
\newblock \showarticletitle{Text-to-image diffusion models in generative ai: A survey}.
\newblock \bibinfo{journal}{\emph{arXiv preprint arXiv:2303.07909}} (\bibinfo{year}{2023}).
\newblock


\bibitem[Zhang and Aslan(2021)]%
        {zhang2021ai}
\bibfield{author}{\bibinfo{person}{Ke Zhang} {and} \bibinfo{person}{Ayse~Begum Aslan}.} \bibinfo{year}{2021}\natexlab{}.
\newblock \showarticletitle{AI technologies for education: Recent research \& future directions}.
\newblock \bibinfo{journal}{\emph{Computers and Education: Artificial Intelligence}}  \bibinfo{volume}{2} (\bibinfo{year}{2021}), \bibinfo{pages}{100025}.
\newblock


\bibitem[Zhang et~al\mbox{.}(2023d)]%
        {zhang2023memory}
\bibfield{author}{\bibinfo{person}{Kai Zhang}, \bibinfo{person}{Fubang Zhao}, \bibinfo{person}{Yangyang Kang}, {and} \bibinfo{person}{Xiaozhong Liu}.} \bibinfo{year}{2023}\natexlab{d}.
\newblock \showarticletitle{Memory-augmented llm personalization with short-and long-term memory coordination}.
\newblock \bibinfo{journal}{\emph{arXiv preprint arXiv:2309.11696}} (\bibinfo{year}{2023}).
\newblock


\bibitem[Zhang et~al\mbox{.}(2023b)]%
        {Zhang2023RetrieveAT}
\bibfield{author}{\bibinfo{person}{Peitian Zhang}, \bibinfo{person}{Shitao Xiao}, \bibinfo{person}{Zheng Liu}, \bibinfo{person}{Zhicheng Dou}, {and} \bibinfo{person}{Jian-Yun Nie}.} \bibinfo{year}{2023}\natexlab{b}.
\newblock \showarticletitle{Retrieve Anything To Augment Large Language Models}.
\newblock \bibinfo{journal}{\emph{ArXiv}}  \bibinfo{volume}{abs/2310.07554} (\bibinfo{year}{2023}).
\newblock
\urldef\tempurl%
\url{https://api.semanticscholar.org/CorpusID:263835099}
\showURL{%
\tempurl}


\bibitem[Zhang et~al\mbox{.}(2023a)]%
        {zhang2023siren}
\bibfield{author}{\bibinfo{person}{Yue Zhang}, \bibinfo{person}{Yafu Li}, \bibinfo{person}{Leyang Cui}, \bibinfo{person}{Deng Cai}, \bibinfo{person}{Lemao Liu}, \bibinfo{person}{Tingchen Fu}, \bibinfo{person}{Xinting Huang}, \bibinfo{person}{Enbo Zhao}, \bibinfo{person}{Yu Zhang}, \bibinfo{person}{Yulong Chen}, {et~al\mbox{.}}} \bibinfo{year}{2023}\natexlab{a}.
\newblock \showarticletitle{Siren's song in the AI ocean: a survey on hallucination in large language models}.
\newblock \bibinfo{journal}{\emph{arXiv preprint arXiv:2309.01219}} (\bibinfo{year}{2023}).
\newblock


\bibitem[Zhang et~al\mbox{.}(2022a)]%
        {zhangStoryBuddyHumanAICollaborative2022a}
\bibfield{author}{\bibinfo{person}{Zheng Zhang}, \bibinfo{person}{Ying Xu}, \bibinfo{person}{Yanhao Wang}, \bibinfo{person}{Bingsheng Yao}, \bibinfo{person}{Daniel Ritchie}, \bibinfo{person}{Tongshuang Wu}, \bibinfo{person}{Mo Yu}, \bibinfo{person}{Dakuo Wang}, {and} \bibinfo{person}{Toby Jia-Jun Li}.} \bibinfo{year}{2022}\natexlab{a}.
\newblock \showarticletitle{{{StoryBuddy}}: {{A Human-AI Collaborative Chatbot}} for {{Parent-Child Interactive Storytelling}} with {{Flexible Parental Involvement}}}. In \bibinfo{booktitle}{\emph{Proceedings of the 2022 {{CHI Conference}} on {{Human Factors}} in {{Computing Systems}}}} \emph{(\bibinfo{series}{{{CHI}} '22})}. \bibinfo{publisher}{Association for Computing Machinery}, \bibinfo{address}{New York, NY, USA}, \bibinfo{pages}{1--21}.
\newblock
\showISBNx{978-1-4503-9157-3}
\urldef\tempurl%
\url{https://doi.org/10.1145/3491102.3517479}
\showDOI{\tempurl}


\bibitem[Zhong et~al\mbox{.}(2022)]%
        {zhong2022should}
\bibfield{author}{\bibinfo{person}{Baichang Zhong}, \bibinfo{person}{Xiaofan Liu}, \bibinfo{person}{Zehui Zhan}, \bibinfo{person}{Qingchao Ke}, {and} \bibinfo{person}{Fulai Wang}.} \bibinfo{year}{2022}\natexlab{}.
\newblock \showarticletitle{What should a Chinese top-level design in STEM Education look like?}
\newblock \bibinfo{journal}{\emph{Humanities and Social Sciences Communications}} \bibinfo{volume}{9}, \bibinfo{number}{1} (\bibinfo{year}{2022}), \bibinfo{pages}{1--8}.
\newblock


\bibitem[Zhong et~al\mbox{.}(2024)]%
        {zhong2024memorybank}
\bibfield{author}{\bibinfo{person}{Wanjun Zhong}, \bibinfo{person}{Lianghong Guo}, \bibinfo{person}{Qiqi Gao}, \bibinfo{person}{He Ye}, {and} \bibinfo{person}{Yanlin Wang}.} \bibinfo{year}{2024}\natexlab{}.
\newblock \showarticletitle{Memorybank: Enhancing large language models with long-term memory}. In \bibinfo{booktitle}{\emph{Proceedings of the AAAI Conference on Artificial Intelligence}}, Vol.~\bibinfo{volume}{38}. \bibinfo{pages}{19724--19731}.
\newblock


\end{thebibliography}

\balance
\appendix
\clearpage
\onecolumn

\section{Questions for Children's Use Experience of \toolname}
\label{app:question}
\begin{enumerate}
    \item \jiaju{Do you like this tool, and why?}
    \item \jiaju{Do you find using this tool interesting, and why?}
    \item \jiaju{What do you think about the approach of learning while reading a story?}
    \item \jiaju{What do you think about this talking robot? Would you like to have a partner like this?}
\end{enumerate}

\section{Example of Conversation Snippets Between Children and \toolname}
\label{example}
This section provides several examples of children's interactions with \toolname. The Greeting phase is shown in Table~\ref{tab:greet_dialogue_1} and Table~\ref{tab:greet_dialogue_2} (see Appendix~\ref{app: eg_greeting}). For the Dialogic Reading phase (Appendix~\ref{app: eg_reading}), Table~\ref{tab:chatbot_water_dialogue} and Table~\ref{tab:chatbot_sunlight_dialogue} illustrate conversations that promote active thinking, enriched with relevant real-world knowledge. Finally, Table~\ref{tab:chatbot_summary_dialogue} presents the conversation that follows after the child finishes reading the story.

\subsection{Greeting}
\label{app: eg_greeting}
\begin{table*}[h]
    \centering
    \begin{tabular}{c|p{0.88\textwidth}}
    \toprule
    \multirow{2}{*}{\parbox{1.5cm}{\centering\includegraphics[width=0.8cm]{imgs/penguin.png}}} & \textit{Hi there, little friend! My name is Sparky, and I’m your reading companion. I can tell you many interesting stories. Can you tell me your name?} \\
    \arrayrulecolor[HTML]{778899}
    \midrule
    \arrayrulecolor{black}
        Child: & \textit{My name is ...} \\
    \arrayrulecolor[HTML]{778899}
    \midrule
    \arrayrulecolor{black}
        \parbox{1.5cm}{\centering\includegraphics[width=0.8cm]{imgs/penguin.png}} & \textit{Hello, (child's name). It’s so nice to meet you! How old are you?} \\
    \arrayrulecolor[HTML]{778899}
    \midrule
    \arrayrulecolor{black}
        Child: & \textit{I’m six years old.} \\
    \arrayrulecolor[HTML]{778899}
    \midrule
    \arrayrulecolor{black}
    \multirow{2}{*}{\parbox{1.5cm}{\centering\includegraphics[width=0.8cm]{imgs/penguin.png}}} & \textit{Wow, being six years old must mean you’re very smart! Do you have any favorite topics? Like space, princesses, dinosaurs, or cars? You can tell me about anything you like!} \\
        \arrayrulecolor[HTML]{778899}
    \midrule
    \arrayrulecolor{black}
        Child: & \textit{I like Princess My Little Pony and Elsa.} \\
    \arrayrulecolor[HTML]{778899}
    \midrule
    \arrayrulecolor{black}
    \multirow{4}{*}{\parbox{1.5cm}{\centering\includegraphics[width=0.8cm]{imgs/penguin.png}}} & \textit{Wow, you like Princess My Little Pony and Elsa! They’re both very brave and kind-hearted princesses!}
    \textbf{[Introduction of reading activity]} \textit{Next, we’ll switch to story-reading mode. You can also switch to dialogic reading mode by clicking the button at the bottom right of the screen. In dialogic reading mode, we’ll explore the story and knowledge together, and you can answer questions by clicking the buttons on the screen. Are you ready? Let’s start reading!} \\
    \bottomrule
    \end{tabular}
    \caption{Greeting dialogue between \toolname\ and a child.}
    \label{tab:greet_dialogue_1}
\end{table*}

\begin{table*}[h]
    \centering
    \begin{tabular}{c|p{0.88\textwidth}}
    \toprule
    \multirow{2}{*}{\parbox{1.5cm}{\centering\includegraphics[width=0.8cm]{imgs/penguin.png}}} & \textit{Hi there, little friend! My name is Sparky, and I’m your reading companion. I can tell you many interesting stories. Can you tell me your name?} \\
    \arrayrulecolor[HTML]{778899}
    \midrule
    \arrayrulecolor{black}
        Child: & \textit{My name is ...} \\
    \arrayrulecolor[HTML]{778899}
    \midrule
    \arrayrulecolor{black}
        \parbox{1.5cm}{\centering\includegraphics[width=0.8cm]{imgs/penguin.png}} & \textit{Hi, (child's name)! So nice to meet you! Can you tell me what's your age now?} \\
    \arrayrulecolor[HTML]{778899}
    \midrule
    \arrayrulecolor{black}
        Child: & \textit{I’m six years old.} \\
    \arrayrulecolor[HTML]{778899}
    \midrule
    \arrayrulecolor{black}
    \multirow{2}{*}{\parbox{1.5cm}{\centering\includegraphics[width=0.8cm]{imgs/penguin.png}}} & \textit{Wow, six-year-old (child's name) must know a lot of things already! Do you have any favorite topics? Like space, princesses, dinosaurs, or cars? You can tell me about anything you like!} \\
    \arrayrulecolor[HTML]{778899}
    \midrule
    \arrayrulecolor{black}
        Child: & \textit{I like Peppa Pig.} \\
    \arrayrulecolor[HTML]{778899}
    \midrule
    \arrayrulecolor{black}
    \parbox{1.5cm}{\centering\includegraphics[width=0.8cm]{imgs/penguin.png}} & \textit{Peppa Pig is such a fun cartoon character! (child's name), it must be fun to like her! \textbf{[Introduction of reading activity]}} \\
    \bottomrule
    \end{tabular}
    \caption{Greeting dialogue between \toolname\ and a child.}
    \label{tab:greet_dialogue_2}
\end{table*}

\newpage
\subsection{Dialogic Reading}
\label{app: eg_reading}

%Interactive dialogue between a chatbot and a child about water forms
\begin{table*}[!htbp]
    \centering
    \begin{tabular}{>{\raggedright\arraybackslash}p{3cm}|p{0.8\textwidth}} % Left-align for first 5 rows
    \toprule
    \multirow{3}{*}{\textbf{Story Text}} & The friends in Dinosaur Valley all admired David the T-Rex because he had seen the ocean. This summer, David became the leader and led his friends on a journey to the \textcolor[HTML]{F94A29}{seaside}. They passed through the Colorful Mountain, the Rocky Mountain, and the Monster Mountain before finally arriving at the \textcolor[HTML]{F94A29}{ocean}. \\
    \arrayrulecolor[HTML]{778899}
    \midrule
    \arrayrulecolor{black}
     \makecell[l]{\textbf{\colorbox[HTML]{86A7FC}{Child's Age}}} & \textcolor[HTML]{86A7FC}{Eight years old} \\
     \makecell[l]{\textbf{\colorbox[HTML]{FF9843}{Child's Interest}}} & \textcolor[HTML]{FF9843}{Little animals} \\
    \arrayrulecolor[HTML]{778899}
    \midrule
    \arrayrulecolor{black}
    \makecell[l]{\textbf{\colorbox[HTML]{708FC2}{Matched Knowledge}}} & \textcolor[HTML]{3468C0}{Water is found in the ocean, rivers, lakes, and ponds. Water exists as solid ice and in liquid form.} \\
     \makecell[l]{\textbf{\colorbox[HTML]{86A7FC}{Knowledge Level}}} & \textcolor[HTML]{86A7FC}{Second Grade} \\
    \midrule
    \midrule
    \end{tabular}
% From this point onward, center-align the first column
    \begin{tabular}{>{\centering\arraybackslash}p{3cm}|p{0.8\textwidth}} % Center-align for remaining rows
    \multirow{3}{*}{\parbox{1.5cm}{\centering\includegraphics[width=0.8cm]{imgs/penguin.png}}} & \textbf{[Opening]} \textit{Hello, (child's name)! I’m Sparky, and today we’re going to read a fantastic story together.} \textbf{[Story Context]} \textit{In the story, T-Rex David leads his friends to the \textcolor[HTML]{F94A29}{seaside}.} \textbf{[Integrating Child's Interest]} \textit{\textcolor[HTML]{3468C0}{Do you know the difference between the water \textcolor[HTML]{FF9843}{little animals} drink and the water in the} \textcolor[HTML]{F94A29}{ocean}?} \\
    \arrayrulecolor[HTML]{778899}
    \midrule
    \arrayrulecolor{black}
        Child: & \textit{I think the water in the ocean is salty.} \\
    \arrayrulecolor[HTML]{778899}
    \midrule
    \arrayrulecolor{black}
    \multirow{2}{*}{\parbox{1.5cm}{\centering\includegraphics[width=0.8cm]{imgs/penguin.png}} }& \textit{Absolutely correct! \textcolor[HTML]{FF9843}{The water little animals drink is usually freshwater}, while ocean water is salty because it contains salt. Now, let's think together: \textcolor[HTML]{3468C0}{What would happen if the ice cubes in the freezer melted?}} \\
    \arrayrulecolor[HTML]{778899}
    \midrule
    \arrayrulecolor{black}
        Child: & \textit{Turn into water.} \\
    \arrayrulecolor[HTML]{778899}
    \midrule
    \arrayrulecolor{black}
    \multirow{2}{*}{\parbox{1.5cm}{\centering\includegraphics[width=0.8cm]{imgs/penguin.png}}} & \textit{Great job, (child's name)! Yes,} \textbf{[Extending to Real-World Knowledge]} \textit{\textcolor[HTML]{3468C0}{ice is the solid form of water, and when it melts and warms up, it becomes the liquid water we commonly see. Do you know that there is ice in the ocean too?}} \\
    \arrayrulecolor[HTML]{778899}
    \midrule
    \arrayrulecolor{black}
        Child: & \textit{I don't know.} \\
    \arrayrulecolor[HTML]{778899}
    \midrule
    \arrayrulecolor{black}
    \multirow{3}{*}{\parbox{1.5cm}{\centering\includegraphics[width=0.8cm]{imgs/penguin.png}}} & \textbf{[Scaffolding]} \textit{No worries, (child's name), we can learn together. \textcolor[HTML]{3468C0}{At Earth's two poles, the seawater freezes into ice because of the cold, so there is indeed ice in the ocean}. We’ve learned so much about water already! Let’s continue reading!} \\
    \bottomrule
    \end{tabular}
    \caption{Interactive dialogue between a chatbot and a child about water forms}
    \label{tab:chatbot_water_dialogue}
\end{table*}

%Interactive dialogue between a chatbot and a child about sunlight and shade
\begin{table*}[!htbp]
    \centering
    \begin{tabular}{>{\raggedright\arraybackslash}p{3cm}|p{0.8\textwidth}} % Left-align for first 5 rows
    \toprule
    \multirow{2}{*}{\textbf{Story Text}} & ``The \textcolor[HTML]{F94A29}{sunset} is so beautiful!'' the friends couldn't help but admire. Under the afterglow of the \textcolor[HTML]{F94A29}{sunset}, they happily set up their tents. \\
     \arrayrulecolor[HTML]{778899}
    \midrule
    \arrayrulecolor{black}
     \textbf{\colorbox[HTML]{86A7FC}{Child's Age}} & \textcolor[HTML]{86A7FC}{Six years old} \\
    \textbf{\colorbox[HTML]{FF9843}{Child's Interest}} & \textcolor[HTML]{FF9843}{Peppa Pig} \\
     \arrayrulecolor[HTML]{778899}
    \midrule
    \arrayrulecolor{black}
     \textbf{\colorbox[HTML]{708FC2}{Matched Knowledge}} & \textcolor[HTML]{3468C0}{Sunlight warms Earth’s surface.} \\
     \textbf{\colorbox[HTML]{86A7FC}{Knowledge Level}} & \textcolor[HTML]{86A7FC}{Kindergarten} \\
    \midrule
    \midrule
    \end{tabular}

    \begin{tabular}{>{\centering\arraybackslash}p{3cm}|p{0.8\textwidth}}
    \multirow{3}{*}{\parbox{2cm}{\centering\includegraphics[width=0.8cm]{imgs/penguin.png}}} & \textbf{[Opening]} \textit{Hello, (child's name)!} \textbf{[Story Context]} \textit{In the story, they pitched their tent under the \textcolor[HTML]{F94A29}{sunset} glow.} \textit{\textcolor[HTML]{3468C0}{Do you know why they chose this time to set up the tent?}} \textbf{[Integrating Child's Interest]} \textit{\textcolor[HTML]{FF9843}{Just like Peppa Pig and her friends, they also love building houses outdoors for fun.}} \\
    \arrayrulecolor[HTML]{778899}
    \midrule
    \arrayrulecolor{black}
        Child: & \textit{They set up the tent to sleep at night.} \\
    \arrayrulecolor[HTML]{778899}
    \midrule
    \arrayrulecolor{black}
    \multirow{4}{*}{\parbox{1.5cm}{\centering\includegraphics[width=0.8cm]{imgs/penguin.png}}} & \textit{You’re partly correct, (child's name). Great answer!} \textbf{[Extending to Real-World Knowledge]} \textit{They also set up the tent to \textcolor[HTML]{3468C0}{avoid direct sunlight, much like how we use sun umbrellas.} Let’s explore this question together: \textcolor[HTML]{FF9843}{Just like Peppa Pig and her friends playing in different places,} \textcolor[HTML]{3468C0}{if you put your hand in a sunny spot and a shady spot, what do you feel?} } \\
    \arrayrulecolor[HTML]{778899}
    \midrule
    \arrayrulecolor{black}
        Child: & \textit{In the sun, it feels hot; in the shade, it feels cool.} \\
    \arrayrulecolor[HTML]{778899}
    \midrule
    \arrayrulecolor{black}
    \multirow{2}{*}{\parbox{1.5cm}{\centering\includegraphics[width=0.8cm]{imgs/penguin.png}}} & \textbf{[Encouraging Feedback]} \textit{That’s absolutely correct, (child's name). Well done! \textcolor[HTML]{3468C0}{Yes, sunlight warms the ground and makes it feel warm to us.}} \\
    \bottomrule
    \end{tabular}
    \caption{Interactive dialogue between a chatbot and a child about sunlight and shade}
    \label{tab:chatbot_sunlight_dialogue}
\end{table*}

%Interactive dialogue between a chatbot and a child after completing a story
\begin{table*}[!htbp]
    \centering
    \begin{tabular}{>{\raggedright\arraybackslash}p{2.3cm}|p{0.83\textwidth}} % Left-align for first 5 rows
    \toprule
    \multirow{6}{*}{\textbf{Story Summary}} & The friends in Dinosaur Valley, led by David the T-Rex, went camping by the sea. Through their journey, they learned about the tidal phenomenon. When the tide went out, they happily collected seashells and caught marine creatures, but they almost found themselves in danger as the tide came in. It was only thanks to Nan Nan’s reminder that they realized the danger of the incoming tide and quickly retreated. They then enjoyed a rich seafood feast together. This story teaches children to observe natural phenomena, pay attention to safety warnings, and work together with others.\\
     \arrayrulecolor[HTML]{778899}
    \midrule
    \arrayrulecolor{black}
     \textbf{\colorbox[HTML]{86A7FC}{Child's Age}} & \textcolor[HTML]{86A7FC}{Six years old} \\
    \textbf{\colorbox[HTML]{FF9843}{Child's Interest}} & \textcolor[HTML]{FF9843}{Fairies} \\
    \midrule
    \midrule
    \end{tabular}

    \begin{tabular}{>{\centering\arraybackslash}p{2.3cm}|p{0.83\textwidth}}
        \multirow{2}{*}{\parbox{1.5cm}{\centering\includegraphics[width=0.8cm]{imgs/penguin.png}}} & \textbf{[Opening]} \textit{Hello, (child's name)! You’re amazing! Now that we’ve finished the story, let’s think about it together.} \textbf{[Story Context]} \textit{The story mentions the ebb and flow of tides. When did the little dinosaurs collect seashells?} \\
    \arrayrulecolor[HTML]{778899}
    \midrule
    \arrayrulecolor{black}
        Child: & \textit{It seems like it was when the tide was going out.} \\
    \arrayrulecolor[HTML]{778899}
    \midrule
    \arrayrulecolor{black}
        \multirow{3}{*}{\parbox{1.5cm}{\centering\includegraphics[width=0.8cm]{imgs/penguin.png}}} & \textit{Well done! When the tide goes out, the sea level lowers, and the little dinosaurs were able to collect seashells.} \textbf{[Integrating Child's Interest]} \textit{\textcolor[HTML]{FF9843}{It's like fairy's magic!} Let's explore this phenomenon together: Which areas get flooded during high tide? Try to think about it.} \\
    \arrayrulecolor[HTML]{778899}
    \midrule
    \arrayrulecolor{black}
        Child: & \textit{I am not sure.} \\
    \arrayrulecolor[HTML]{778899}
    \midrule
    \arrayrulecolor{black}
        \multirow{3}{*}{\parbox{1.5cm}{\centering\includegraphics[width=0.8cm]{imgs/penguin.png}}} & \textbf{[Scaffolding]} \textit{It's ok, (child's name). We can think together! During high tide, the seawater floods the beach. Just like the little dinosaurs in the story, observing tides is important. \textcolor[HTML]{FF9843}{Imagine a fairy casting magic in the sea. Would it make the tides more magical?}} \\
    \arrayrulecolor[HTML]{778899}
    \midrule
    \arrayrulecolor{black}
        Child: & \textit{The seawater might change into other colors.} \\
    \arrayrulecolor[HTML]{778899}
    \midrule
    \arrayrulecolor{black}
        \multirow{2}{*}{\parbox{1.5cm}{\centering\includegraphics[width=0.8cm]{imgs/penguin.png}}} & \textit{Wow, that’s so interesting! The fairy's magic might make the seawater become colorful! Thank you for sharing these imaginative ideas with me!} \\
    \bottomrule
    \end{tabular}
    \caption{Interactive dialogue between a chatbot and a child after completing a story.}
    \label{tab:chatbot_summary_dialogue}
\end{table*}

\newpage
\section{Technical Details of \toolname}
\label{app: technical process}
\begin{figure*}[h]
    \centering
    \twiceinclude[width=0.78\linewidth]{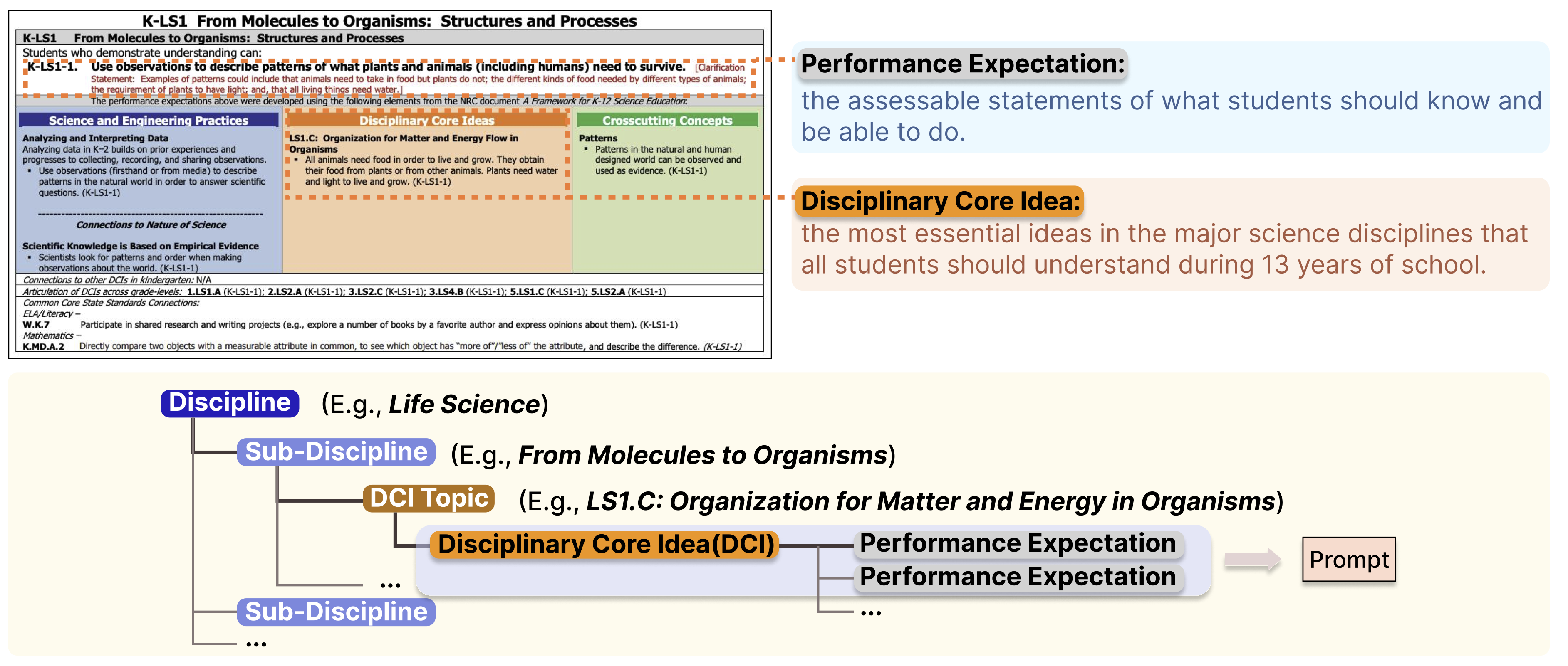}
    \caption{An example of the architecture of the Next Generation Science Standards (NGSS)~\cite{national2013next}. 
    % The sentences in the upper part of the table are \textit{Performance Expectations}, and the orange box represents the \textit{Disciplinary Core Ideas (DCIs)}. 
    We manually organize the \textit{DCIs} alongside their corresponding \textit{Performance Expectations} to build our knowledge source.}
    \vspace{-16px}
    \label{fig:NGSS}
\end{figure*}

\begin{figure*}
    \centering
    \twiceinclude[width=0.7\linewidth]{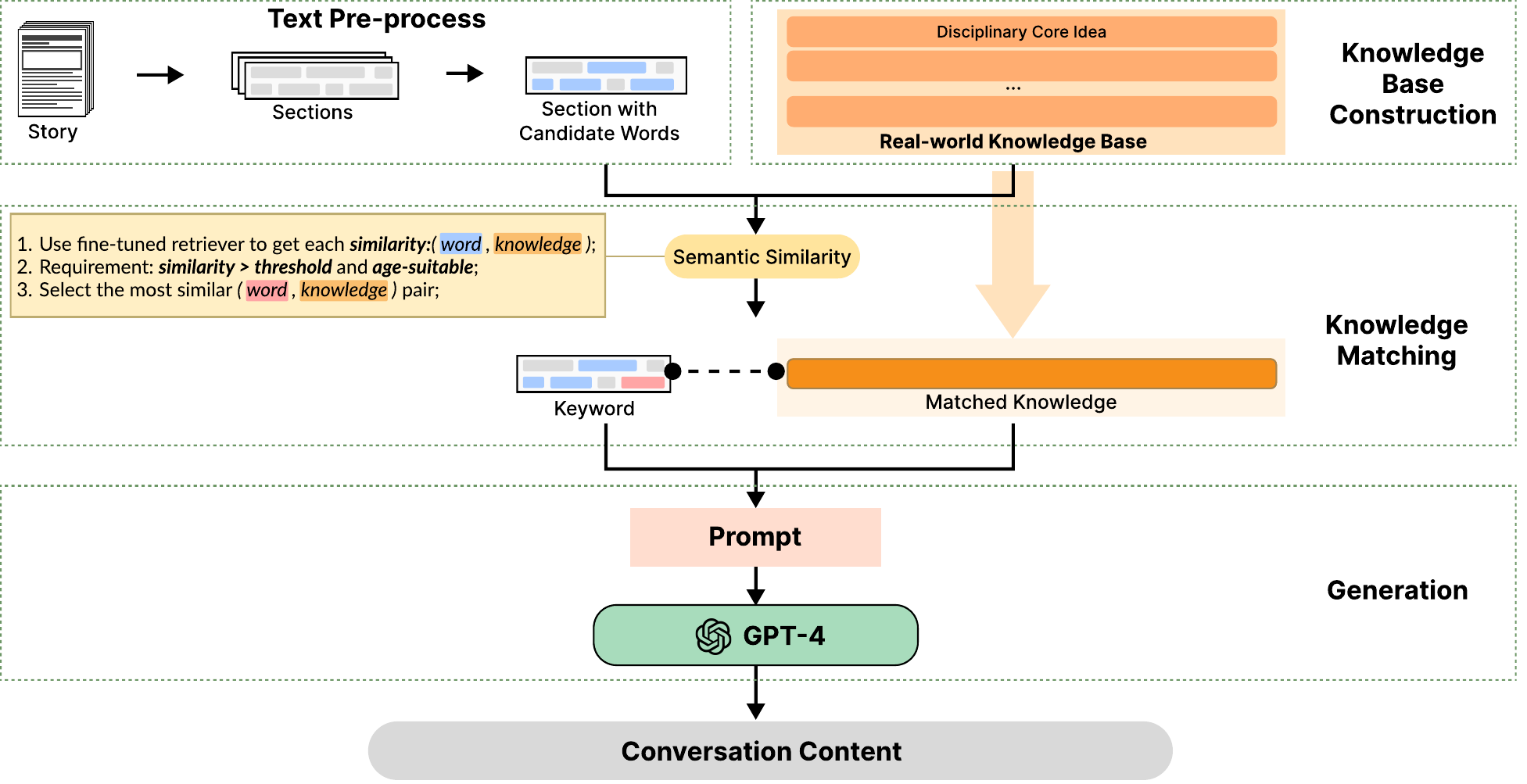}
    \caption{The technical process behind \toolname's knowledge infusion. We use a fine-tuned retriever to identify the most educationally appropriate piece of knowledge and keyword. 
    The word-knowledge pair is then integrated into the prompt. }
    \label{fig:rag}
\end{figure*}

Technically, the knowledge infusion in \toolname\ is achieved using Retrieval-Augmented Generation (RAG). 
\jiaju{Due to the lack of a standardized curriculum for this age group in China~\cite{zhong2022should}, and considering many parents' interest in U.S. educational content~\cite{hsbc2017value}, we chose the Next Generation Science Standards (NGSS) as the knowledge base for our study. NGSS's age-specific content is well-aligned with our educational goals. To adapt it for use in China, we translated the content and conducted multiple rounds of internal evaluations with experts to ensure its educational suitability.}
We constructed a knowledge graph using the Disciplinary Core Ideas (DCIs) and Performance Expectations from NGSS. The detailed structure of the DCIs and Performance Expectations we utilized can be found in Figure~\ref{fig:NGSS}.
Then, we use our knowledge-matching algorithm to retrieve the most educationally appropriate piece of knowledge for each story section.
Specifically, we use a BGE retriever~\cite{Zhang2023RetrieveAT} fine-tuned on StorySparkQA~\cite{chen2023fairytalecqa} to retrieve educationally appropriate pieces of knowledge for each keyword in story sections. 
To ensure that children are not overwhelmed by too much story-associated knowledge, we conducted several rounds of internal evaluations to set a threshold to control the relevance and amount of associated knowledge with keywords.
Finally, for each identified keyword and matched knowledge pair, we input them into GPT-4 for further conversation generation. This process is illustrated in Figure~\ref{fig:rag}.

For story-based conversations that do not incorporate external knowledge, we prompt GPT-4 to summarize the story narrative, providing an overall context. This summarization is then used as additional input to generate conversations focused on story comprehension.

\clearpage
\section{Prompts for \toolname}
\label{app:prompt}

\begin{figure*}[h]
    \centering
    \twiceinclude[width=0.8\linewidth]{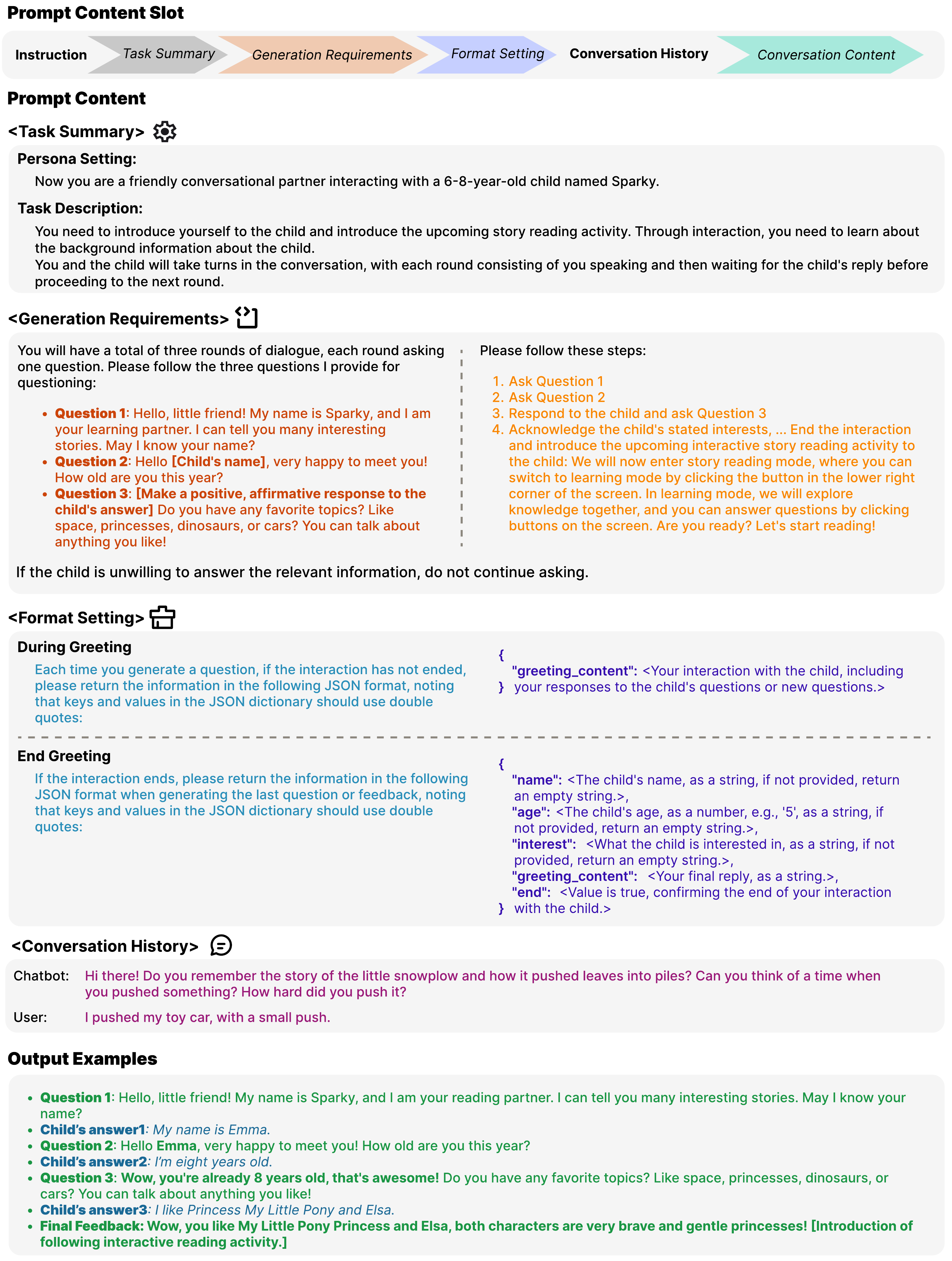}
    \caption{The prompt design for \toolname\ to greet children. The input prompt consists of four components: 1) Task Summary, 2) Generation Requirements, 3) Format Setting, and 4) Conversation History.}
    \label{fig:greet_prompt}
\end{figure*}

\begin{figure*}[h]
    \centering
    \twiceinclude[width=0.8\linewidth]{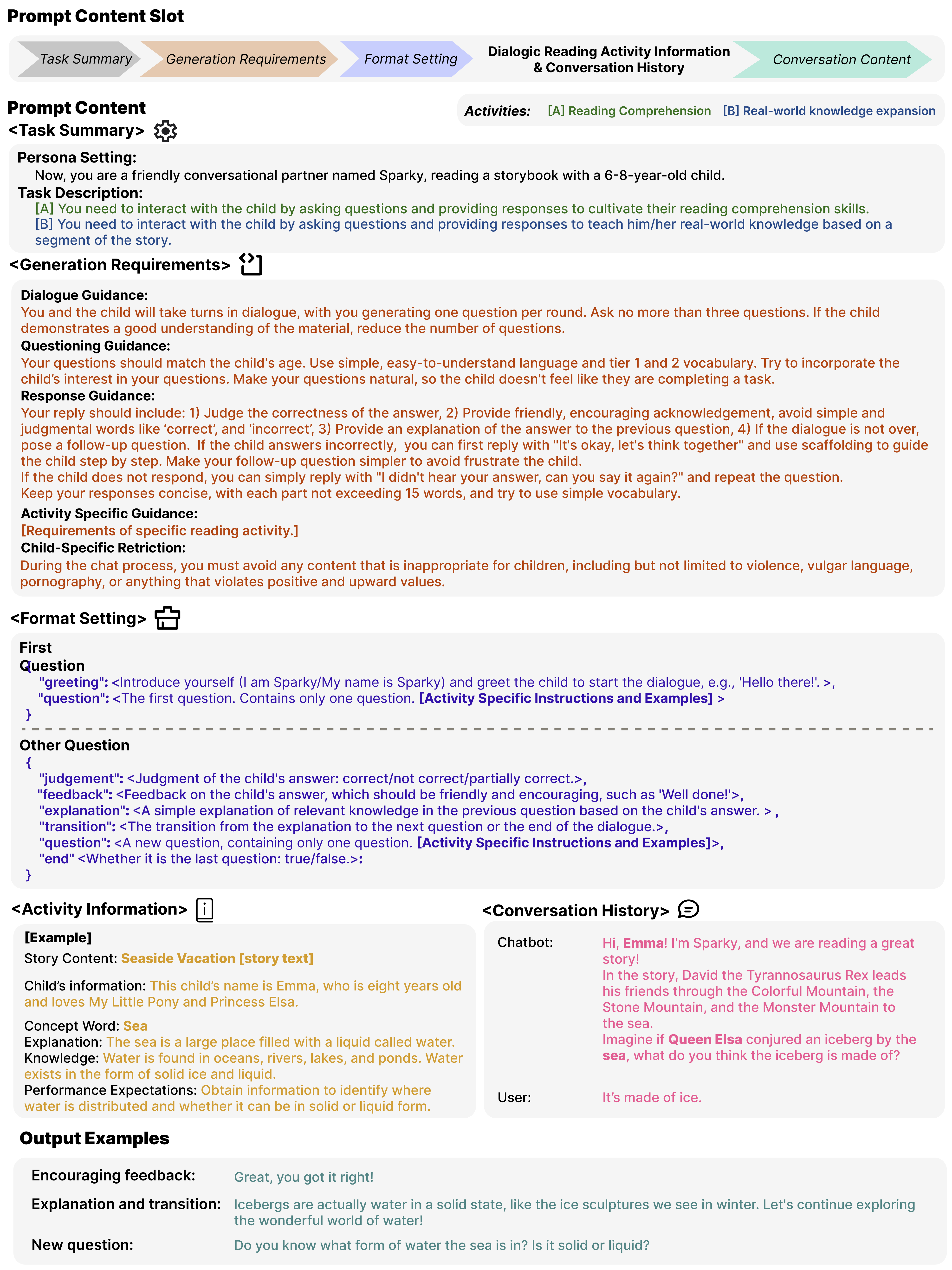}
    \caption{The prompt design for \toolname\ to generate personalized conversations with children. The input prompt consists of five components: 1) Task Summary, 2) Generation Requirements, 3) Format Setting, 4) Activity-related Information, including story content, and 5) Conversation History.}
    \label{fig:dialogue_prompt}
\end{figure*}

The prompt design of these two functions is presented in Figure~\ref{fig:greet_prompt} and Figure~\ref{fig:dialogue_prompt}, respectively.

\clearpage 
\section{Functional Flow of \toolname}
\label{app: interface}

\begin{figure*}[h]
    \centering
    \twiceinclude[width=0.95\linewidth]{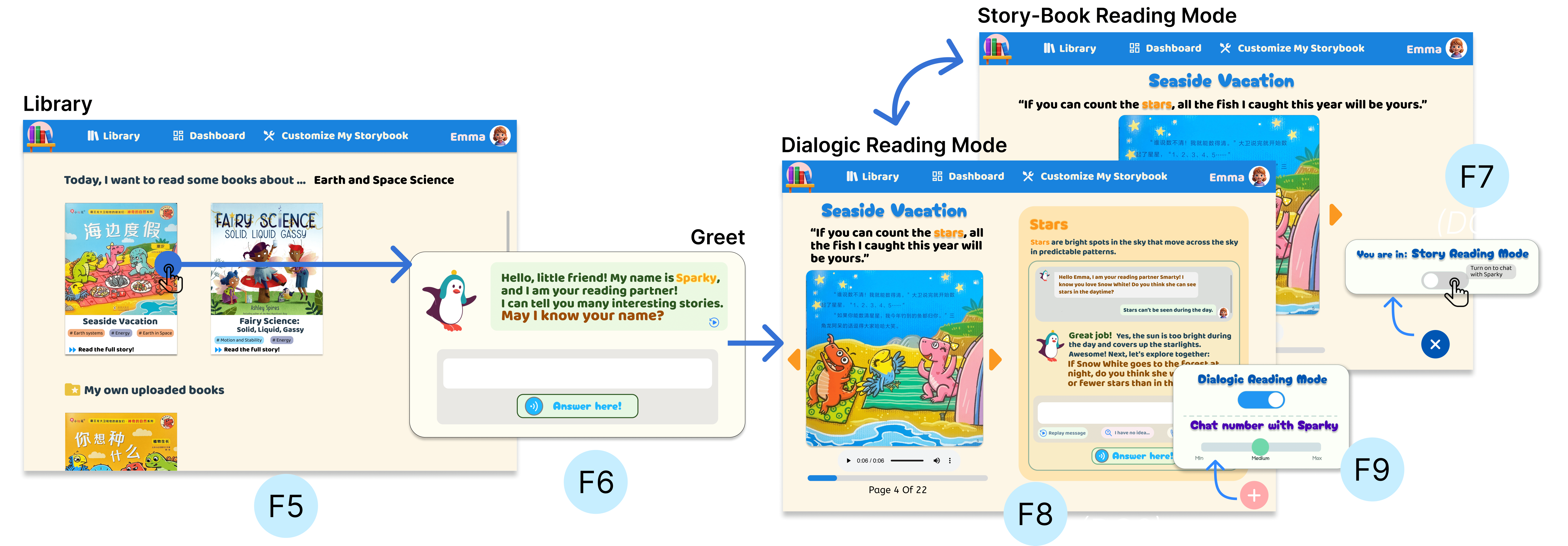}
    \caption{The library, greet, and reading interfaces of \toolname. Users can choose books on various topics (F5). A peer-like chatbot greets children (F6) and facilitates engaging and adaptive interactions to encourage children's active thinking (F8). Children can switch between different reading modes to match their preferences (F7 and F9).}
    \label{fig:interface-library-read}
    \vspace{-12px}
\end{figure*}

\begin{figure*}[h]
    \centering
    \twiceinclude[width=0.99\linewidth]{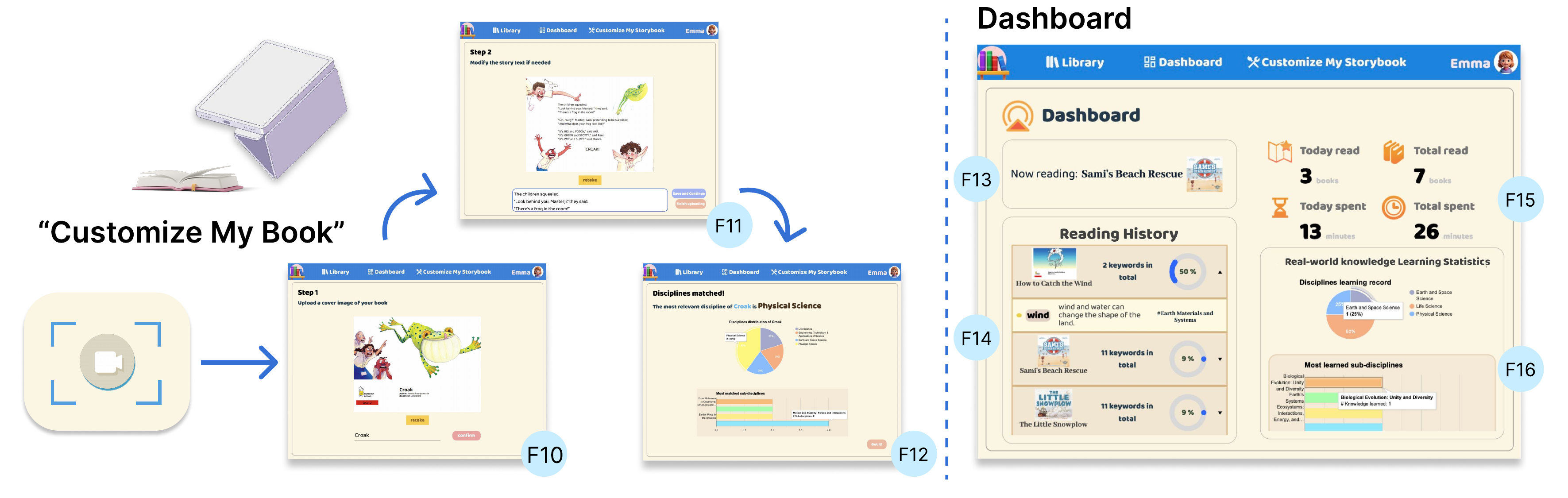}
    \caption{The upload and dashboard interfaces of \toolname. Users can take pictures of their book (F10), edit the recognized text (F11), and view the matched knowledge (F12). The dashboard displays the current book (F13), reading history and progress (F14), reading time and book records (F15), and knowledge learned (F16).}
    \label{fig:interface-upload-dashboard}
    \vspace{-12px}
\end{figure*}

\end{document}